\documentclass{iopart}
\usepackage{graphicx}
\usepackage{bm}

\usepackage{amsbsy}
\usepackage{amssymb}
\usepackage{iopams}  
\usepackage{lineno}

\def\dd{\mbox{d}}
\def\ve{\varepsilon}

\def\D{{\bf D}}
\def\E{{\bf E}}

\begin{document}

\title[NESM: A paradigm and applications]{Non-equilibrium
  statistical mechanics: From a paradigmatic model to biological transport} 
\author{T. Chou$^{1}$, K. Mallick$^{2}$,
  and R. K. P. Zia$^{3}$} 
\address{$^{1}$Depts. of Biomathematics and Mathematics, UCLA,  Los Angeles, CA 90095-1766, USA\\
$^{2}$Institut de Physique Th\'eorique, C. E. A.  Saclay, 91191 Gif-sur-Yvette Cedex, France\\
$^{3}$Department of Physics,  Virginia Tech, Blacksburg, VA 24061, USA}

\date{today}

\begin{abstract}
Unlike equilibrium statistical mechanics, with its well-established
foundations, a similar widely-accepted framework for non-equilibrium
statistical mechanics (NESM) remains elusive. Here, we review some of
the many recent activities on NESM, focusing on some of the
fundamental issues and general aspects. Using the language of
stochastic Markov processes, we emphasize general properties of the
evolution of configurational probabilities, as described by master
equations. Of particular interest are systems in which the dynamics
violate detailed balance, since such systems serve to model a wide
variety of phenomena in nature. We next review two distinct approaches
for investigating such problems. One approach focuses on models
sufficiently simple to allow us to find exact, analytic, non-trivial
results. We provide detailed mathematical analyses of a
one-dimensional continuous-time lattice gas, the totally asymmetric
exclusion process (TASEP). It is regarded as a paradigmatic model for
NESM, much like the role the Ising model played for equilibrium
statistical mechanics.  It is also the starting point for the second
approach, which attempts to include more realistic ingredients in
order to be more applicable to systems in nature. Restricting
ourselves to the area of biophysics and cellular biology, we review a
number of models that are relevant for transport phenomena.  Successes
and limitations of these simple models are also highlighted.

\end{abstract}


\runninglinenumbers

\section{Introduction}

What can we expect of a system which consists of a large number of
simple constituents and evolves according to relatively simple rules?
To answer this question and bridge the micro-macro connection is the
central goal of statistical mechanics. About a century ago, Boltzmann
made considerable progress by proposing a bold hypothesis: When an
\emph{isolated} system eventually settles into a state of equilibrium,
all its microstates are equally likely to occur over long periods of
time. Known as the microcanonical ensemble, it provides the basis for
computing averages of macroscopic observables: (a) by assuming (time
independent) ensemble averages can replace time averages in such an
equilibrium state, (b) by labeling each microstate ${\cal C}$, (a
configuration of the constituents which can be reached via the rules
of evolution) as a member of this ensemble, and (c) by assigning the
same weight to every member ($P^{*}\left({\cal C}\right) \propto
1$). This simple postulate forms the foundation for equilibrium
statistical mechanics (EQSM), leads to other ensembles for systems in
thermal equilibrium, and frames the treatment of thermodynamics in
essentially all textbooks. The problem of answering the question posed
above shifts, for systems in equilibrium, to computing averages with
Boltzmann weights.

By contrast, there is no similar stepping stone for non-equilibrium
statistical systems (NESM), especially ones far from equilibrium. Of
course, given a set of rules of stochastic evolution, it is possible
to write down equations which govern the time dependent weights,
$P\left( {\cal C},t\right)$. But that is just the starting point of
NESM, as little is known, in general, about the solutions of such
equations. Even if we have a solution, there is an added complication:
the obvious inequivalence of time- and ensemble-averages \`{a} la
Boltzmann. Instead, since our interest is the full dynamic behavior of
such a statistical system, we must imagine (a) performing the same
experiment many times, (b) collecting the data to form an ensemble of
trajectories through configuration space, and (c) computing \emph{time
  dependent} averages of macroscopic observables from this
ensemble. The results can then be compared to averages obtained from
$P\left({\cal C},t\right)$. Despite these daunting tasks, there are
many studies \cite{Young98,DG10,HenkelPleimling10} with the goal of
understanding such far-from-equilibrium phenomena.

Here, we focus on another aspect of NESM, namely, systems which evolve
according rules that violate detailed balance. In general, much less
is known about their behavior, though they are used to model a much
wider range of natural phenomena. Examples include the topic in this
review -- transport in biological systems, as well as epidemic
spreading, pedestrian/vehicular traffic, stock markets, and social
networks. A major difficulty with such systems is that, even if such a
system is known (or assumed) to settle eventually in a
time-independent state, the appropriate stationary weights are not
generally known. In other words, there is no overarching principle, in
the spirit of Boltzmann's fundamental hypothesis, which provides the
weights for such \emph{non-equilibrium} steady states (NESS). We
should emphasize that, if the dynamics is Markovian, then these
weights can be constructed formally from the rules of evolution
\cite{Hill66}. However, this formal solution is typically far too
intractable to be of practical use. As a result, such NESS
distributions are explicitly known only for a handful of model
systems. Indeed, developing a fundamental and comprehensive
understanding of physics far from equilibrium is recognized to be one
of the `grand challenges' of our time, by both the US National Academy
of Sciences \cite{CMMP2010} and the US Department of Energy
\cite{DOE07}. Furthermore, these studies point out the importance of
non-equilibrium systems and their impact far beyond physics, including
areas such as computer science, biology, public health, civil
infrastructure, sociology, and finance.

One of the aims of this review is to provide a framework in which
issues of NESM are well-posed, so that readers can appreciate why NESM
is so challenging. Another goal is to show that initial steps in this
long journey have been taken in the form of a few mathematically
tractable models. A good example is the totally asymmetric simple
exclusion process (TASEP). Like the Ising model, TASEP also consists
of binary constituents, but evolves with even simpler rules. Unlike
the scorn Ising's model faced in the 1920's, the TASEP already enjoys
the status of a paradigmatic model. Fortunately, it is now recognized
that seemingly simplistic models can play key roles in the
understanding fundamental statistical mechanics and in formulating
applied models of real physical systems. In this spirit, our final aim
is to provide potential applications of the TASEP, and its many
relatives, to a small class of problems in biology, namely, transport
at molecular and cellular levels.

This article is organized along the lines of these three goals. The
phrase `non-equilibrium statistical mechanics' has been used in many
contexts, referring to very different issues, in a wide range of
settings. The first part of the next section will help readers discern
the many facets of NESM. A more specific objective of section
\ref{General} is to review a proposal for characterizing all
stationary states by a \emph{pair} of time independent distributions,
$\left\{ P^{*}\left( \mathcal{C}\right),K^{*}\left(
\mathcal{C}\rightarrow \mathcal{C}^{\prime }\right) \right\} $, where
$K^{*}$ is the probability current `flowing' from $\mathcal{C}$ to
$\mathcal{C}^{\prime }$ \cite{ZiaSchmittmann07}. In this scheme,
ordinary equilibrium stationary states (EQSS) appear as the restricted
set $\left\{P^{*},0\right\} $, whereas states with $K^{*}\neq 0$ are
identified as NESS. Making an analogy with electromagnetism, this
distinction is comparable to that of electrostatics \emph{vs.}
magnetostatics, as the hallmark of the latter is the presence of
steady and persistent currents. Others (e.g., \cite{JQQ04}) have also
called attention to the importance of such current loops for NESS and
the key role they play in the understanding of fluctuations and
dissipation. Two examples of NESS phenomena, which appear contrary to
the conventional wisdom developed from EQSM, will be provided here.

Section \ref{KM} will be devoted to some details on how to `solve' the
TASEP, for readers who are interested in getting involved in this type
of study. In particular, we will present two complementary techniques,
with one of them exploiting the relationship between two-dimensional
systems in equilibrium and one-dimensional systems in NESM.  While
TASEP was introduced in 1970 \cite{Spitzer70} for studying interacting
Markov processes, it gained wide attention two decades later in the
statistical physics community \cite{Spohn91,Liggett99,Schutz00}. In a
twist of history, two years before its formal introduction, Gibbs and
his collaborators introduced \cite{MGP68,MG69} a more complex version
of TASEP to model mRNA translation in protein synthesis. As the need
for modeling molecular transport in a biological setting provided the
first incentives for considering such NESM systems, it is fitting that
we devote the next part, section \ref{TC}, to potential applications
for biological transport. In contrast to the late 60's, much more
about molecular biology is known today, so that there is a large
number of topics, even within this restricted class of biological
systems. Though each of which deserves a full review, we will limit
ourselves to a few paragraphs for each topic. The reader should regard
our effort here as a bird's eye view `tour guide', pointing to more
detailed, in-depth coverages of specific avenues within this rich
field. Finally, we should mention that TASEP naturally lends itself to
applications in many other areas, e.g., traffic flow \cite{CSS00} and
surface growth \cite{KPZ86,WT90}, etc. All are very interesting, but
clearly beyond the scope of this review, as each deserves a review of
its own. In the last section, \ref{sum}, we conclude with a brief
summary and outlook.


\section{General aspects of non-equilibrium statistical mechanics}
\label{General}

In any quantitative description of a system in nature, the first step
is to specify the degrees of freedom to focus our attention, while
ignoring all others. For example, in Galileo's study of the motion of
balls dropped from a tower, the degrees of freedom associated with the
planets is ignored.  Similarly, the motion of the atoms within the
balls plays no role. The importance of this simple observation is to
recognize that all investigations are necessarily limited in scope and
all quantitative predictions are approximations to some degree. Only
by narrowing our focus to a limited window of length- and/or
time-scales can we make reasonable progress towards quantitative
understanding of natural phenomena.  Thus, we must start by specifying
a set of configurations, $\left\{{\cal C}\right\}$, which accounts for
all the relevant degrees of freedom of the system. For example, for
the traditional kinetic theory of gases (in $d$ spatial dimensions),
${\cal C}$ is a point in a $2dN$ dimensional phase space: $\left\{
\vec{x}_i,\vec{p}_i\right\} ,\,\,i=1,...,N$.  For an Ising model with
spins $s=\pm 1$, the set $\left\{ {\cal C}\right\} $ is the $2^N$
vertices of an $N$ dimensional cube: $\left\{s_i\right\}
,\,\,i=1,...,N$. In the first example, which should be suitable for
describing Argon at standard temperature and pressure, the window of
length scales certainly excludes Angstr\"{o}ms or less, since the
electronic and hadronic degrees of freedom within an atom are
ignored. Similarly, the window in the second example also excludes
many details of solid state physics. Yet, the Ising model is
remarkably successful at predicting the magnetic properties of several
physical systems \cite{WangLu85,LCLELVS89,Wolf00}.

Now, as we shift our focus from microscopic to macroscopic lengths,
both $\left\{{\cal C}\right\}$ and the description also
change. Keeping detailed accounts of such changes is the key idea
behind renormalization, the application of which led to the extremely
successful prediction of non-analytic thermodynamic properties near
second order phase transitions \cite{Wilson83}. While certain aspects
of these different levels of description change, other aspects --
e.g., fundamental symmetries of the system -- remain. One particular
aspect of interest here is time reversal. Although physical laws at
the atomic level respect this symmetry\footnote{Strictly speaking, if
  we accept CPT as an exact symmetry, then time reversal is violated
  at the subatomic level, since CP violation has been observed. So
  far, there is no direct observation of T violation. For a recent
  overview, see, e.g., reference \cite{Quinn09}.}, `effective
Hamiltonians' and phenomenological descriptions at more macroscopic
levels often do not. One hallmark of EQSM is that the dynamics,
effective for whatever level of interest, retain this symmetry. Here,
the concept and term `detailed balance' is often used as well as `time
reversal.' By contrast, NESM provides a natural setting for us to
appreciate the significance of this micro-macro connection and the
appearance of time-irreversible dynamics. We may start with a system
with many degrees of freedom evolving with dynamics obeying detailed
balance. Yet, when we focus on a \emph{subsystem} with far fewer
degrees of freedom, it is often reasonable to consider a dynamics that
violates detailed balance.  Examples of irreversible dynamics include
simple friction in solid mechanics, resistance in electrical systems,
and viscosity in fluid flows.


Before presenting the framework we will use for discussing fundamental
issues of NESM, let us briefly alert readers to the many settings
where this term is used. Starting a statistical system in some initial
configuration, ${\cal C}_0$, and letting it evolve according to rules
which respect detailed balance, it will eventually wind up in an EQSS
(precise definitions and conditions to be given at the beginning of
section \ref{General} below). To be explicit, let us denote the
probability to find the system in configuration ${\cal C}$ at time $t$
by $P\left( {\cal C},t\right)$ and start with $P\left( {\cal
  C},0\right) =\delta \left( {\cal C}-{\cal C}_0\right) $. Then,
$P({\cal C},t\rightarrow \infty)$ will approach a stationary
distribution, $P^{*}({\cal C})$, which is recognized as a Boltzmann
distribution in equilibrium physics. Before this `eventuality', many
scenarios are possible and all of them rightly deserve the term
NESM. There are three important examples from the literature. Physical
systems in which certain variables change so slowly that reaching
$P^{*}$ may take many times the age of the universe. For time scales
relevant to us, these systems are always `far from equilibrium'. To
study the statistics associated with fast variables, these slow ones
might as well be considered frozen, leading to the concept of
`quenched disorder'.  The techniques used to attack this class of
problems are considerably more sophisticated than computing Boltzmann
weights \cite{Young98,DG10}, and are often termed NESM. At the other
extreme, there is much interest in behavior of systems near
equilibrium, for which perturbation theory around the EQSS is quite
adequate. Linear response is the first step in such approaches
\cite{Onsager31,Green54,Kubo57,dGM62}, with a large body of well
established results and many textbooks devoted to them. Between these
extremes are system which evolve very slowly, yet
tractably. Frequently, these studies come under the umbrella of NESM
and are found with the term `aging' in their titles
\cite{HenkelPleimling10}.

Another frequently encountered situation is the presence of
time-dependent rates. Such a problem corresponds to many experimental
realizations in which control parameters, e.g., pressure or
temperature, are varied according to some time-dependent protocol. In
the context of theoretical investigations, such changes play central
roles in the study of work theorems \cite{Jarzynski,CrooksJSP,Crooks,
  HatanoSasa,Hummer,Seifert,seifert:2006,SeifertSpeck2010,ChetriteGawedzki2}.
In general these problems are much less tractable and will not be
considered here.

By contrast, we will focus on systems which evolve according to
dynamics that \emph{violate} detailed balance. The simplest context
for such a system is the coupling to two or more reservoirs (of the
same resource, e.g., energy) in such a way that, when the system
reaches a stationary state, there is a steady flux \emph{through}
it. A daily example is stove-top cooking, in which water in a pot
gains energy from the burner and loses it to the room. At a steady
simmer, the input balances the heat loss and our system reaches a
non-equilibrium steady state (NESS). That these states differ
significantly from ordinary EQSS's has been demonstrated in a variety
of studies of simple model systems coupled to thermal reservoirs at
two different temperatures. Another example, at the global scale, is
life on earth, the existence of which depends on a steady influx of
energy from the sun and re-radiation to outer space. Indeed, all
living organisms survive (in relatively steady states) by balancing
input with output -- of energy and matter of some form. Labeling these
reservoirs as `the medium' in which our system finds itself, we see
the following scenario emerging. Though the medium+system combination
is clearly evolving in time, the system may be small enough that it
has arrived at a time-independent NESS. While the much larger,
combined system may well be evolving according to a time-reversal
symmetric dynamics, it is quite reasonable to assume that this
symmetry is violated by the effective dynamics appropriate for our
smaller system with its shorter associated time scales. In other
words, when we sum over the degrees of freedom associated with the
medium, the dynamics describing the remaining configurations ${\cal
  C}$'s of our system should in general violate detailed balance.

In general, it is impossible to derive such effective dynamics for
systems at the mesoscopic or macroscopic scales from well-known
interactions at the microscopic, atomic level. There are proposals to
derive them from variational principles, based on postulating some
quantity to be extremized during the evolution, in the spirit of least
action in classical mechanics.  The most widely known is probably
`maximum entropy production'. The major challenge is to identify the
constraints appropriate for each NESM system at hand. None of these
approaches has achieved the same level of acceptance as the maximum
entropy principle in EQSM (where the constraints are well established,
e.g., total energy, volume, particle number, etc.). In particular, the
NESS in the uniformly driven lattice gas is known to differ from the
state predicted by this principle \cite{KLS84}. Readers interested in
these approaches may study a variety of books and reviews which
appeared over the years
\cite{Prigogine67,Gyarmati70,Jaynes80,Gallavotti04,Attard06,MS06,OttenStock10,Monthus11}.
How an effective dynamics (i.e., a set of rules of evolution) arise is
not the purpose of our review. Instead, our goal is to explore the
nature of stationary states, starting from a \emph{given} dynamics
that violates detailed balance.  Specifically, in modeling biological
transport, the main theme of the applications section, it is
reasonable to postulate a set of transition rates for the system of
interest.

\subsection{Master equation and other approaches to statistical mechanics}

Since probabilities are central to statistical mechanics, our starting
point for discussing NESM is $P\left( {\cal C},t\right) $ and its
evolution.  Since much of our review will be devoted to models
well-suited for computer simulations, let us restrict ourselves to
discrete steps ($\tau =0,1,...$ of time $\delta t$) as well as a
discrete, finite set of ${\cal C}$'s (${\cal C}_1$, ${\cal C}_2$, ..,
${\cal C}_{{\cal N}}$). Also, since we will be concerned with time
reversal, we will assume, for simplicity, that all variables are even
under this operation (i.e., no momenta-like variables which change
sign under time reversal).  To simplify notation further, let us write
$P_i\left( \tau \right)$, interchangeably with $P\left({\cal C}_i,\tau
\delta t\right)$. Being conserved ($\sum_{i} P_{i}\left(\tau
\right)=1$ for all $\tau$), $P$ must obey a continuity equation, i.e.,
the vanishing of the time rate of change of a conserved density plus
the divergence of the associated current density. Clearly, the
associated currents here are \emph{probability currents}. Since we
restrict our attention to a discrete configuration space, each of
these currents can be written as the flow from ${\cal C}_j$ to ${\cal
  C}_i$, i.e., $K_i^j\left( \tau \right)$ or $K\left( {\cal
  C}_j\rightarrow {\cal C}_i,\tau \delta t\right)$. As a net current,
$K_j^i\left(\tau \right) $ is by definition $-K_i^j\left( \tau
\right)$, while its `divergence' associated with any ${\cal C}_i$ is
just $\sum_{j} K_{i}^{j}\left(\tau \right)$. In general, $K$ is a new
variable and how it evolves must be specified. For example, in quantum
mechanics, $P$ is encoded in the amplitude of the wavefunction $\psi$
only, while $K$ contains information of the phase in $\psi$ as well
(e.g., $K\propto \psi ^{*}\stackrel{\leftrightarrow }{\nabla }\psi$
for a single non-relativistic particle). In this review, as well as
the models used in essentially all simulation studies, we follow a
much simpler route: the master equation or the Markov chain. Here, $K$
is assumed to be proportional to $P$, so that $P_i\left(\tau
+1\right)$ depends only on linear combinations of the probabilities of
the previous step, $P_j\left( \tau \right) $. In a further
simplification, we focus on time-homogeneous Markov chains, in which
the matrix relating $K$ to $P$ is constant in time. Thus, we write
$P_i\left( \tau +1\right) =\sum_jw_i^jP_i\left(\tau \right)$, where
$w_i^j$ are known as the transition rates (from ${\cal C}_j$ to ${\cal
  C}_i$).

As emphasized above, we will assume that these rates are \emph{given}
quantities, as in a mathematical model system like TASEP or in
phenomenologically-motivated models for biological
systems. Probability conservation imposes the constraint
$\sum_iw_i^j=1$ for all $j$, of course. A convenient way to
incorporate this constraint is to write the master equation in terms
of the changes
\begin{eqnarray}
\Delta P_i\left( \tau \right) &\equiv &P_i\left(\tau +1\right)-P_i\left(
\tau \right) \nonumber \\
&=&\sum_{j\neq i}\left[ w_i^jP_j(\tau )-w_j^iP_i(\tau)\right]  
\label{me}
\end{eqnarray}
This equation can be written as 
\begin{equation}
\Delta P_i\left( \tau \right) =\sum_jL_i^jP_j(\tau)  \label{me-L}
\end{equation}
where

\begin{equation}
L_i^j=\left\{ 
\begin{array}{c}
w_i^j \\ 
-\sum_{k\neq j}w_k^j
\end{array}
\right. \ \mathrm{{if}\quad 
\begin{array}{c}
i\neq j \\ 
i=j
\end{array}
}  \label{def-L}
\end{equation}
is a matrix (denoted by $\Bbb{L}$; sometimes referred to as the
Liouvillian) that plays much the same role as the Hamiltonian in
quantum mechanics. Since all transition rates are non-negative,
$w_i^j$ is a stochastic matrix and many properties of the evolution of
our system follow from the Perron-Frobenius theorem \cite{PFT}. In
particular, $\sum_iL_i^j=0$ for all $j$ (probability conservation), so
that at least one of the eigenvalues must vanish. Indeed, we recognize
the stationary distribution, $P^{*}$, as the associated right
eigenvector, since $\Bbb{L} P^{*}=0$ implies $P_i^{*}\left(\tau
+1\right)=P_i^{*}\left(\tau \right)$.  Also, this $P^{*}$ is unique,
provided the dynamics is ergodic, i.e., every ${\cal C}_i$ can be
reached from any ${\cal C}_j$ via the $w$'s. Further, the real parts
of all other eigenvalues must be negative, so that the system must
decay into $P^{*}$ eventually.

Since the right-hand side of equation (\ref{me}) is already cast in
the form of the divergence of a current, we identify

\begin{equation}
K_i^j\left( \tau \right) \equiv w_i^jP_j(\tau )-w_j^iP_i(\tau)
\label{P-current}
\end{equation}
as the (net) probability current from ${\cal C}_j$ to ${\cal C}_i$.
Note that the antisymmetry $K_i^j=-K_j^i$ is manifest here. When a
system settles into a stationary state, these time-independent
currents are given simply by

\begin{equation}
K_i^{*j}\equiv w_i^jP_j^{*}-w_j^iP_i^{*}  \label{K*}
\end{equation}
As we will present in the next subsection, a reasonable way to
distinguish EQSS from NESS is whether the $K^{*}$ vanish or
not. Before embarking on that topic, let us briefly remark on two
other common approaches to time dependent statistical mechanics. More
detailed presentations of these and related topics are beyond the
scope of this review, but can be found in many books
\cite{vanKampen07,Reichl09,Risken89,SCN11}.

Arguably the most intuitive approach to a stochastic process is the
Langevin equation. Originating with the explanation of Brownian motion
\cite{RBrown1828} by Einstein and Smoluchowski
\cite{Einstein05,Smoluchowski06}, this equation consists of adding a
random drive to an otherwise deterministic evolution. The
deterministic evolution describes a single trajectory through
configuration space: ${\cal C}\left(\tau \right)$ (starting with
${\cal C}\left( 0\right) ={\cal C}_0$), governed by say, an equation
of the form $\Delta {\cal C}\left(\tau \right) ={\cal F}\left[ {\cal
    C}\left( \tau \right) \right]$. In a Langevin approach, ${\cal F}$
will contain both a deterministic part and a noisy component. Of
course, a trajectory (or history, or realization) will depend on the
specific noise force appearing in that run. Many trajectories are
therefore generated, each depending on a particular realization of the
noise and the associated probability. Although each trajectory can be
easily understood, the fact that many of them are possible means the
system at time $\tau$ can be found at a collection of ${\cal C}$'s. In
this sense, the evolution is best described by a probability
distribution, $P\left( {\cal C},\tau \right)$, which is controlled by
both the deterministic and the noisy components in ${\cal F}$.
Historically, such considerations were first provided for a classical
point particle, where $\Delta {\cal C} ={\cal F}$ would be Newton's
equation, $\partial_t^2\vec{x}\left(t\right) =\vec{F}/m$, with
continuous time and configuration variables. How the deterministic and
noisy parts of $\vec{F}$ are connected to each other for the Brownian
particle is the celebrated Einstein-Smoluchowski relation.  Of course,
$P\left({\cal C},\tau \right) $ becomes $P\left(\vec{x}, t\right) $ in
this context, while the Langevin approach can be reformulated as a PDE
for $P\left(\vec{x}, t\right)$

\begin{equation}
\partial _tP\left( \vec{x},t\right) =\frac{\partial ^2}{\partial x_\alpha
\partial x_\beta }D_{\alpha \beta }\left( \vec{x}\right) P\left( \vec{x}
,t\right) -\frac \partial {\partial x_\alpha }V_\alpha \left( \vec{x}\right)
P\left( \vec{x},t\right)  \label{FP}
\end{equation}
Here, $D_{\alpha \beta }$ and $V_\alpha $ are the diffusion tensor and
the drift vector, respectively, and are related to the noisy and
deterministic components of the drive\footnote{Note that
  $D_{\alpha\beta}(\vec{x})$ here derives from the rate of change of
  the variance of the distribution and is different from the diffusion
  coefficient used to define Fick's law. Here, both spatial
  derivatives operate on $D_{\alpha\beta}(\vec{x})$.}. This PDE is
referred to as the Fokker-Planck equation, although it was first
introduced for the velocity distribution of a particle
\cite{Risken89}.

An experienced reader will notice that the master equation (\ref{me})
for $P\left({\cal C},\tau \right) $ and the Fokker-Planck equation
(\ref{FP}) for $P\left(\vec{x},t\right)$ are both linear in $P$ and
first order in time, but that the right-hand sides are quite
different. Let us comment briefly on their similarities and
differences. Despite the simpler appearance, equation (\ref{me}) is
the more general case, apart from the complications associated with
discrete \emph{vs.} continuous variables.  Thus, let us facilitate the
comparison by considering a discrete version of equation (\ref{FP}),
i.e., $t\rightarrow \tau \delta t$ and $\vec{x}\rightarrow
\vec{\zeta}\delta x$, so that $P\left(\vec{x},t\right) \rightarrow
P(\vec{\zeta},\tau)$. In this light, it is clear that the derivatives
on the right correspond to various linear combinations of
$P(\zeta_\alpha \pm 1,\zeta_\beta \pm 1;\tau)$. In other words, only a
handful of the `nearest configurations' are involved in the evolution
of $P(\vec{\zeta},\tau)$. By contrast, the range of $w_i^j$, as
written in equation (\ref{me}), is not restricted.

Let us illustrate by a specific example. Consider a system with $N$
Ising spins (with any interactions between them) evolving according to
random sequential Glauber spin-flip dynamics \cite{Glauber63}. In a
time step, a random spin is chosen and flipped with some
probability. Now, as noted earlier, the configuration space is the set
of vertices of an $N$ dimensional cube. Therefore, the only
transitions allowed are moves along an edge of the cube, so that the
range of $w_i^j$ is `short'. In this sense, field theoretic
formulations of the Ising system evolving with Glauber dynamics are
possible, taking advantage of Fokker-Planck like equations, cast in
terms of path integrals. On the other hand, consider updating
according to a cluster algorithm, e.g., Swendsen-Wang \cite{SW87}, in
which a large cluster of spins (say, $M$) are flipped in a single
step. Such a move clearly corresponds to crossing the body diagonal of
an $M$ dimensional cube. Since $M$ is conceivably as large as $N$,
there is no limit to the range of this set of $w$'s. It is hardly
surprising that field theoretic approach for such systems are yet to
be formulated, as they would be considerably more complex.

\subsection{Non-equilibrium \emph{vs.} equilibrium stationary states,
persistent probability currents}

Following the footsteps of Boltzmann and Gibbs, we study statistical
mechanics of systems in thermal equilibrium by focusing on time
independent distributions such as $P^{*}\left( {\cal C}\right) \propto
1$ or $\exp \left[ -\beta {\cal H}\left( {\cal C}\right) \right] $
(where ${\cal H}$ is the total energy associated with ${\cal C}$ and
$\beta $ is the inverse temperature). Apart from a few model systems,
it is not possible to compute, analytically and in general, averages
of observable quantities, i.e.,

\begin{equation}
\left\langle {\cal O}\right\rangle \equiv \sum_j{\cal O}\left( 
{\cal C}_j\right) P^{*}\left( {\cal C}_j\right).  \label{average}
\end{equation}
Instead, remarkable progress over the last fifty years was achieved
through computer simulations, in which a small subset of $\left\{{\cal
  C}\right\} $ is generated -- with the appropriate (relative) weights
-- and used for computing the desired averages. This approach is an
advanced art \cite{LandauBinder09}, far beyond the scope (or purpose)
of this review. Here, only some key points will be mentioned and
exploited -- for highlighting the contrast between the stationary
distributions of Boltzmann-Gibbs and those in NESS.

In a classic paper \cite{MRRTT53}, Metropolis, \emph{et.al.}
introduced an algorithm to generate a set of configurations with
relative Boltzmann weights. This process also simulates a dynamical
evolution of the system, in precisely the sense of the master
equation. Starting from some initial ${\cal C}\left( 0\right) ={\cal
  C}_0$, a new one, ${\cal C}_k$, is generated (by some well defined
procedure) and accepted with probability $w_k^0$. Thus, ${\cal
  C}\left( 1\right) $ is ${\cal C}_k$ or ${\cal C}_0$. with relative
probability $w_k^0/\left( 1-w_k^0\right) $. After some transient
period, the system is expected to settle into a stationary state,
i.e., the frequencies of ${\cal C}_i$ occurring in the run are
proportional to a time independent $P^{*}\left( {\cal C}_i\right)
$. To implement this Monte-Carlo method, a set of transition rates,
$w_j^i$, must be fixed. Further, if the desired outcome is
$P^{*}\propto e^{-\beta {\cal H}}$, then $w_j^i$ cannot be
arbitrary. A sufficient (but not necessary) condition is referred to,
especially in the simulations community \cite{LandauBinder09}, as
`detailed balance':

\begin{equation}
w_k^iP^{*}\left( {\cal C}_i\right) =w_i^kP^{*}\left( {\cal C}_k\right).  
\label{db}
\end{equation}
In other words, it suffices to constrain the ratio $w_k^i/w_i^k$ to be
$\exp \left[ -\beta \Delta {\cal H}\right]$, where $\Delta {\cal
  \ H}\equiv {\cal H}\left( {\cal C}_k\right)- {\cal H}\left( {\cal
  C}_i\right)$ is just the difference between the configurational
energies. A common and simple choice is $w_k^i=\min \left\{
1,e^{-\beta \Delta {\cal H}}\right\}$.

Of course, constraining the ratios still leaves us with many
possibilities.  To narrow the choices, it is reasonable to regard a
particular set of $w$'s as the simulation of a physical
dynamics\footnote{In this light, the sequence of configurations
  generated $\left({\cal C}_{j_1}, {\cal C}_{j_2}, \ldots, {\cal
    C}_{j_\tau}\right)$ can be regarded as a history, or trajectory,
  of the system. By collecting many ($M$) such sequences, $\left\{
  {\cal \ C}_j^\alpha \right\} ,\alpha =1,...,M$, time dependent
  averages $\left\langle {\cal O}\right\rangle_\tau \equiv \sum_j
  {\cal \ O}\left( {\cal C}_j\right) P_j\left(\tau \right) $ are
  simulated by $M^{-1}\sum_\alpha {\cal O}\left( {\cal C}_{j_\tau
  }^\alpha \right) $.}. In that case, other considerations will guide
our choices. For example, the Lenz-Ising system is used to model spins
in ferromagnetism \cite{Ising25} as well as occupations in binary
alloys \cite{YangLee52,Huang87}. In the former, individual spins can
be flipped and it is quite appropriate to exploit Glauber
\cite{Glauber63} dynamics, in which the $w$'s connect ${\cal C}$'s
that differ by only one spin. In the latter however, a Zn atom, say,
cannot be changed into a Cu atom, so that exchanging a neighboring
pair of different `spins' -- Kawasaki \cite{Kawasaki66,Kawasaki70}
dynamics -- is more appropriate. In terms of the $N$ dimensional cube
representation of $\left\{ {\cal C}\right\} $, these $w$'s connect two
vertices along an edge (Glabuer) or across the diagonal of a square
face or plaquette (Kawasaki). Both dynamics involve $w$'s that only
connect ${\cal C}$'s with one or two different spins. The idea is
that, in a short $\delta t$, exchanging energy with the thermal
reservoir can randomly affect only one or two spins.  Also, in this
sense, we can regard the $w$'s as how the system is coupled to the
surrounding medium. Clearly, $\Delta {\cal H}$ measures the energy
exchanged between the two. Another important quantity is entropy
production, whether associated with the system or the medium, in which
the $w$'s will play a crucial role.

It is significant that regardless of the details of the associated
dynamics, a set of $w$'s that obey detailed balance (\ref{db})
necessarily leads the system to a state in which \emph{all} stationary
currents vanish. This follows trivially from the definition
(\ref{K*}). By contrast, transition rates that violate detailed
balance necessarily lead to some \emph{non-vanishing} stationary
currents. To appreciate this statement, let us provide a better
criterion, due to Kolmogorov \cite{Kolmogorov36}, for rates that
respect/violate detailed balance. In particular, while equation
(\ref{db}) gives the wrong impression that detailed balance is defined
with respect to a given ${\cal H}$, the Kolmogorov criterion for
detailed balance is applicable to all Markov processes, whether an
underlying Hamiltonian exists or not.

Consider a closed loop in configuration space, e.g., ${\cal L}$
$\equiv {\cal C}_i\rightarrow {\cal C}_j\rightarrow {\cal
  C}_k\rightarrow ...\rightarrow {\cal C}_n\rightarrow {\cal
  C}_i$. Define the product of the associated rates in the `forward'
direction by $\Pi \left[{\cal L}\right] \equiv w_j^iw_k^j\,...\,w_i^n$
and also for the `reverse' direction: $\Pi \left[{\cal L}_{rev}\right]
\equiv w_i^jw_j^k\,...\,w_n^i$. The set of rates are said to satisfy
detailed balance if and only if

\begin{equation}
\Pi \left[ {\cal L}\right] =\Pi \left[ {\cal L}_{rev}\right] \quad
\label{loops}
\end{equation}
for \emph{all loops.} If this criterion is satisfied, then we can show
that a (single valued) functional in configuration space can be
constructed simply from the set of ratios $w_k^i/w_i^k$, and that it
is proportional to $P^{*}\left( {\cal C}\right)$. If this criterion is
violated for certain loops, these will be referred to here as
`irreversible rate loops' (IRLs).  Despite the lack of detailed
balance, $P^{*}\left({\cal C}\right) $ exists and can still be
constructed from the $w$'s, though much more effort is
required. Established some time ago
\cite{Hill66,Schnakenberg76,Haken83}, this approach to $P^{*}$ is
similar to Kirchhoff's for electric networks
\cite{Kirchhoff1847}. More importantly, this construction provides the
framework for showing that, in the stationary state, some $K^{*}$'s
must be non-trivial \cite{ZiaSchmittmann06,ZiaSchmittmann07}. Since
the divergence of these currents must vanish, they must form current
loops. As time-independent current loops, they remind us of
magnetostatics. The distinction between this scenario and the case
with detailed balance $w$'s is clear: The latter resembles
electrostatics. In this light, it is reasonable to label a stationary
state as an equilibrium one if and only if \emph{all} its
(probability) currents vanish, associated with a set of $w$'s with no
IRLs. Similarly, a non-equilibrium steady state -- NESS -- would be
associated with non-trivial current loops, generated by detailed
balance-violating rates with IRLs
\cite{ZiaSchmittmann06,ZiaSchmittmann07}. Our proposal is that all
stationary states should be characterized by the pair $\left\{
P^{*},K^{*}\right\}$.  In this scheme, `equilibrium states' correspond
to the subset $\left\{ P^{*},0\right\}$, associated with a dynamics
that respect detailed balance and time reversal.

The presence of current loops and IRLs raises a natural and intriguing
question: Is there a intuitively accessible and simple relationship
between them? Unfortunately, the answer remains elusive so
far. Venturing further, it is tempting to speculate on the existence
of a gauge theory, along the lines of that in electromagnetism, for
NESM. If such a theory can be formulated, its consequences may be
far-reaching.

Time-independent probability current loops also carry physical
information about a NESS. Referring readers to a recent article
\cite{ZiaSchmittmann07} for the details, we provide brief summaries
here for a few key points.

\vspace{2mm}

\noindent (i) In particular, it is shown how the $K$'s can be used to
compute currents associated with physical quantities, such as energy
or matter. In addition, we have emphasized that a signature of NESS is
the existence of a steady flux (of, e.g., energy) \emph{through} the
system. All aspects of such through-flux, such as averages and
correlation, can also be computed with the $K$'s.

\vspace{2mm}

\noindent (ii) Following Schnakenberg \cite{Schnakenberg76}, we may
define the \emph{total} entropy production $\mathbf{\Sigma}_{\rm tot}$
as a quantity associated with the rates $\left\{ w_j^i\right\} $. This
$\mathbf{\Sigma}_{\rm tot}$ can be written as the sum of two terms,
$\mathbf{\Sigma}_{\rm sys}+\,\mathbf{\Sigma}_{\rm med}$. The first is
associated with entropy production within our system (recognizable as
the derivative of the Gibbs' entropy, $-\sum_iP_i\ln P_i$, in the
continuous time limit):

\begin{equation}
\mathbf{\Sigma}_{\rm sys}(\tau)\equiv \frac 12\sum_{i,j}K{}_i^j(\tau)\ln 
\frac{P_j(\tau )}{P_i(\tau)}\,.\,  \label{S-sys}
\end{equation}
It is straightforward to show that for a NESS with $K^{*}\neq 0$,
$\mathbf{\Sigma}_{\rm sys}$ vanishes as expected. However, 
a second contribution to entropy production is
associated with the medium:

\begin{equation}
\mathbf{\Sigma}_{\rm med}(\tau)\equiv \frac 12\sum_{i,j}K_i^j(\tau)\ln
\frac{w_i^j}{w_j^i}, \label{S-med}
\end{equation}
where the positivity of $\sum K_{i}^{*j}\ln(w_i^j/w_j^i)$ has been
demonstrated. This result is entirely consistent with
our description of a NESS, namely, a system coupled to surroundings which
continue to evolve and generate entropy.

\vspace{2mm}

\noindent (iii) The following inverse question for NESS is also
interesting.  As we noted, given a Boltzmann distribution, a well
known route to generate it is to use a dynamics which obey detailed
balance (\ref{db}). If we accept that a NESS is characterised not only
by the stationary distribution, but by the pair
$\left\{P^{*},K^{*}\right\} $, then the generalized condition is
$w_k^iP_i^{*}=w_i^kP_k^{*}+K_k^{*i}$, or more explicitly,

\begin{equation}
w_k^iP^{*}\left( {\cal C}_i\right) =w_i^kP^{*}\left({\cal C}_k\right)
+K^{*}\left( {\cal C}_i\rightarrow {\cal C}_k\right).  \label{db+K}
\end{equation}
It is possible to phrase this condition more elegantly (perhaps offering a
little insight) by performing a well-known similarity transform on $w_k^i$:
Define the matrix $\Bbb{U}$, the elements of which are 
\[
U_k^i\equiv \left( P_k^{*}\right) ^{-1/2}w_k^i\left( P_i^{*}\right) ^{1/2}. 
\]
The advantage of this form of `coding' the dynamics is that $\Bbb{U}$ is
symmetric if rates obey detailed balance. Meanwhile, since $K^{*}$ is a
current, we can exploit the analog $J=\rho v$ to define the `velocity
matrix' $\Bbb{V}$, the elements of which are 
\[
V_k^i\equiv \left( P_k^{*}\right) ^{-1/2}K_k^{*i}\left(P_i^{*}\right)
^{-1/2},
\]
associated with the flow from ${\cal C}_i$ to ${\cal C}_k$. Our
generalized condition (\ref{db+K}) now reads simply: The antisymmetric
part of $\Bbb{U}$ is $\Bbb{V}/2$. Similar ideas have also been pursued
recently \cite{ChetriteGupta11}.

To summarize, if a dynamics is to lead a system to a desired $\left\{
P^{*},K^{*}\right\}$, then the associated antisymmetric part of
$\Bbb{R}$ is completely fixed by $K^{*}$. By contrast, its symmetric
part is still unconstrained, corresponding to dynamics that
takes us to the same $\left\{ P^{*},K^{*}\right\} $.

\vspace{2mm}

\noindent (iv) As long as $K^{*}\neq 0$ for a transition, we can focus
on the direction with positive current (say, $K_k^{*i}>0 $) and choose
the maximally asymmetric dynamics, namely, $w_i^k\equiv 0$ and
$w_k^i=K_k^{*i}/P_i^{*}$. (Understandably, such choices are impossible
for systems in thermal equilibrium, except for $T=0$ cases.) Whether
systems with such apparently unique dynamics carry additional
significance remains to be explored. Certainly, TASEP -- the
paradigmatic model of NESS, to be presented next -- belongs in this
class. Before embarking on the next section, let us briefly comment on
some typical features of NESS which are counter-intuitive, based on
our notions of EQSM.

\subsection{Beyond expectations of equilibrium statistical mechanics}

Equilibrium statistical mechanics has allowed us to develop physical
intuition that can be valuable guides when we are faced with new
problems in unfamiliar settings. A good example is energy-entropy
competition, which tends to serve us well when we encounter novel
phase transitions: The former/latter `wins' for systems at low/high
temperatures, so that it displays order/disorder phenomena. Another
example is `positive response': To ensure thermodynamic stability, we
expect the system to respond in a certain manner (positively) when its
control parameters are changed. Thus, it is reasonable to expect,
e.g., positive specific heat and compressibility for systems in
thermal equilibrium. A final example is long-range correlations, which
are generically absent in equilibrium systems with short-range
interactions and dynamics. There are exceptions, of course, such as in
critical phenomena associated with second order phase
transitions. When we encounter systems in NESS however, we should be
aware that such physical intuition often leads us astray. At present,
we are not aware of another set of overarching principles which are
generally applicable for NESS. Instead, in the following, we will
provide two specific circumstances in which our expectations fail.

\vspace{2mm}

\noindent \emph{Negative response.} In order for a system to be in
thermal equilibrium, it must be stable against small
perturbations. Otherwise, fluctuations will drive it into another
state. Such consistent behaviors of a system may be labeled as
`positive response'. Related to the positivity of certain second
derivatives of the free energy, elementary examples include positive
specific heat and compressibility. By contrast, a surprisingly common
hallmark of NESS is `negative response.' For example, imagine a room
in which the internal energy decreases when the thermostat is turned
up! One of the first systems in NESS where this type of surprising
behavior surfaced is the driven Ising lattice gas
\cite{KLS84}. Referring the reader to, for example, reference
\cite{SZ95} for details, the key ingredients are the following. An
ordinary Ising system (with nearest neighbor ferromagnetic
interactions on a square lattice) is subjected to an external drive,
and observed to undergo the phase transition at a temperature
\emph{higher} than that expected from Onsager's solution. Since the
external drive tends to break bonds, its effect is similar to coupling
the system to another energy reservoir with a \emph{higher}
temperature. Nevertheless, this NESS system displays more order than
its equilibrium counterpart. In other words, despite the fact that
more energy appears to be `pumped into' the system, the internal
energy decreases. A more direct manifestation of this form of
`negative response' has been observed in the two-temperature Ising
lattice gas, in which particle hops in the $x$ or $y$ direction are
updated by Metropolis rates appropriate to exchanging energy with a
thermal reservoir set at temperature $T_x$ or $T_y$. Changing $T_x$,
with $T_y$ held fixed, the average internal energy, $U$ (i.e.,
$\left\langle {\cal H} \right\rangle $), is found to vary with
$\partial U/\partial T_y<0$ \cite{PSZ00}! Such surprising negative
response is so generic that it can be found in exceedingly simple,
exactly solvable cases \cite{ZPM02}.

Of course, we should caution the reader that `negative response' may
be simply a misnomer, poor semantics, or careless interpretation of an
observed phenomenon. After all, fluctuations of observables in a
stationary state must be positive and if the appropriate conjugate
variable is used, then the response to that variable will again be
positive. In particular, for any observable ${\cal O}$, we can always
define the cumulant generating function $\Xi \left(\omega \right)
\equiv \ln \left\langle e^{\omega {\cal O}}\right\rangle $ and its
derivative $X\left( \omega \right) \equiv \Xi^{\prime }\left(\omega
\right)$.  Of course, the average $\left\langle {\cal O}\right\rangle
$ is $\Xi ^{\prime }\left( 0\right) $, while $X\left(\omega \right)$
is, in general, the average of ${\cal O}$ in the {\it modified
  distribution} $\tilde{P}^{*}\left( {\cal C} \right) \propto
e^{\omega {\cal O}\left( {\cal C} \right)} P^{*}\left({\cal C}
\right)$.  Then, we are guaranteed `positive response:' $\partial
X/\partial \omega >0$. However, unlike internal energy and temperature
for systems in thermal equilibrium, simple physical interpretations of
these mathematical manipulations for NESS's are yet to be
established. Clearly, this issue is related to the
fluctuation-dissipation theorem in EQSM.
General results valid for NESM have been derived during the last two
decades; we refer the reader to the seminal articles
\cite{EvansCohenMorriss,EvansSearles,Gallavotti,Jarzynski,CrooksJSP}.
The generalization of the fluctuation-dissipation theorem to NESM is a
major topic \cite{BramWynants,SeifertSpeck2010} and lies outside our
scope.
Here, let us remark that the foundations of this theorem
lies in the time reversal properties of the underlying dynamics
\cite{Kurchan,LeboSpohn}, which control the nature of the fluctuations
of the random variables. To characterize these fluctuations
quantitatively, large-deviation functions (LDF) have been
introduced. They play a crucial role in NESM, akin to that of
thermodynamic potentials in EQSM \cite{DerrReview,Touchette}. Valid
for systems far from equilibrium, the fluctuation theorem can be
stated as a symmetry property of the LDF (see the next section for a
explicit example in the case of TASEP). Near an equilibrium state this
theorem implies the fluctuation-dissipation relation, previously
derived from linear response theory \cite{Kubo,Rondoni}. A related set
of significant results is the non-equilibrium work relations
\cite{Jarzynski,Seifert,Crooks,CrooksJSP,HatanoSasa,ChetriteGawedzki2,seifert:2006,SeifertSpeck2010,Hummer},
also a topic worthy of its own review (see, e.g.,
\cite{GallEPJB,JarzEPJB} and references therein). Here, in the context
of the exact solution of TASEP (section \ref{KM}), another result
along this theme -- the macroscopic fluctuation theory developed by
Jona-Lasinio {\it et al.} \cite{Bertini} -- plays an important role.

\noindent \emph{Generic long-range correlations.} For systems with
short-range interactions in thermal equilibrium, the static (i.e.,
equal time) correlations of the fluctuations are generically
short-ranged. This is true even if the dynamics were to contain long
ranged components (e.g., the `artificial' Swendsen-Wang \cite{SW87}
dynamics): As long as it obeys detailed balance, we are assured of
$P^{*}\propto e^{-\beta {\cal H}}$.  Of course, time-dependent
correlations are \emph{not} similarly constrained.  An excellent
example is diffusive dynamics obeying certain conservation laws (e.g.,
hydrodynamics or Kawasaki \cite{Kawasaki66,Kawasaki70} dynamics
modeling Cu-Zn exchange), where long time tails (power law decays) are
well known phenomena: The autocorrelation function, $G\left(
r=0,t\right) $, decays as $t^{-d/2}$ in $d$ dimensions, even for a
single particle. As pointed out by Grinstein \cite{Grinstein91}, a
simple scaling argument, along with $r\thicksim t^{1/z}$, would lead
to \emph{generic} long-range correlations, $G\left(r,t=0\right)
\rightarrow Ar^{-zd/2}$ (i.e., $r^{-d}$ in the case of random walkers
subjected to short-range interactions, where $z=2$). Yet,
$G\left(r,t=0\right)$ generally decays as an exponential in
equilibrium systems! These seemingly contradictory statements can be
reconciled when the amplitude $A$ is examined. In the equilibrium
case, detailed balance (or the fluctuation-dissipation theorem)
constrains $A$ to vanish. For NESS, there is no such constraint,
leaving us with long-range correlations generically. For further
details of these considerations, the reader may consult references
\cite{SZ95,DKS94}.

To end the discussion on such correlations, we should caution the
readers on a subtle point. Although we emphasized how long-range
correlations can emerge from a short-range dynamics that violates
detailed balance, the latter is not \emph{necessary}. An excellent
example is a driven diffusive system with three species -- the ABC
model \cite{EKKM98,CDE03} -- in one dimension. The system evolves with
a short-range dynamics which generally violate detailed balance, and
displays long-range \emph{order} (as well as
correlations). Remarkably, for a special set of parameters, detailed
balance is restored and so, an exact $P^{*}$ was easily found. When
interpreted as a Boltzmann factor (i.e., $P^{*}\propto e^{-\beta {\cal
    H}}$), the Hamiltonian contains inter-particle interactions which
range over the entire lattice! Despite having an ${\cal H}$ with
long-range interactions, it is possible to construct a short-range
dynamics (e.g., $w$'s that involve only nearest neighbor particle
exchanges) that will lead the system to an EQSS: $\left\{ P^{*}\propto
e^{-\beta {\cal H}},K^{*}\equiv 0\right\} $. To appreciate such a
counter-intuitive situation, consider $\Delta {\cal H}={\cal H}\left(
{\cal C}_i\right)-$ ${\cal H}\left( {\cal C}_j\right)$ for a pair of
configurations that differ in some local variables. The presence of
long-range terms in ${\cal H}$ typically induce similar terms in
$\Delta {\cal H}$, leading to a long-range dynamics. If such terms
conspire to cancel, then $\Delta {\cal H}$ becomes short-ranged and it
is simple to choose $w$'s with no long range components. We believe it
is important to investigate whether such examples belong to a class of
mathematical curiosities or form the basis for a wide variety of
physical/biological systems. With these two examples, we hope to have
conveyed an important lesson we learned from our limited explorations
of NESS: Expect the unexpected, when confronted with a novel system
evolving according to detailed balance violating dynamics.

In this section, we attempted to provide a bird's eye view of the many
facets of non-equilibrium statistical mechanics, and then to focus on
a particular aspect: stationary states associated with dynamics that
violate detailed balance.  We emphasized the importance of this class
of problems and pointed out significant features of NESS that run
counter to the physical intuition learned from equilibrium statistical
mechanics. In the next two sections, we will turn our attention to
specific systems. As common in all theoretical studies, there are two
typically diverging goals associated with the models we pursue. One
goal is to account for as many features found in nature as
realistically possible. The other is to consider models with just one
or two features so that they are simple enough to be solved
analytically and exactly. These goals obviously diverge since more
realistic models are typically mathematically intractable while
exactly solvable models generally lack essential aspects of physical
systems (or those in chemistry, biology, finance, sociology,
etc.). Nevertheless, we believe is it worthwhile to devote our
attention to both of them, albeit separately. In this spirit, we will
next present a simple model: the exclusion process, with the TASEP
being an extreme limit. Arguably the simplest possible model with a
nontrivial NESS, the TASEP is not only amenable to exact and
approximate solution strategies, but it also has shed considerable
light on problems in the real world. In section (\ref{TC}), we turn to
a number of generalizations of this model, each taking into account
new physical features required for modeling various aspects of
transport in biological systems.


\section{A paradigmatic model: The asymmetric simple exclusion process (ASEP)}
\label{KM}

Building a simple representation for complex phenomena is a common
procedure in physics, leading to the emergence of paradigmatic models:
the harmonic oscillator, the random walker, the Ising magnet. All
these beautiful models often display wonderful mathematical structures
\cite{Baxter,Sutherland}. In the field of NESM, and for investigating
the role of detailed balance violating dynamics in particular, the
asymmetric simple exclusion process (ASEP) is reaching the status of
such a paradigm. A model with possibly the simplest of rules, it
nevertheless displays a rich variety of NESS's. Further, it is
sufficiently simple to allow us to exploit rigorous mathematical
methods and, over the last two decades, to arrive at a number of exact
results. In this way, valuable insights into the intricacies of NESM
have been garnered. In this section, we delve into some details of two
such methods, in the hope that readers unfamiliar with these
techniques can join in these efforts. Of course, as we consider models
which are more suited for physical/biological systems, we will
encounter more complex ingredients than in ASEPs. As a result, these
models are not exactly solvable at present. In these cases,
approximations are necessary for further progress. One successful
scheme is the mean field approximation, leading to hydrodynamics
equations (PDE's) for density fields. Since this strategy is the only
one to offer some quantitative understanding of the more
realistic/complex models, we will devote the last subsection
\ref{HMFT} here to this approach.

In the previous section, we presented the master equation (\ref{me})
in discrete time, which is clearly the most appropriate description
for computer simulations. On the other hand, for analytic studies, it
is often far easier to use continuous variables (or infinite systems,
in a similar vein). Thus, all the discussions in this section will be
based on continuous time, $t$. As discussed in the context of the
Fokker-Planck equation (\ref{FP}), we introduced this connection:
$t=\tau \delta t$. Here, let us be explicit and write the continuous
version of equation (\ref{me-L}) as

\begin{equation}
\partial_t P_i\left( t\right) =\sum_j M_{i}^{j} P_j(t)  
\label{me-M}
\end{equation}
where the matrix on the right, $\Bbb{M}$, is just 
$\Bbb{L}/\delta t$. Taking the limit $\delta t\rightarrow 0$ 
in the formal solution,
$P\left( \tau \right) =\left[ \Bbb{I}+\Bbb{L}\right] ^\tau P\left(
0\right) $ ($\Bbb{I}$ being the identity), leads then to 
$P\left( t\right) =\exp \left[ \Bbb{M} t\right] P\left( 0\right)$.

\subsection{Motivation and definition of ASEP and TASEP}

The ASEP is a many-body dynamical system, consisting of particles
located on a discrete lattice that evolves in continuous time. The
particles hop randomly from a site on a lattice to one of its
immediate neighbors, provided the target site is empty. Physically,
this constraint mimics short-range interactions amongst particles. In
order to drive this lattice gas out of equilibrium, non-vanishing
currents must be established in the system. This can be achieved by
various means: by starting from non-uniform initial conditions, by
coupling the system to external reservoirs that drive currents
\cite{Krug} through the system (transport of particles, energy, heat),
or by introducing some intrinsic bias in the dynamics that favors
motion in a privileged direction. Then, each particle is an asymmetric
random walker that drifts steadily along the direction of an external
driving force. Due to its simplicity, this model has appeared in
different contexts. As noted in the Introduction, it was first
proposed by Gibbs, \emph{et al.} in 1968 \cite{MGP68,MG69} as a
prototype to describe the dynamics of ribosomes translating along an
mRNA. In the mathematical literature, Brownian processes with
hard-core interactions were defined in 1970 by Spitzer
\cite{Spitzer70} who coined the term `exclusion process' (see also
\cite{Harris,Liggett1,Liggett99,Spohn91}). In addition to motivation
from issues in molecular biology -- the main focus of section
\ref{TC}, the ASEP has also been used to describe transport in
low-dimensional systems with strong geometrical constraints
\cite{MartinRev1} such as macromolecules transiting through
anisotropic conductors, or quantum dots, where electrons hop to vacant
locations and repel each other via Coulomb interaction
\cite{VonOppen}, while many of its variants are ubiquitous in modeling
traffic flow \cite{MartinShock,CSS00}. More generally, the ASEP
belongs to the class of driven diffusive systems defined by Katz,
Lebowitz and Spohn in 1984
\cite{KLS84,SZ95,Schutz00,DerridaRep,MartinRev1}. We emphasize that
the ASEP is defined through dynamical rules, while no energy is
associated with a microscopic configuration. More generally, the
kinetic point of view seems to be a promising and fruitful approach to
non-equilibrium systems (see e.g., \cite{PaulK}).

To summarize, the ASEP is a minimal model to study non-equilibrium
behavior.  It is simple enough to allow analytical studies, however it
contains the necessary ingredients for the emergence of a non-trivial
phenomenology:

\begin{itemize}
\item  \textit{Asymmetric}: The external driving breaks detailed balance and
creates a stationary current. The system settles into a NESS.

\item  \textit{Exclusion}: The hard-core interaction implies that there is
at most one particle per site. The ASEP is a genuine many-body problem, with
arguably the simplest of all interactions.

\item  \textit{Process}: With no underlying Hamiltonian, the dynamics is
stochastic and Markovian.
\end{itemize}

Having a general picture of an ASEP, let us turn to a complete
definition of the model. Focusing on only exactly solvable cases, we
restrict ourselves here to a one dimensional lattice, with sites
labeled by $i=1,...,L$ (here, we will use $i$ to label a site rather
than a configuration).  The stochastic evolution rules are the
following: A particle located at site $i$ in the bulk of
the system jumps, in the interval $\left[t,t+\dd t\right] $, with
probability $p\dd t$ to site $i+1$ and with probability $q \dd t$ to site
$i-1$, provided the target site is empty (\emph{exclusion rule}). The
rates $p$ and $q$ are the parameters of our system. By rescaling time,
$p$ is often set to unity, while $q$ is left as a genuine control
parameter. In the totally asymmetric exclusion process (TASEP), the
jumps are totally biased in one direction (e.g., $q=0$). On the other
hand, the \emph{symmetric} exclusion process (SEP) corresponds to the
choice $p=q$. The physics and the phenomenology of the ASEP is
extremely sensitive to the boundary conditions. We will mainly discuss
three types of boundary conditions:

\vspace{2mm}
\noindent {\it Periodic boundary conditions:} 

\begin{figure}[th]
\begin{center}
\includegraphics[width=4.4in]{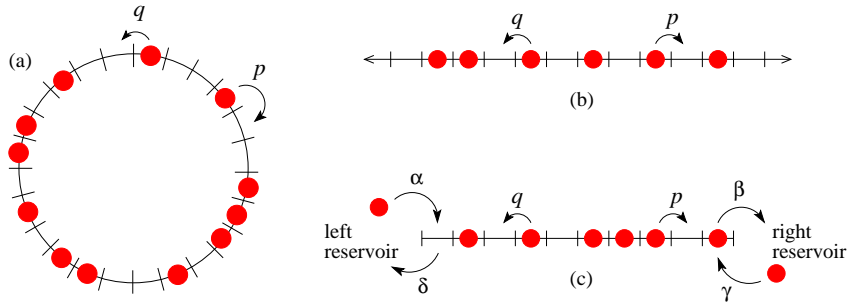}
\end{center}
\caption{(a) The asymmetric simple exclusion process on a ring of $L$
  sites filled with $N$ particles ($L=24,\,N=12$ here). The total
  number of configurations is just $\Omega = {L \choose N}$.  (b) ASEP
  on an infinite lattice. The particles perform asymmetric random
  walks $(p\neq q)$ and interact through the exclusion constraint.
  (c) A schematic of a ASEP with open boundaries on a finite lattice
  with $L=10$ sites.}
\label{fig:ASEPgeneral}
\end{figure}

The exclusion process is defined on a ring, so that sites $i$ and
$L+i$ are identical. The system is filled with $N\leq L$ particles
(figure \ref{fig:ASEPgeneral}(a)). Of course, the dynamics conserves
$N$.

\vspace{2mm}
\noindent {\it Infinite lattice:} 

Here, the boundaries are sent to $\pm \infty $.  Boundary conditions
are here of a different kind. This system always remains sensitive to
the initial conditions. Therefore, an initial configuration, or a
statistical set of configurations, must be carefully
specified. Figure~\ref{fig:ASEPgeneral}(b) is an illustration of this
system.

\vspace{2mm}
\noindent {\it Open boundaries:}

Here, the boundary sites $i=$ $1,L$ play a special role, as they are
coupled to particles in reservoirs just outside the system. Thus, if
site 1 is empty, a particle can enter (from the `left reservoir') with
rate $\alpha$. If it is occupied, the particle can exit (into this
reservoir) with rate $\gamma $. Similarly, the interactions with the
`right reservoir' are: If site $L$ is empty/occupied, a particle can
enter/exit the system with rate $\delta/\beta $ respectively. These
rates can be regarded as the coupling of our finite system with
infinite reservoirs set at different `potentials'.
Figure~\ref{fig:ASEPgeneral}(c) illustrates this system. A much
simpler limiting case is the TASEP, with $q=\gamma =\delta =0$, in
which particles injected from the left, hopping along the lattice only
to the right, and finally exiting to the right.

We emphasize that there are very many variants of the ASEP. The
dynamical rules can be modified, especially in computer simulation
studies using discrete-time updates (e.g., random sequential,
parallel, or shuffle updates). The hopping rates can be modified to be
either site- or particle-dependent, with motivations for such
additions from biology provided below. In modeling vehicular traffic,
the former is suitable, e.g., for including road work or traffic
lights. The latter can account for the range of preferred driving
velocities, while a system can also be regarded as one with many
`species' of particles. In addition, these disorders can be dynamic or
quenched
\cite{Barma92,Barma1,Barma2,Barma3,HS04,SETHNA2004,SHAW2004,DONG2011}. Further,
the exclusion constraint can be relaxed, so that fast cars are allowed
to overtake slower ones, which are known as `second class' or `third
class' particles \cite{DJLS,km,Speer,MMR,FM,EFM,KMCairns}. Finally,
the lattice geometry itself can be generalized to multi-lanes, higher
dimensions, or complex networks. All these modifications drastically
alter the collective behavior displayed in the system, as hundreds of
researchers discovered during the last two decades. Though more
relevant for applications, more realistic models cannot, in general,
be solved exactly. As the primary focus of this section is exact
solutions, we will focus only on the homogeneous systems presented
above. These problems are amenable to sophisticated mathematical
analysis thanks to a large variety of techniques: Bethe Ansatz,
quadratic algebras, Young tableaux, combinatorics, orthogonal
polynomials, random matrices, stochastic differential equations,
determinantal representations, hydrodynamic limits etc. Each of these
approaches is becoming a specific subfield that has its own links with
other branches of theoretical physics and mathematics. Next, we will
present some of these methods that have been developed for these three
ideal cases.

\subsection{Mathematical setup and fundamental issues}

The evolution of the system is given by equation (\ref{me-M}) and controlled
by the Markov operator $\Bbb{M}$ as follows. In order distinguish the two
uses of $i$ -- label for configurations and for sites on our lattices, let us
revert to using ${\cal C}$ for configurations. Then, this master equation
reads

\begin{equation}
\frac{dP({\cal C},t)}{dt}=\sum_{{\cal C}^{\prime }}M({\cal C},
{\cal C}^{\prime })P({\cal C}^{\prime },t)\,.  \label{Eq:Markov}
\end{equation}
As a reminder, the off-diagonal matrix elements of $\Bbb{M}$ represent
the transition rates, which the diagonal part $M({\cal C},{\cal
  C})=-\sum_{ {\cal C}^{\prime }\neq {\cal C}}M({\cal
  C}^{\prime},{\cal C} ^{\prime })$ accounts for the exit rate from
${\cal C}$. Thus, the sum of the elements in any given column
vanishes, ensuring probability conservation. For a finite ASEP,
$\left\{ {\cal C}\right\}$ is finite and the Markov operator $\Bbb{M}$
is a matrix. For the infinite system, $\Bbb{M}$ is an operator and its
precise definition needs more sophisticated mathematical tools than
linear algebra, namely, functional analysis
\cite{Liggett1,Spohn91}. Unless stated otherwise, we will focus here
on the technically simpler case of finite $L$ and deduce results for
the infinite system by taking $L\to \infty $ limit formally. An
important feature of the finite ASEP is ergodicity: Any configuration
can evolve to any other one in a finite number of steps. This property
ensures that the Perron-Frobenius theorem applies (see, e.g.,
\cite{PFT,Gantmacher}). Thus, $E=0$ is a non-degenerate eigenvalue of
$\Bbb{M}$, while all other eigenvalues have a strictly negative real
part, $\mathrm{Re}(E) < 0$.  The physical interpretation of the
spectrum of $\Bbb{M}$ is clear: The right eigenvector associated with
the eigenvalue $E=0$ corresponds to the stationary state of the
dynamics. Because all non-zero eigenvalues have strictly negative real
parts, these eigenvectors correspond to relaxation modes, with
$-1/\mathrm{Re}(E)$ being the relaxation time and $\mathrm{Im}(E)$
controlling the oscillations.

We emphasize that the operator $\Bbb{M}$ encodes all statistical
information of our system, so that any physical quantity can be traced
ultimately to some property of $\Bbb{M}$. We will list below generic
mathematical and physical properties of our system that motivate the
appropriate calculation strategies:

\begin{enumerate}
\item Once the dynamics is properly defined, the basic question is to
  determine the steady-state, $P^{*}$, of the system, i.e.,
  the right eigenvector of $\Bbb{M}$ with eigenvalue 0. Given a
  configuration ${\cal C}$, the value of the component $P^{*}({\cal
    C})$ is known as the measure (or stationary weight) of ${\cal C}$
  in the steady-state, i.e., it represents the frequency of
  occurrence of ${\cal C}$ in the stationary state.

\item The vector $P^{*}$ plays the role of the Boltzmann factor in
  EQSM, so that all steady-state properties (e.g., all equal-time
  correlations) can be computed from it. Some important questions are:
  what is the mean occupation $\rho_i$ of a given site $i$? What does
  the most likely density profile, given by the function $i\to \rho
  _i$, look like? Can one calculate density-density correlation
  functions between different sites? What is the probability of
  occurrence of a density profile that differs significantly from the
  most likely one? The last quantity is known as the large deviation
  of the density profile.

\item The stationary state is a dynamical state in which the system
  constantly evolves from one micro-state to another. This microscopic
  evolution induces macroscopic fluctuations (the equivalent of the
  Gaussian Brownian fluctuations at equilibrium). How can one
  characterize such fluctuations? Are they necessarily Gaussian? How
  are they related to the linear response of the system to small
  perturbations in the vicinity of the steady-state? These issues can
  be tackled by considering tagged-particle dynamics, anomalous
  diffusion, time-dependent perturbations of the dynamical rules
  \cite{Kavita}.

\item As expected, the ASEP carries a finite, non-zero, steady-state
  particle current, $J$, which is clearly an important physical
  observable.  The dependence of $J$ on the external parameters of the
  system allows us to define different phases of the system.

\item The existence of a non-zero $J$ in the stationary state implies
  the physical transport of an extensive number of particles, $Q$,
  through the lattice. The total number of particles $Q_{t}$,
  transported up to time $t$ is a random quantity. In the
  steady-state, the mean value of $Q_t$ is just $Jt$, while in the
  long time limit, the distribution of the random variable $Q_t/t-J$
  represents exceptional fluctuations (i.e., large deviations) of the
  mean-current. This LDF is an important observable that provides
  detailed properties of the transport through the system. While
  particle current is the most obvious quantity to study in an ASEP,
  similar questions can be asked of other currents and the transport
  of their associated quantities, such as mass, charge, energy, etc.,
  in more realistic NESM models.

\item The way a system relaxes to its stationary state is also an
  important characteristic. The typical relaxation time of the ASEP
  scales with the system size as $L^z$, where $z$ is the dynamical
  exponent. The value of $z$ is related to the spectral gap of the
  Markov matrix $\Bbb{M}$, i.e., to the largest
  $-\mathrm{Re}\left(E\right)$. For a diffusive system, $z=2$. For the
  ASEP with periodic boundary condition, an exact calculation leads to
  $z=3/2$. More generally, the transitory state of the model can be
  probed using correlation functions at different times.

\item The matrix $\Bbb{M}$ is generally a non-symmetric matrix and,
  therefore, its right eigenvectors differ from its left
  eigenvectors. For instance, a right eigenvector $\psi_E$
  corresponding to the eigenvalue $E$ is defined as

\begin{equation}
\Bbb{M}\psi =E\psi. 
\label{eq:mpsi=epsi}
\end{equation}
Because $\Bbb{M}$ is real, its eigenvalues (and eigenvectors) are
either real numbers or complex conjugate pairs. However, M is
generally asymmetric, so that its left eigenvectors are different from
its right eigenvectors. Powerful analytical techniques, such as the
Bethe Ansatz, can be exploited to diagonalize $\Bbb{M}$ in some
specific cases, providing us with crucial information on its spectrum.

\item Solving the master equation~(\ref{Eq:Markov}) analytically would
  allow us to calculate exactly the evolution of the system. A
  challenging goal is to determine the finite-time Green function (or
  transition probability) $P({\cal C},t|{\cal C}_0,0)$, the
  probability for the system to be in configuration ${\cal C}$ at time
  $t$, given that the initial configuration at time $t=0$ was ${\cal
    C}_0$. Formally, it is just the ${\cal C},{\cal C}_0$ element of
  the matrix $\exp \left[\Bbb{M}t\right] $ here. In principle, it
  allows us to calculate all the correlation functions of the
  system. However, explicit results for certain quantities will
  require not only the knowledge of the spectrum and eigenvectors of
  $\Bbb{M}$, but also explicit evaluations of matrix elements
  associated with the observable of interest.
\end{enumerate}

The following sections are devoted to a short exposition of several
analytical techniques that have been developed to answer some of these
issues for the ASEP.

\subsection{Steady state properties of the ASEP}
\label{SSASEP}

Given a stochastic dynamical system, the first question naturally
concerns the stationary measure. We will briefly discuss the ASEP with
periodic boundary conditions and the infinite line case. More details
will be provided for the highly non-trivial case of the open ASEP.

\hfill\break
\noindent \textit{Periodic boundary conditions:}

This is the simplest case, with the stationary measure being flat
\cite{DerridaRep}. That the uniform measure is also stationary can be
understood as follows. A given configuration consists of $k$ clusters
of particles separated by holes. A particle that leads a cluster can
hop in the forward direction with rate $p$ whereas a particle that
ends a cluster can hop backwards with rate $q$; thus the total rate of
leaving a configuration consisting of $k$ clusters is
$k(p+q)$. Similarly, the total number of \textit{ancestors} of this
configuration (i.e., of configurations that can evolve into it) is
also given by $k(p+q)$. The fact that these two quantities are
identical suffices to show that the uniform probability is
stationary. To obtain the precise value of $P^{*}$, $1/\Omega $, is
also elementary. Since $N$ is a constant, we only need the total
number of configurations for $N$ particles on a ring with $L$ sites,
which is just $\Omega =L!/[N!(L-N)!]$.

\hfill\break
\noindent \textit{Infinite lattice:}

For the exclusion process on an infinite line, the stationary measures
are studied and classified in \cite{Liggett1,Liggett99}. There are two
one-parameter families of invariant measures. One family, denoted by
$\nu_\rho$, is a product of local Bernoulli measures of constant
density $\rho$: this means that each site is occupied with probability
$\rho$.  The other family is discrete and is concentrated on a
countable subset of configurations. For the TASEP, this second family
corresponds to \textit{blocking measures}, which are point-mass
measures concentrated on step-like configurations (i.e.,
configurations where all sites to the left/right of a given site are
empty/occupied).

\hfill\break
\noindent \textit{Open boundaries:}

Turning to the case of the ASEP on a finite lattice with open
boundaries, we note that the only knowledge we have, without detailed
analysis, is the existence of a unique stationary measure (thanks to
the Perron-Frobenius theorem), i.e., the vector $P^{*}$ with $2^L$
components. We emphasize again that finding $P^{*}$ is a non-trivial
task because we have no \emph{a priori} guiding principle at our
disposal. With no underlying Hamiltonian and no temperature, no
fundamental principles of EQSM are relevant here. The system is far
from equilibrium with a non-trivial steady-state current that does not
vanish for even large $L$.

To simplify the discussion, we focus on the TASEP where particles
enter from the left reservoir with rate $\alpha$, hop only to the
right and leave the system from the site $L$ with rate $\beta$. A
configuration ${\cal C} $ can be represented by the binary string,
$(\sigma_1,\ldots ,\sigma_L)$, of occupation variables: $\sigma_i=1$
if the site $i$ is occupied and $\sigma_i=0$ otherwise. Our goal is to
determine $P^{*}\left({\cal C}\right) $, with which the steady-state
current $J$ can be expressed simply and \emph{exactly}:

\begin{equation}
J=\alpha \left(1-\langle \sigma_1\rangle \right) =\langle 
\sigma_1(1-\sigma_2)\rangle =\ldots =
\langle \sigma_i(1-\sigma_{i+1})\rangle =\beta \langle\sigma_L\rangle.
\label{CourantCorrelTASEP}
\end{equation}
Here, the brackets $\langle \,\, \rangle $ denote expectation values
as defined in equation (\ref{average}). Even if $J$ were known
somehow, this set of equations is not sufficient for fixing the
(unknown) density profile $\rho_i=\langle \sigma_i\rangle$ and the
nearest neighbor (NN) correlations $\langle
\sigma_i\sigma_{i+1}\rangle$.  Typical in a truly many-body problem,
there is a hierarchy \cite {BBGKY1,BBGKY2,BBGKY3,Reichl09} of
equations that couple $k$-site and $\left(k+1\right)$-site
correlations.  A very important approximation, which often provides
valuable insights, is the \textit{mean-field assumption} in which the
hierarchy is simply truncated at a given level. If the correlations
beyond this level are small, this approximation can be quite
good. Applying this technique here consists of replacing two-sites
correlations by products of single-site averages:

\begin{equation}
\langle \sigma_i\sigma_j\rangle \rightarrow \langle \sigma_i\rangle \langle
\sigma_j\rangle =\rho_i\rho_j.
\label{MF-2spins}
\end{equation}
Thus, equation (\ref{CourantCorrelTASEP}) becomes

\begin{equation}
J\simeq \alpha \left(1-\rho_1\right) =\rho _1(1-\rho_2)=\ldots
=\rho_i(1-\rho_{i+1})=\beta \rho_L.  
\label{MeanFieldTASEP}
\end{equation}
and we arrive at a closed recursion relation between $\rho_{i+1}$ and
$\rho_i,$ namely $\rho_{i+1}=1-J/\rho_i$. Of course, $J$ is an
unknown, to be fixed as follows. Starting with $\rho_1=1-J/\alpha $,
the recursion eventually leads us to $\rho_L$ as a rational function
of $J$ (and $\alpha $). Setting this to $J/\beta $ gives a polynomial
equation for $J$. Solving for $J$, we obtain the desired dependence of
the steady-state current on the control parameters: $J\left( \alpha
,\beta ,L\right)$. This approach to an approximate solution was known
to Gibbs, \emph{et al.} \cite{MGP68,MG69} (within the context of a
more general version of TASEP) and explored further in \cite{DeDoMuk}
recently.

Analyzing $J\left(\alpha, \beta, L\right) $ gives rise to the phase
diagram of the TASEP (see figure \ref{fig-DIAGPHASE}). For studying
these aspects of the TASEP, the mean-field method provides us with a
reasonably good approximation. Indeed, the correct phase diagram of
the model was obtained in \cite{Krug}
\footnote{See also \cite{Spohn91} and \cite{Janowski} for a
  pedagogical example.} through physical reasoning by using such
mean-field arguments along with the hydrodynamic limit. Since this
strategy is quite effective and more widely applicable, we will
briefly discuss how it is applied to ASEP in section \ref{HMFT}
below. Despite many effective MFT-based strategies, exact solutions to
ASEP are desirable, especially for an in-depth analysis.  In
particular, MFT cannot account for correlations, fluctuations, or rare
events. In fact, the mean-field density profile (from solving the
recursion relation (\ref{MeanFieldTASEP}) numerically) does not agree
with the exact profile (from evaluating the expression
(\ref{ExactProfile}) below). Of course, it is rare that we are able to
calculate the stationary measure for a non-equilibrium interacting
model. Not surprisingly, the exact steady-state solution of the TASEP
\cite{DeDoMuk,DEHP,SchutzDomany} was considered a major breakthrough.

\subsection{The matrix representation method for the open TASEP}
\label{TASEP-MR}

The exact calculation of the stationary measure for the TASEP with
open boundaries and the derivation of its phase diagram triggered an
explosion of research on exactly solvable models in NESM. The
fundamental observation \cite{DeDoMuk} is the existence of recursion
relations for the stationary probabilities between systems of
different sizes. These recursions are particularly striking when
$\alpha =\beta =1$ \cite{DeDoMuk}; they can be generalized to
arbitrary values of $\alpha$ and $\beta$ \cite{SchutzDomany,DEHP} and
also to the more general case in which backward jumps are allowed
(also known as a partially asymmetric exclusion process, or
PASEP). The most elegant and efficient way to encode these recursions
is to use the matrix Ansatz \cite{DEHP}. We caution readers that the
matrices to be presented in this subsection have nothing to do with
the transition matrices $\Bbb{M}$ discussed above.  They do not
represent the intrinsic physics of TASEP, but instead, provide a
convenient framework for representing the algebra that arises from the
recursion relations. In particular, we need only two matrices ($\D$
and $\E$) and two `eigenvectors' (a bra $\langle W|$ and a ket
$|V\rangle $, normalized by $\langle W|V\rangle =1$) here. The algebra
we need is

\begin{eqnarray}
\D \E &=& \D+\E  \nonumber  \\
\D \,|V\rangle  &=&\frac 1\beta |V\rangle \,\,  \nonumber \\
\langle W|\,\E &=&\frac 1\alpha \langle W|\,\,.  \label{DEHPAlgebra}
\end{eqnarray}
We emphasize that in general the operators $\D$ and $\E$ do not commute
with each other, so that an explicit representation would be typically
infinite dimensional.

Remarkably, it was shown that the stationary $P^{*}\left( {\cal
  C}\right)$ for TASEP can be written as the scalar

\begin{equation}
P^{*}(\sigma_1,\ldots, \sigma_L)=\frac 1{Z_L}\langle W|\prod_{i=1}^L\left(
\sigma_i\D+(1-\sigma_i)\E\right) |V\rangle \,.  \label{MPA}
\end{equation}
where
\begin{equation}
Z_L=\langle W|\left(\D+\E\right) ^L|V\rangle 
\label{def:Z_L}
\end{equation}
is a normalization constant. Each operation of the matrices $\D$ or
$\E$ is associated with a filled or empty site on the lattice. For
example, the weight of the configuration shown in figure
\ref{fig:ASEPgeneral}(c) is $\langle W| \E \D \D \E \D \E \D \E \D
\E|V\rangle/Z^{10}$. To obtain explicit values for (\ref{MPA}), one
method is to find an explicit representation of this algebra. Another
possibility is to use systematically the algebraic
rules~(\ref{DEHPAlgebra}) without referring to any
representation. Indeed, a product of $\D$'s and $\E$'s in an arbitrary
order can be decomposed as a linear combination of monomials of the
type $\E^n \D^m$ by using repeatedly the rule $\D \E=\D+\E$. Then,
using $\langle W|\E^n \D^m\,|V\rangle =\alpha^{-n}\beta^{-m}$, we find
that any matrix element of the type (\ref{MPA}) is a polynomial in
$1/\alpha$ and $1/\beta $. In particular, we find the following
general formula \cite{DEHP} for $Z_L$:

\begin{equation}
\begin{array}{rl}
Z_L & =\langle W|\left(\D+\E\right) ^L|V\rangle  \\[13pt]
\: & \displaystyle =\sum_{p=1}^{L}\frac{p\,(2L-1-p)!}{L!\,(L-p)!}\frac{\beta
^{-p-1}-\alpha ^{-p-1}}{\beta ^{-1}-\alpha ^{-1}}.\label{def:Z_LBIS}
\end{array}
\end{equation}
When $\alpha =\beta=1$, $Z_L$ is a Catalan number \cite{DEHP}. Another
special case is $\alpha =1-\beta $, for which the scalar
representation, $\D=1/\beta, \E=1/\alpha $, suffices and $P^{*}$
simplifies greatly. In all cases, quantities of physical interest
(current, profile, correlations) can be explicitly computed using the
algebra~(\ref{DEHPAlgebra}). In this sense, the TASEP is `solved:' All
equal time correlations in the stationary state are known.

The matrix method may look puzzling at first sight. One informal way
to motivate it is the following: the steady state weight of a
configuration cannot be expressed in general as a product of single
site occupancy or vacancy probabilities because in the presence of
interactions there are non-zero correlations (i.e., mean-field theory
is not exact).  However, by writing the stationary measure as a
product of matrices, a sort of factorization still holds and at the
same time correlations are taken into account by the fact that the
matrices do not commute.

\subsection{Phase diagram of the open TASEP}
\label{TASEP-PD}

Thanks to the matrix representation method, exact expressions for the
phase diagram of the TASEP, as well as stationary equal-time
correlations and density profiles, can be readily obtained.  For
example, the mean occupation of a site $i$ (with $ 1 \leq i \leq L$)
is given by

\begin{equation}
 \langle \sigma_i\rangle = \frac{1}{Z_L} \langle W | \left(\D + \E
 \right)^{i-1}\D \left(\D +\E \right)^{L-i} | V \rangle \, .
\label{ExactProfile}
\end{equation}
This expression is obtained by summing over all configurations in
which site $i$ is occupied.

Similarly, the average value of the steady state current $J$ through
the ``bond'' connecting site $i$ and $i+1$ can be calculated as

\begin{equation}
\begin{array}{rl} 
J(\alpha,\beta, L) & =  \langle \sigma_i (1 - \sigma_{i+1}) \rangle  \\[13pt]
\: & \displaystyle = \frac{1}{Z_L} 
 \langle W | \left(\D + \E \right)^{i-1} \D \E  
  \left(\D + \E \right)^{L-i-1}  | V  \rangle \\[13pt]
\: & \displaystyle =  \frac{Z_{L-1}}{Z_L} \, ,
\label{CURRENTJ}
\end{array}
\end{equation}
where we have used the algebra~(\ref{DEHPAlgebra}).  We note that this
value does not depend on the specific bond considered.  This was
expected because particles are neither created nor destroyed in the
system.  \hfill\break

\begin{figure}[ht]
\begin{center}
\includegraphics[width=4.8in]{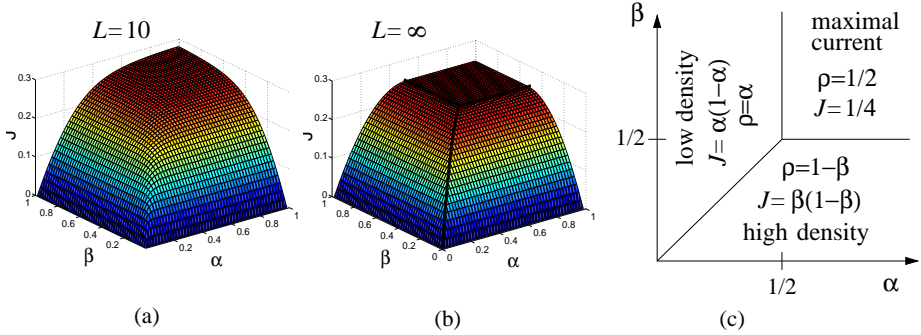}
\caption{(a) Exact solution (derived from eqs.~(\ref{CURRENTJ}) and
  (\ref{def:Z_LBIS}) for the steady-state current $J$ of a TASEP with
  $L=10$ sites.  (b) The steady-state current $J$ plotted in the $L\to
  \infty$ limit. (c) The phase diagram for the current of an infinite
  lattice ($L=\infty$) TASEP as a function of the injection and
  extraction rates.}
\label{fig-DIAGPHASE}
\end{center}
\end{figure}

In the limit of large system sizes, the phase diagram
(figure~\ref{fig-DIAGPHASE}) consists of three main regions

\begin{itemize}
  \item In the {\it Low-Density Phase}, when $\alpha <
    \min\{\beta, 1/2\}$, the bulk-density is $\rho = \alpha$
    and the average current $J = \alpha(1 -\alpha)$ is a function only
    of the rate-limiting injection process.

\item In the {\it High Density phase}, when $\beta < \min\{\alpha,
  1/2\}$, the typical density is characterized by $\rho = 1 -
  \beta$ and the steady-state current, ${J} = \beta(1 -\beta)$, is a
  function only of the rate-limiting extraction step.

\item In the {\it Maximal Current Phase,} when $\alpha >1/2$ and
  $\beta >1/2$, the bulk behavior is independent of the boundary
  conditions and one finds $\rho = 1/2$ and $J = 1/4$. In this phase,
  particle-particle correlations decay only algebraically away from
  the boundaries, in contrast to exponentially decaying correlations
  in both the low and high density phases.

\item The low and high density phases are separated by the
  `shock-line', $\alpha = \beta \leq 1/2$, across which the
  bulk-density is discontinuous. In fact, the profile on this line is
  a mixed-state of shock-profiles interpolating between the lower
  density $\rho = \alpha$ and the higher density $\rho = 1 - \beta$.
\end{itemize}

\noindent Detailed properties of the phase diagram are reviewed in
\cite{DerridaRep,Schutz,MartinRev2}.

The particle density profiles in the large $L$ limit, described in
each phase above, are only approximately uniform in that they are
asymptotically accurate only in the bulk, but must vary near the
boundaries in order to satisfy conditions determined by the
steady-state particle injection and extraction \cite{LAKATOS2006}.  It
turns out that the MFT approaches, recursion and hydrodynamic
equations, reproduce the exact $L \to \infty$ phase diagram; however,
the matrix product approach finds the correct density profile which is
not obtained by mean-field approximations.

\subsection{Some extensions of the matrix Ansatz}

\noindent {\it Extension to the general ASEP model:} 

If we allow jumps in both directions and entrance/exit at both ends of
the system, the matrix technique can still be applied. The
algebra~(\ref{DEHPAlgebra}) must be replaced by the more general rules

\begin{eqnarray}
             p\,  \D \, \E  - q\,  \E \, \D  & = & \D \, + \, \E    \nonumber \\
           (\beta  \D  \, - \delta \E)  \, | V  \rangle   & = & 
  | V  \rangle    \nonumber   \\
              \langle W | \, (\alpha\, \E  - \gamma \, \D)  & = &  \langle W |  \,\, . 
 \label{DEHPGenerale}
\end{eqnarray}
This new algebra allows one to calculate the general phase diagram of the
open ASEP using orthogonal polynomials
\cite{AlcarazRitt,MartinRev2,Sasamoto99,Sasamoto2000}. The phase
diagram of the ASEP turns out to be identical to that of the TASEP in
a two-dimensional slice across effective parameters that are functions
of the intrinsic rates $\alpha, \beta, \gamma, \delta, p, q$.

\vspace{2mm}

\noindent {\it Multispecies exclusion processes:} 

The matrix method can be used to represent the stationary measure of
periodic systems with defects or with different classes of particles
\cite{DJLS,km,Speer,MMR,FM,EFM,KMCairns}. In particular, this allows
us to investigate shock profiles that appear in the ASEP and to prove
that these shocks, predicted by Burgers' equation, do exist at the
microscopic level and are not artifacts of the hydrodynamic limit
\cite{DJLS,F92,FF94,FFK94}.
\vspace{2mm}

\noindent {\it Macroscopic density profiles:} 

Knowing exactly the microscopic invariant measure allows us to
rigorously coarse-grain the problem and to determine the probability
of occurrence of a density profile that differs from the average
profile. The calculation of such `free energy functionals' is an
important step in understanding how non-equilibrium macroscopic
behavior emerges from the microscopic model. For a review of these
very important issues and for relevant references, we refer the reader
to \cite{DerrReview,DerridaCAIRNS}.
\vspace{2mm}

\noindent {\it Other models:} 

We emphasize that the matrix product representation method has proved
to be very fruitful for solving many one-dimensional systems; a very
thorough review of this method can be found in \cite{MartinRev2}.

\subsection{Time-dependent properties: the Bethe Ansatz}
\label{BA}

In order to investigate the behavior of the system away from
stationarity, the spectrum of the Markov matrix is needed.  For an
arbitrary stochastic system, the evolution operator cannot be
diagonalized analytically for any system sizes.  However, the ASEP
belongs to a very special class of models: it is {\it integrable} and
it can be solved using the Bethe Ansatz as first noticed by D. Dhar in
1987 \cite{Dhar}.  Indeed, the Markov matrix that encodes the
stochastic dynamics of the ASEP can be rewritten in terms of Pauli
matrices; in the absence of a driving field, the {\it symmetric}
exclusion process can be mapped exactly into the Heisenberg spin
chain.  The asymmetry due to a non-zero external driving field breaks
the left/right symmetry and the ASEP becomes equivalent to a
non-Hermitian spin chain of the XXZ type with boundary terms that
preserve the integrable character of the model.  The ASEP can also be
mapped into a six vertex model \cite{Baxter,Kandel,Rajesh}.  These
mappings suggest the use of the Bethe Ansatz to derive spectral
information about the evolution operator, such as the spectral gap
\cite{Dhar,Gwa,Kim1,Kim2,OGKMgap,ogkmrev} and large deviation
functions \cite{DLeb,evans}.

Here, we will apply the Bethe Ansatz to the ASEP on a ring.  A
configuration can also be characterized by the positions of the $N$
particles on the ring, $(x_1, x_2, \dots, x_N)$ with $1 \leq x_1 < x_2
< \dots < x_N \leq L$. With this representation, the eigenvalue
equation (\ref{eq:mpsi=epsi}) becomes
\begin{equation}
\begin{array}{l}
\fl E \psi(x_1,\dots, x_N) =  \\[13pt]
\fl \: \hspace{1cm} \sum_i  p \left[ 
             \psi(x_1, \dots, x_{i-1},\ x_i-1,\ x_{i+1}, \dots, x_N) 
 - \psi(x_1,\dots, x_n) \right] \,\, + \\[13pt]
\fl \: \hspace{1cm}  \sum_j  q \left[ \psi(x_1, \dots, x_{j-1},\ x_j+1,\ x_{j+1}, \dots, x_N)
- \psi(x_1,\dots, x_N) \right] \, , 
\label{Eq:VPASEP}
\end{array}
\end{equation}
where the sum runs over the indices $i$ such that $ x_{i-1} < x_i-1$
and over the indices $j$ such that $ x_j + 1 < x_{j+1} \,;$ these
conditions ensure that the corresponding jumps are allowed.

We observe that equation~(\ref{Eq:VPASEP}) is akin to a discrete
Laplacian on a $N$-dimensional lattice: the major difference is that
the terms corresponding to forbidden jumps are absent. Nevertheless,
this suggests that a trial solution, or {\it Ansatz}, in the form of
plane waves may be useful.  This is precisely the idea
underlying the Bethe Ansatz, originally developed to
study the Heisenberg spin chain model of quantum magnetism \cite{Bethe}.
Following Bethe's method, we will also refer to $\psi$ as a
`wavefunction,' but the reader should not confuse our problem with one
in quantum mechanics. In our opinion, the ASEP is the one of the
simplest system with which one can learn the Bethe Ansatz. The next
subsection is devoted to such a tutorial.

\subsection{Bethe Ansatz for ASEP: a crash-course}

Our aim is to solve the linear eigenvalue problem~(\ref{Eq:VPASEP})
which corresponds to the relaxation modes of the ASEP with $N$
particles on a ring of $L$ sites.  We will study some special cases
with small $N$ to illustrate the general structure of
the solution.

\hfill\break
\noindent {\bf The single particle case:} For $N =1$,
equation~(\ref{Eq:VPASEP}) reads
\begin{equation}
 E \psi(x) = p  \psi(x -1)  +  q  \psi(x + 1) - (p+q) \psi(x) \, , 
 \label{N=1PASEP}
\end{equation}
 with $ 1 \le x \le L$ and where periodicity is assumed
\begin{equation}
  \psi(x + L ) =  \psi(x)  \, .
\end{equation}
  
Equation~(\ref{N=1PASEP}) is simply a linear recursion of order 2 that
is solved as
 \begin{eqnarray}
  \psi(x) = A z_{+}^x + B z_{-}^x \,  ,
 \end{eqnarray}
 where $r=z_{\pm}$ are the two  roots of the characteristic equation
\begin{eqnarray}
    q r^2 -(E +p +q) r + p = 0  \, .
 \end{eqnarray}
The periodicity condition imposes that at least one of the two
characteristic values is a $L$-th root of unity (Note that because
$z_{+} z_{-} = p/q$, only one of them can be a root of unity when
$p \neq q$). The general solution is given by

\begin{eqnarray}
  \psi(x) = A z^x  \quad \quad \hbox{ with }  \quad z^L = 1  \,  ,
\label{1bodyEigenP}
 \end{eqnarray}
i.e., simple {\it plane waves} with `momenta' being integer 
multiples of $2\pi /L$ and associated with eigenvalue
\begin{equation}
  E  =  \frac{p}{z} + q z - (p+q)  \, .
\label{1bodyEnergy}
\end{equation}

\hfill\break
\noindent {\bf The two-particle case:} The case $N=2$ where
two particles are present is more interesting because when the
particles are located on adjacent sites the exclusion effect plays a
role.  Indeed, the general eigenvalue equation~(\ref{Eq:VPASEP}) can
be split into two different cases:

\begin{itemize}
  \item In the generic case, $x_1$ and $x_2$ are separated by at least
    one empty site:

\begin{equation}
\begin{array}{l}
\fl E \psi(x_1,x_2) = p \left[ \psi(x_1 -1 ,x_2) + \psi(x_1,x_2 -1) \right]\\[13pt]
\: + q\left[ \psi(x_1 +1,x_2) + \psi(x_1,x_2 +1) \right]- 2(p+q) \psi(x_1,x_2) \, .
\label{2-generic}
\end{array}
\end{equation}

\item In the (special) adjacency case, $x_2 = x_1 +1$, some jumps are
  forbidden and the eigenvalue equation reduces to:

\begin{equation}
\fl E \psi(x_{1}, x_{1}+1) = p \psi(x_{1}-1 ,x_{1}+1) + q \psi(x_{1},
x_{1}+2) - (p+q) \psi(x_{1},x_{1}+1).
\label{2-special}
\end{equation}
This equation differs from the generic equation~(\ref{2-generic}) in
which we substitute $x_2 = x_1 +1$. An equivalent way to take into
account the adjacency case is to impose that the generic
equation~(\ref{2-generic}) is valid {\it for all values} of $x_1$ and
$x_2$ and add to it the following {\it cancellation boundary
  condition:}
\begin{equation}
\fl p \psi(x_1 ,x_1)    +  q  \psi(x_1+1 ,x_1+ 1)   - (p+q) \psi(x_1,x_1+1) = 0 \, .
\label{2-annulation}
\end{equation}
\end{itemize}
We now examine how these equations can be solved.  In the generic case
particles behave totally independently (i.e., they do not
interact).  The solution of the generic equation~(\ref{2-generic}) can
therefore be written as a product of plane waves $\psi(x_1,x_2) = A
z_1^{x_1} z_2^{x_2}$, with the eigenvalue
 \begin{equation}
  E  =  p \left(\frac{1}{z_1} +  \frac{1}{z_2} \right)
  + q \left(z_1 + z_2 \right)  -2 (p+q)  \, .
\label{2-Energy}
\end{equation}
However, the simple product solution cannot be the full answer:
the cancellation condition for the adjacency
case~(\ref{2-annulation}) also has to be satisfied.  The first crucial
observation, following H. Bethe \cite{Bethe}, is that the eigenvalue
$E$, given in~(\ref{2-Energy}) is invariant by the permutation $z_1
\leftrightarrow z_2$.  In other words, there are two plane waves $ A
z_1^{x_1} z_2^{x_2} $ and $ B z_2^{x_1} z_1^{x_2} $ with the same
eigenvalue $E$ which has a two-fold degeneracy. The full eigenfunction
corresponding to $E$ can thus be written as
 \begin{equation}
\psi(x_1,x_2) = A_{12} z_1^{x_1} z_2^{x_2} + A_{21} z_2^{x_1} z_1^{x_2} \, ,
\label{2-Bethe}
\end{equation}
where the amplitudes $A_{12} $ and $A_{21} $ are yet arbitrary.  The
second key step is to understand that these amplitudes can now be
chosen to fulfill the adjacency cancellation condition: substitution
of (\ref{2-Bethe}) into equation~(\ref{2-annulation}), we obtain
\begin{equation}
   \frac{  A_{21} } {  A_{12} } =  
 - \frac{ q z_1  z_2 - (p +q)  z_2  + p}{ q z_1  z_2 - (p +q)  z_1  + p} \, .
\label{2-BetheAmplitudes}
\end{equation}
The eigenfunction~(\ref{2-Bethe}) is therefore determined, but for an
overall multiplicative constant.  We now implement the periodicity
condition that takes into account the fact that the system is defined
on a ring. This constraint can be written as follows for $ 1 \le x_1 <
x_2 \le L $
\begin{equation}
\psi(x_1, x_2) =  \psi( x_2, x_1 + L) \, .
\label{2-Period}
\end{equation}
This relation plays the role of a quantification condition for the
scalars $z_1$ and $z_2$, which we will call {\it Bethe roots}.  If we
now impose the condition that the expression~(\ref{2-Bethe}) satisfies
equation~(\ref{2-Period}) for all generic values of the positions
$x_1$ and $ x_2$, new relations between the amplitudes arise:
 \begin{equation}
   \frac{A_{21}} {A_{12}} = z_2^L = \frac{1}{z_1^L} \, . 
\label{2-BetheAmpliBIS}
\end{equation}
Comparing equations~(\ref{2-BetheAmplitudes})
and~(\ref{2-BetheAmpliBIS}) leads to a set of algebraic equations
obeyed by the Bethe roots $z_1$ and $z_2$:
 \begin{eqnarray}
        z_1^L &=& -  \frac{ q z_1  z_2 - (p +q)  z_1  + p}{ q z_1  z_2 - (p +q)  z_2  + p} \, \\
       z_2^L &=& -   \frac{ q z_1  z_2 - (p +q)  z_2  + p}{ q z_1  z_2 - (p +q)  z_1  + p}
\end{eqnarray}

These equations are known as the {\it Bethe Ansatz Equations}.
Finding the spectrum of the matrix $\Bbb{M}$ for two particles on a
ring of size $L$ is reduced to solving these two coupled polynomial
equations of degree of order $L$ with unknowns $z_1$ and
$z_2$. Surely, this still remains a very challenging task but the
Bethe equations are explicit and very symmetric.  Besides, we
emphasize that the size of the matrix $\Bbb{M}$ (and the degree of its
characteristic polynomial) is of order $L^2$.

\hfill\break {\bf The three-particle case:} We are now ready
to consider the case $N=3$.  For a system containing three particles,
located at $x_1 \le x_2 \le x_3$, the generic equation, valid when the
particles are well separated, can readily be written using
equation~(\ref{Eq:VPASEP}).  But now, the special adjacency cases are
more complicated:
\vspace{2mm}

\noindent (i) Two particles are next to each other and the third one is far
 apart; such a setting is called a 2-body collision and the boundary
 condition that results is identical to the one obtained for the case
 $N=2$.  There are now two equations that correspond to the cases
 $x_1=x \le x_2 = x +1 \ll x_3 $ and $x_1 \ll x_2 =x \le x_3= x +1$:
\begin{eqnarray}
\fl p \psi(x ,x , x_3)    +  q  \psi(x+1 ,x + 1, x_3 )   - (p+q) \psi(x ,x +1 , x_3 ) &=&  0
\label{3-annulation} \\[13pt]
\fl p\psi(x_1, x ,x )    +  q  \psi(x_1, x+1 ,x + 1)   - (p+q) \psi(x_1,x ,x +1  ) &=&  0.
\label{3-annulationbis}
\end{eqnarray}
We emphasize again that these equations are identical to
equation~(\ref{2-annulation}) because the third particle, located far
apart, is simply a {\it spectator} ($x_3$ is a spectator in the first
equation; $x_1$ in the second one).

\vspace{2mm}

\noindent (ii) There can be 3-body collisions, in which the three particles are
 adjacent, with $x_1 =x, x_2 = x +1, x_3 = x +2$.  The resulting
 boundary condition is then given by
\begin{eqnarray}
\fl p \left[\psi(x ,x , x+2) + \psi(x , x+1 ,x + 1) \right] + q \left[
   \psi(x+1 ,x + 1, x +2 ) + \psi(x , x + 2 ,x + 2) \right] \nonumber \\
   \: \hspace{1cm} - 2 (p+q) \psi(x ,x +1 , x +2 ) = 0 \, .
\label{3-bodyCollision}
\end{eqnarray}
   
The fundamental remark is that {\it 3-body collisions do not lead to
  an independent constraint}.  Indeed,
equation~(\ref{3-bodyCollision}) is simply a linear combination of the
constraints~(\ref{3-annulation}) and~(\ref{3-annulationbis}) imposed
by the 2-body collisions. To be precise:
equation~(\ref{3-bodyCollision}) is the sum of
equation~(\ref{3-annulation}), with the substitutions $x \to x$ and
$x_3 \to x+2$, and of equation~(\ref{3-annulationbis}) with $x_1 \to
x$ and $x \to x+1$. Therefore, {\it it is sufficient to fulfill the
  2-body constraints because then the 3-body conditions are
  automatically satisfied.}  The fact that 3-body collisions decompose
or `factorise' into 2-body collisions is the {\it crucial property}
that lies at the very heart of the Bethe Ansatz. If it were not true,
the ASEP would not be exactly solvable or `integrable'.

For $N=3$, the plane wave $\psi(x_1,x_2,x_3) = A z_1^{x_1} z_2^{x_2}
z_3^{x_3}$ is a solution of the generic equation with the eigenvalue

\begin{equation}
  E  =  p \left( \frac{1}{z_1} +  \frac{1}{z_2 + z_{3}^{-1}} \right)
  + q \left(z_1 + z_2 + z_3 \right)  -3 (p+q)  \, .
\label{3-Energy}
\end{equation}
However, such a single plane wave does not satisfy the boundary
conditions (\ref{3-annulation}) and~(\ref{3-annulationbis}).
Again, we note that the eigenvalue $E$ is invariant under the
permutations of $z_1, z_2$ and $z_3$. There are 6 such permutations,
that belong to $S_3$, the permutation group of 3 objects.  The
Bethe wave-function is therefore written as a sum of the 6 plane
waves, corresponding to the same eigenvalue $E$, with unknown
amplitudes:

\begin{eqnarray}
\psi(x_1,x_2,x_3) & = &  A_{123} \, z_1^{x_1} z_2^{x_2} z_3^{x_3}
   +  A_{132}\, z_1^{x_1} z_3^{x_2} z_2^{x_3}
 +  A_{213} \, z_2^{x_1} z_1^{x_2} z_3^{x_3} \nonumber  \\    
\: & \: &  +   A_{231}\,  z_2^{x_1}  z_3^{x_2} z_1^{x_3}  +
   A_{312}\,  z_3^{x_1} z_1^{x_2} z_2^{x_3}  + A_{321}\,  z_3^{x_1} z_2^{x_2}  z_1^{x_3} 
\nonumber \\[12pt]
\:   &=&  \displaystyle \sum_{s \in S_3} A_{s}z_{s(1)}^{x_1} z_{s(2)}^{x_2} z_{s(3)}^{x_3}.
\label{3-Bethe}
\end{eqnarray}
The 6 amplitudes $A_{s}$ are uniquely and unambiguously
determined (up to an overall multiplicative constant) by the 2-body
collision constraints.  It is therefore absolutely crucial that
3-body collisions do not bring additional independent constraints that
the Bethe wave function could not satisfy.  We encourage the
reader to perform the calculations (which are very similar to the
$N=2$ case) of the amplitude ratios.

Finally, the Bethe roots $z_1$, $z_2$ and $z_3$ are quantized through
the periodicity condition
\begin{equation}
 \psi( x_1, x_2, x_3) =  \psi( x_2, x_3,  x_1 + L) \, , 
\label{3-Period}
\end{equation}
for $ 1 \le x_1 < x_2 < x_3 \leq L $.  This condition leads to the
Bethe Ansatz equations (the equations for general $N$ are given
below).

\hfill\break
\noindent {\bf The general case:} Finally, we briefly discuss the
general case $ N > 3$.  Here one can have $k$-body collisions with
$k=2,3,\ldots N$. However, all multi-body collisions `factorize' into
2-body collisions and ASEP can be diagonalized using the Bethe Wave
Function
 \begin{eqnarray}
\psi(x_1,x_2, \ldots, x_N) =   \sum_{s \in S_N} A_{s}  \,
  z_{s(1)}^{x_1} z_{s(2)}^{x_2} \cdots  z_{s(N)}^{x_N}  \, , 
\label{N-Bethe}
\end{eqnarray}
where $S_N$ is the permutation group of $N$ objects.  The $N!$
amplitudes $A_{s}$ are fixed (up to an overall multiplicative
constant) by the 2-body collisions constraints.  The corresponding
eigenvalue is given by
 \begin{equation}
  E  =  p  \sum_{i=1}^N  \frac{1}{z_i} 
  + q   \sum_{i=1}^N z_i   -N (p+q)  \, .
\label{N-Energy}
\end{equation}
The periodicity  condition 

\begin{equation}
\psi(x_1, x_2,\ldots, x_N) =  \psi(x_2, x_3, \ldots, x_N,  x_1 + L) \, ,
\label{N-Period}
\end{equation}
where $1 \leq x_1 < x_2 <\ldots < x_N \leq L$, leads to a set of
algebraic equations satisfied by the Bethe roots $z_1,
z_2,\ldots,z_N$.  The Bethe Ansatz equations are given by

\begin{equation}
        z_i^L = (-1)^{N-1} \prod_{j \neq i} \frac{ q z_i z_j - (p +q)
          z_i + p}{ q z_i z_j - (p +q) z_j + p} \, ,
 \label{EQ:BetheAnsatz}
\end{equation}
for $i=1, \ldots N$. The Bethe Ansatz thus provides us with a set of
$N$ coupled algebraic equations of degree of order $L$ (Recall that
the size of the matrix $\Bbb{M}$ is of order $2^L$, when $N \simeq
L/2$).  Although the degree of the polynomials involved are extremely
high, a large variety of methods have been developed to analyze them
\cite{Baxter,Gaudin,Langlands}.

We remark that for $p = q = 1$ the Bethe equations are the same as the
ones derived by H. Bethe in 1931. Indeed, the symmetric exclusion
process is mathematically equivalent to the isotropic Heisenberg
spin chain (although the two systems describe totally different
physical settings).

\subsection{Some applications of the Bethe Ansatz}
\label{BAapplications}

The Bethe Ansatz allows us to derive information about the spectrum of
the Markov matrix that governs the evolution of ASEP. Below, we review
some applications and refer to the original works.
 
\vspace{2mm}

\noindent {\it Structure of the Bethe wave function:} 

In the totally asymmetric case (TASEP), the Bethe equations simplify
and it is possible to perform analytical calculations even for finite
values of $L$ and $N$. Besides, the TASEP Bethe wave function takes
the form of a determinant:

\begin{equation}
  \psi(\xi_1,\dots,\xi_N) = \det(\Bbb{R})  \, , 
   \label{eq:psidet}
\end{equation}
where $\Bbb{R}$ is a $N \times N$ matrix with elements
\begin{equation}
   R_{ij} = \frac{z_{i}^{\xi_j}}{(1-z_i)^j}
  \ \ \mbox{for } 1 \le i,j \le N  \, ,
   \label{eq:r}
\end{equation}
$(z_1, \dots, z_N)$ being the Bethe roots. By expanding the
determinant, one recovers the generic form~(\ref{N-Bethe}) for the
Bethe wave function. The formulas~(\ref{eq:psidet}) and~(\ref{eq:r})
can be verified directly by proving that the eigenvalue
equation~(\ref{Eq:VPASEP}) and all the boundary conditions are
satisfied \cite{KMCairns,OGKMdet}. As we will see in the next
subsection, determinantal representations of the eigenvectors and of
the exact probability distribution at finite time play a very
important role in the study of the TASEP
\cite{Bogoliubov,SchutzBA,Priezzhev,PriezzhevInde}.

\vspace{2mm}

\noindent {\it Spectral gap and dynamical exponent:} 

The Bethe Ansatz allows us to determine the spectral gap which amounts
to calculating the eigenvalue $E_{1}$ with the largest real part.  For
a density $\rho = N/L$, one obtains for the TASEP

\begin{equation}
    E_1 = \underbrace{{-2 \sqrt{\rho( 1 - \rho)}
        \frac{6.509189337\ldots}{L^{3/2}}}}_{\textrm{relaxation}} \pm
    \underbrace{{\frac{ 2 i \pi (2 \rho
          -1)}{L}}}_{\textrm{oscillations}}.
\end{equation}
The first excited state consists of a pair of conjugate complex
numbers when $\rho$ is different from 1/2. The real part of $E_1$
describes the relaxation towards the stationary state: we find that
the largest relaxation time scales as $T \sim L^z$ with the dynamical
exponent $z=3/2$ \cite{VanBSpohn,Dhar,Gwa,Kim1,OGKMgap}.  This value
agrees with the dynamical exponent of the one-dimensional
Kardar-Parisi-Zhang equation that belongs to the same universality
class as ASEP (see the review of Halpin-Healy and Zhang 1995
\cite{HHZ}). The imaginary part of $E_1$ represents the relaxation
oscillations and scales as $L^{-1}$; these oscillations correspond to
a kinematic wave that propagates with the group velocity $2 \rho -1$;
such traveling waves can be probed by dynamical correlations
\cite{Majumdar,Barma1}.  The same procedure also allows us to classify
the higher excitations in the spectrum \cite{deGier2}.  For the
partially asymmetric case ($p, q > 0, p\neq q$), the Bethe equations do not
decouple and analytical results are much harder to obtain. A
systematic procedure for calculating the finite size corrections of
the upper region of the ASEP spectrum was developed by Doochul Kim
\cite{Kim1,Kim2}.

\vspace{2mm}
\noindent {\it Large deviations of the current:} 

The exclusion constraint modifies the transport properties through the
ASEP system. For instance, if one considers the ASEP on a ring and
tags an individual particle (without changing its dynamics), the
particle displays diffusive behavior in the long time limit, but with
a tracer diffusion coefficient ${\mathcal D}$ that depends on the size
of the system. It is different from the free, non-interacting particle
diffusion constant: ${\mathcal D}_{\rm ASEP} \neq {\mathcal D}_{\rm
  free}$.  In a ring of size $L$, $\mathcal{D}_{\rm ASEP}$ scales as
$L^{-1/2}$ in presence of an asymmetric driving. In the case of
symmetric exclusion, one has $\mathcal{D}_{\rm SEP} \sim L^{-1}$. These
scaling behaviors indicate that in the limit of an infinite system the
diffusion constant vanishes and fluctuations become anomalous
\cite{DeMasiFerrari}. In the classic result
\cite{Harris,Liggett1,PaulK} of the SEP on an infinite line, root-mean
square displacement of a tagged particle scales as $t^{1/4}$. In the
ASEP, subtleties associated with the initial conditions arise. For a
\textit{fixed initial condition}, the position of the tagged particle
spreads as $t^{1/3}$. However, if an average is carried out by
choosing random initial conditions (with respect to the
stationary measure), then a normal $t^{1/2}$ diffusion is
recovered. The latter is a trivial effect due to the local density
fluctuations that result from varying the initial condition; these
fluctuations completely overwhelm and mask the intrinsic dynamical
fluctuations. Another feature that one expects is a non-Gaussian
behavior, i.e., cumulants beyond the second are present.

An alternative observable that carries equivalent information (and is
amenable to analytical studies) is the total current transported
through the system. Consider the statistics of $Y_t$, the total
distance covered by all the particles between the time 0 and $t$. It
can be shown (see e.g., \cite{KMCairns} and references therein) that
in the long time limit the cumulant generating function of the random
variable $Y_t$ behaves as

\begin{equation}
  {  \left\langle  e^{\mu Y_t}   \right\rangle \simeq 
    e^{ E(\mu) t} }  \, . 
 \label{eq:limF}
\end{equation}
This equation implies that all the cumulants of $Y_t$ grow linearly
with time and can be determined by taking successive derivatives of
the function $E(\mu)$ at $\mu = 0$. Another way to characterize the
statistics of $Y_t$ is to consider the {\it large-deviation function}
of the current, defined as follows

\begin{equation} 
\textrm{Prob}\left(\frac{Y_{t}}{t}=j\right) \sim e^{-tG(j)} \, .\label{def:LDF}
\end{equation}
From equations (\ref{eq:limF},\ref{def:LDF}), we see that $G(j)$ and 
$E(\mu )$ are the \textit{Legendre transforms } of each other.
Large deviation functions
play an increasingly important role in non-equilibrium statistical
physics \cite{Touchette}, especially through application of the Fluctuation
Theorem \cite{LeboSpohn}. Thus, determining exact expressions for these
large deviation functions for interacting particle processes, either
analytically or numerically, is a major challenge in the field
\cite{DerrReview}. Moreover, higher cumulants of the current and large
deviations are also of experimental interest in relation to counting
statistics in quantum systems \cite{Flindt}.

The crucial step in the calculation of the cumulants is to identify
the generating function $E(\mu)$ as the dominant eigenvalue of a
matrix $\Bbb{M}(\mu)$, obtained by the following deformation of the original
Markov matrix $\Bbb{M}$:
\begin{equation}
  \Bbb{M}(\mu) =   \Bbb{M}_0 + e^\mu  \Bbb{M}_1 +e^{-\mu}  \Bbb{M}_{-1}, 
 \label{eq:defMmu}
\end{equation}
where $\Bbb{M}_{0}$ is the matrix of the diagonal of $\Bbb{M}$, and
$\Bbb{M}_{1}$ ($\Bbb{M}_{-1}$) is a matrix containing the transitions
rates of particle hopping in the positive (negative) direction. Hence,
the statistics of the transported mass has been transformed into an
eigenvalue problem.  The deformed matrix $\Bbb{M}(\mu)$ can be
diagonalized by the Bethe Ansatz by solving the following Bethe Ansatz
equations
\begin{equation}
  z_i^L  = (-1)^{N-1} \prod_{j =1}^N
   \frac{q e^{-\mu}  z_i z_j - (p+q) z_i + p e^{\mu}  }
  {qe^{-\mu}  z_i z_j -(p+q) z_j + pe^{\mu} }  \, .
\label{eq:BAmu}
\end{equation}

\noindent The corresponding eigenvalues are given by

\begin{equation}
 E(\mu; z_1, z_2 \ldots  z_N ) =   pe^{\mu}  \sum_{i =1}^N 
   \frac{1}{ z_i}  + 
        q e^{-\mu}  \sum_{i=1}^N   z_i \, - N (p+q)\, .
\label{eq:Emu}
\end{equation}
For $\mu =0$ we know that the largest eigenvalue is 0. For $\mu \neq
0$ the cumulant generating function corresponds to the continuation of
this largest eigenvalue (the existence of this continuation, at least
for small values of $\mu$ is guaranteed by the Perron-Frobenius
theorem).

We remark that equations~(\ref{eq:BAmu}) and (\ref{eq:Emu}) are
invariant under the transformation
 \begin{eqnarray}
                 z &\rightarrow& \frac{1}{z}  \nonumber  \\
                 \mu &\rightarrow& \mu_0- \mu
 \,\,\,\, \hbox{ with } \,\,\,\,   \mu_0 = \log\frac{q}{p}    \, .
  \end{eqnarray}
This symmetry implies that the spectra of $\Bbb{M}(\mu)$ and of $\Bbb{M}(\mu_0
-\mu)$ are identical.  This functional identity is in particular
satisfied by the largest eigenvalue of $\Bbb{M}$ and we have

\begin{equation}
  E(\mu)  =   E(\mu_0- \mu) \, . 
 \label{eq:GC}
  \end{equation}
Using the fact that the LDF $G(j)$ is the
Legendre transform of $E(\mu)$, we obtain the canonical form of the
{\it Fluctuation Theorem} (or Gallavotti-Cohen relation)
\begin{equation}
  G(j) =   G(-j) - \mu_0 j  \, . 
\end{equation} 
We observe that in the present context this relation manifests
itself as a symmetry of the Bethe equations. However, it is satisfied
by a large class of systems far from equilibrium
\cite{EvansCohenMorriss,EvansSearles,Gallavotti}.  The validity of the
Fluctuation Theorem does not rely on the integrability of the system
and it can be proved for Markovian or Langevin dynamical systems using
time-reversal symmetry \cite{Kurchan,LeboSpohn} as well as 
measure-theoretic considerations \cite{QIANBOOK,SHARGEL}.

\begin{figure}[ht]
\begin{center}
\includegraphics[width=5in]{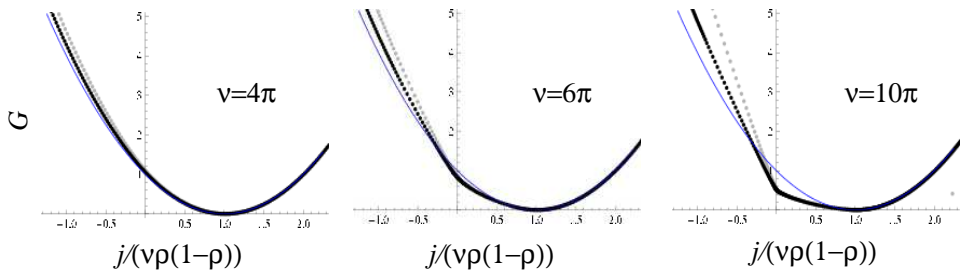}
\caption{Behaviour of the large deviation function $G$ as a function
  of the current $j/(\nu\rho(1-\rho))$ for different values of $\nu$.
  The gray dots correspond to $L=50, N=25$ and the black dots
  correspond to $L=100, N=50$. The curves are formally obtained by
  numerically solving the Bethe Ansatz equations (\ref{eq:BAmu}), and
  Legendre transforming $E(\mu)$. The thin blue curve represents the
  leading Gaussian behavior indicated by equation (\ref{LDFGaussian}).}
\label{Fig:LDF}
\end{center}
\end{figure}

The complete calculation of the current statistics in the ASEP is a
difficult problem that required more than a decade of effort.  The
breakthrough was the solution of the TASEP, by B. Derrida and
J. L. Lebowitz in 1998 \cite{DLeb}.  These authors obtained the
following parametric representation of the function $E(\mu)$ in terms
of an auxiliary parameter $B$:

\begin{equation}
E(\mu) \displaystyle = -N \sum_{k=1}^\infty \left(\begin{array}{c}
  kL-1 \\kN \end{array} \right) \frac{B^k}{kL-1} \, , \label{eq:EofY}
\end{equation}
\noindent where $B(\mu)$ is implicitly defined through 

\begin{equation}
\mu  \displaystyle = - \sum_{k=1}^\infty
\left(\begin{array}{c} kL\\kN \end{array} \right) \frac{B^k}{kL}.
        \label{eq:gamY}
\end{equation}
These expressions allow us to calculate the cumulants of $Y_t$, for
example the mean-current $J$ and a `diffusion constant' $D$:

\begin{equation}
\begin{array}{l}
\fl \displaystyle J =  \lim_{t \to \infty} \frac{\langle Y_t \rangle}{t} = 
 \frac{ {\rm d}  E(\mu)}{{\rm d} \mu} 
 \Big|_{\mu =0} =  \frac{ N(L-N)}{L-1}\, ,    \\[13pt]
\fl  \displaystyle D = \lim_{t \to \infty} 
  \frac{ \langle Y_t^2 \rangle - \langle Y_t \rangle^2 }{t} =  \frac{ {\rm d}^2  E(\mu)}{ {\rm d} \mu^2} 
 \Big|_{\mu =0} = \frac{N^2\; (2L-3)!\; (N-1)!^2 \;(L-N)!^2 }
{ (L-1)!^2 \; (2N-1)! \; (2L-2N-1)! } \, .
 \end{array}
\end{equation}
When $ L \to \infty$,  $\rho = L/N$, and $ |j -
L\rho (1 -\rho) | \ll L$, the large deviation function $G(j)$ can be
written in the following scaling form:
\begin{equation}
 G(j) =     \sqrt{ \frac{\rho (1 -\rho)} {\pi N^3}}
 H \Big(\frac{j -  L\rho (1 -\rho)}{\rho (1 -\rho)}   \Big) 
 \label{eq:scalf}
  \end{equation}
with 
\begin{eqnarray}
   H(y) \simeq  - \frac{ 2 \sqrt{3}}{5 \sqrt{\pi}} y ^{5/2} \,\,\,\,
 &\hbox{ for }&  \,\,\,\,   y \to +\infty \, , \\
     H(y) \simeq -  \frac{ 4 \sqrt{\pi}}{3} |y| ^{3/2} \,\,\,\,
 &\hbox{ for }&  \,\,\,\,   y \to -\infty \, . 
\end{eqnarray}
The shape of the large deviation function is skewed: it decays with a
$5/2$ power as $y\to +\infty$ and with a $3/2$ power as $y\to
-\infty$.  For the general case $q \neq 0$ on a ring, the solution was
found by rewriting the Bethe Ansatz as a functional equation and
restating it as a purely algebraic problem
\cite{Sylvain1,Sylvain2}. For example, this method allows us to
calculate the first two cumulants, $J$ and $D$:

\begin{equation}
\begin{array}{l}
\displaystyle J ={(p-q)\over p}\frac{N(L-N)}{L-1}\sim
     {(p-q)\over p}L\rho(1-\rho) \, \hbox{ for } \,\,
     L\rightarrow\infty  \, ,    \\[14pt]
\displaystyle D = {(p-q)\over p}
     \frac{2L}{L-1}\sum_{k>0} k^{2}
     \frac{C_{L}^{N+k}}{C_{L}^{N}}
     \frac{C_{L}^{N-k}}{C_{L}^{N}}
     \left(\frac{p^{k}+q^{k}}{p^{k}-q^{k}}\right),
 \end{array}
\end{equation}
where $C_{L}^{N}$ are the binomial coefficients. Note that the formula
for $D$ was previously derived using the matrix product representation
discussed above \cite{DMal}.  Higher cumulants can also be found and
their scaling behavior investigated.  For ASEP on a ring, the problem
was completely solved recently by S. Prolhac \cite{Sylvain4}, who
found a parametric representation analogous to
equations~(\ref{eq:EofY}) but in which the binomial coefficients are
replaced by combinatorial expressions enumerating some tree
structures.  A particularly interesting case \cite{Sylvain3} is a
weakly driven limit defined by ${q/p=1-\frac{\nu}{L}}$,
${L\rightarrow\infty}$.  In this case, we need to rescale the fugacity
parameter as ${\mu}/{L}$ and the following asymptotic formula for the
cumulant generating function can be derived

\begin{eqnarray}
\fl E\left(\frac{\mu}{L}, 1- \frac{\nu}{L} \right)\simeq
    \frac{\rho(1-\rho) (\mu^{2} + \mu\nu)}{L}
    -\frac{\rho(1-\rho)\mu^{2}\nu}{2L^{2}} + \frac{1}{L^{2}} \phi [
      \rho(1-\rho)(\mu^{2} + \mu\nu)] \, ,
 \label{eq:series1WASEP} \\
       \hbox{with}   \qquad  \phi(z) = \sum_{k=1}^{\infty} \frac{B_{2k-2}}{k!(k-1)!}z^k  \, ,
 \label{eq:series2WASEP}
\end{eqnarray}
and where ${B_{j}}$'s are Bernoulli Numbers. We observe that the
leading order (in ${1/L}$) is quadratic in $\mu$ and describes
Gaussian fluctuations. It is only in the subleading correction (in
$1/L^{2}$) that the non-Gaussian character arises.  This formula was
also obtained for the symmetric exclusion case $\nu = 0$ in
\cite{Appert}.  We observe that the series that defines the function
$\phi(z)$ has a finite radius of convergence and that $\phi(z)$ has a
singularity for $z = -\pi^2$.  Thus, non-analyticities appear in
${E}(\mu,\nu)$ as soon as $$ \nu \ge \nu_c = \frac{2\pi
}{\sqrt{\rho(1-\rho)}}.$$ By Legendre transform,
non-analyticities also occur in the large deviation function $G(j)$
defined in (\ref{def:LDF}).  At half-filling, the singularity appears
at $\nu_c = {4\pi }$ as can be seen in figure \ref{Fig:LDF}.  For $\nu
< \nu_c$ the leading behavior of $G(j)$ is quadratic (corresponding to
Gaussian fluctuations) and is given by

\begin{equation}
G(j) = \frac{( j -\nu \rho(1-\rho))^2}{4 L \rho(1-\rho) } \, .
\label{LDFGaussian}
\end{equation}
For $\nu > \nu_c$, the series expansions
(\ref{eq:series1WASEP},\ref{eq:series2WASEP}) 
break down and the LDF 
$G(j)$ becomes non-quadratic even at leading order. This phase
transition was predicted by T.  Bodineau and B. Derrida using
macroscopic fluctuation theory \cite{Bodineau,Bodineau1,Bodineau2}.
These authors studied the optimal density profile that corresponds to
the total current $j$ over a large time $t$.  They found that this
optimal profile is flat for $j < j_c$ but it becomes linearly unstable
for $j > j_c$. In fact, when $j > j_c$ the optimal profile is
time-dependent.  The critical value of the total current for which
this phase-transition occurs is $j_c =\rho(1-\rho) \sqrt{ \nu^2 -
  \nu_c^2}$ and therefore one must have $\nu \ge \nu_c$ for this
transition to occur.  One can observe in figure \ref{Fig:LDF} that for
$\nu \ge \nu_c$, the large deviation function $G(j)$ becomes
non-quadratic and develops a kink at a special value of the total
current $j$.

\vspace{2mm}

\noindent {\it Bethe Ansatz for other systems out of equilibrium:} 

We have discussed in this review the Bethe Ansatz for a system on a
periodic ring where the total number of particles is a conserved
quantity. It is possible to extend the Bethe Ansatz for the finite
ASEP with open boundaries in which the total number of particles is
not constant. The resulting Bethe equations have a more complicated
structure; they have been thoroughly analyzed in a series of papers by
J. de Gier and F. Essler \cite{deGier1,deGier2,deGier3} who calculated
the spectral gaps in the different phases of the open ASEP. In
particular, they discovered sub-phases in the phase diagram that could
not be predicted from the steady state measure alone.  We note that
the Bethe Ansatz can be applied to variants of the ASEP, such as
models with impurities \cite{evans}, multispecies exclusion processes
\cite{Alcaraz,Cantini,Chikashi} as the zero-range process
\cite{Povolotsky}, the raise and peel model \cite{AlcarazRitt},
vicious walkers \cite{Dorlas}, or the Bernoulli matching model of
sequence alignment \cite{MMS,PriezSchutz}.
  
\subsection{ASEP on an infinite lattice: Bethe Ansatz and random matrices} 
\label{Infinite}

In this last subsection, we briefly review some important analytical
results that have been obtained for exclusion processes on an infinite
line, especially in connection with the Bethe Ansatz method discussed
above.  this approach will allow us to derive some insight into the
dyanmics of the systems. More detailed results and historical
discussions can be found in the literature cited in the text.

We first consider the case of the ASEP on the infinite lattice
$\mathbb{Z}$ but with only a {\it finite} number $N$ of particles.
Because the particles cannot overtake one another, the ordering of
the particles is conserved by the dynamics, and we can label the
particles from right to left. Suppose that at $t=0$, the particles are
located at positions $y_1> \ldots> y_N$ and at some later time $t$,
they are located at $x_1> \ldots >x_N$.

The transition probability, $P(x_1, \ldots, x_N, t | y_1, \ldots,
y_N,0),$ for reaching the final configuration $x_1, \ldots, x_N$ at $t$
starting from $y_1, \ldots, y_N$ at $t=0$ has the following exact
expression:

\begin{eqnarray}
\fl P(x_1, \ldots, x_N, t| y_1, \ldots, y_N, 0) = 
\!\!\sum_{s \in S_N} \prod_{k=1}^N \oint_{C_R} \frac{d z_k}{2 \pi i z_k}
 \, A_{s}(\{z\}) e^{(\frac{p}{z_k} + q  z_k -1)t}
 \,   z_{s(k)}^{x_k - y_{s(k)}}\!\!\!\!\!. 
 \label{TracyWidomASEP}
\end{eqnarray}
The crucial observation is that there exists a closed formula for the
amplitude $A_{s}$, given by

\begin{eqnarray}
A_{s}(\{z\}) = {\rm sgn}(s) 
\frac
{\prod_{i<j}\left(p + q z_{s(i)}z_{s(j)} - z_{s(i)}\right) }
{\prod_{i<j}\left(p + q z_i z_j -  z_i \right)},
 \label{TracyWidomAmplitude}
\end{eqnarray}
where we use the convention $p +q = 1$.  The
expressions~(\ref{TracyWidomASEP}) and (\ref{TracyWidomAmplitude}) are
reminiscent of the formulae given by the Bethe Ansatz. These results
were initially derived for the TASEP on $\mathbb{Z}$ by G. Sch\"utz in
1997, who constructed them from the Bethe Ansatz \cite{SchutzBA} and
then proved rigorously that equation~(\ref{TracyWidomASEP}) solves
exactly the time-dependent Markov equation with the correct initial
condition (generalized to the periodic case by
V. Priezzhev in \cite{Priezzhev,PriezzhevInde}). The general result
for the ASEP was found by Tracy and Widom in 2008 \cite{Tracy1} and
has motivated many studies in the last few years.
Equation~(\ref{TracyWidomASEP}) is a seminal result: it is an exact
expression for the Green function of the ASEP from which, in
principle, individual particle distributions and correlations
functions can be extracted (this can be an extremely difficult task in
practice).  Finally, we emphasize that the `Bethe roots' $z_i$ are not
quantified in the infinite system: each of them takes a continuous set
of values along the circular contour $C_R$ of radius $R$, so small
that the poles of $A_{s}(\{z\})$ lie outside $C_R$.

To give one specific example, consider the {\it totally} asymmetric
exclusion with unit hopping rate $p=1$ (and $q=0$) and with a {\it
  step initial condition}, where all sites right of the origin ($i >
0$) are empty and all to the left ($i \leq 0$) are occupied.  If we
are interested in the behaviour of {\it only} the right most $N$
particles after time $t$, then we can replace the semi-infinite string
by a finite segment of $N$ particles. The point is, in a TASEP, none
of the particles hops to the left and so, particles to the left of our
$N$-particle string cannot affect their behaviour.  Thus, it is
sufficient to consider $P(x_1, \ldots, x_N, t | y_1 =0, \ldots,
y_N=1-N, 0)$.  Now, suppose we ask a more restricted question: What is
$\tilde{P}(M,N,t)$, the probability that the $N$-th particle
(initially located at $i = 1-N$, has performed at least $M$ hops by
time $t$?  In terms of $P$, we write

\begin{equation}
\fl \hspace{1cm} \tilde{P}(M,N,t) = \!\!\!\!\!\sum_{x_1>\ldots > x_N > M-N}
\!\!\!\!\!P(x_1, \ldots, x_N, t | y_1 =0, \ldots, y_n=1-N, 0)
\end{equation}
We next exploit a determinantal representation,
analogous to (\ref{eq:psidet}) and (\ref{eq:r}) for $\psi$,
and express $P$ as:

\begin{equation}
P(x_1, \ldots, x_N, t | y_1, \ldots, y_N, 0) =
   \det({\Bbb{B}})  \, ,
   \label{eq:Detexpression}
\end{equation}
where ${\Bbb{B}}$ is a $N \times N$ matrix with elements

\begin{equation}
B_{ij} = \oint_{C_R}  \frac{\dd z}{2\pi i z}
z^{x_i - y_j} (1-z)^{j-i}  e^{(\frac{1}{z}  -1)t}
\quad \mbox{for}\quad 1 \le i,j \leq N  \, ,
   \label{eq:gdR}
\end{equation}
After some calculation \cite{Rakos}, these equations allow us to express
$\tilde{P}(M,N,t)$ compactly:

\begin{equation}
\tilde{P}(M,N,t) = \frac{1}{Z_{M,N}} \int_{[0,t]^N}\!\!\!\!\dd^N x \!\!\!\! \prod_{1\le i<j \leq N}
 (x_i - x_j)^{2} \prod_{j=1}^N x_j^{M-N} e^{-x_j}.
\end{equation}
where $Z_{M,N}$ is a normalization factor. The integral on the right
hand side has a direct interpretation in random matrix theory: it is
the probability that the largest eigenvalue of the random matrix
$\Bbb{A} \Bbb{A}^{\dagger}$ is less than or equal to $t$, with
$\Bbb{A}$ being a $N \times M$ matrix of complex random variables with
zero mean and variance 1/2 (unitary Laguerre random matrix ensemble).

Finally, we note that $\tilde{P}(N,N,t)$ can also be interpreted as
the probability that the integrated current $Q_t$ through the bond
(0,1) is at least equal to $N$ ($Q_t$ is the number of
particles that have crossed the bond (0,1) between time 0 and
$t$). From the classical Tracy-Widom laws for the distribution of the
largest eigenvalue of a random matrix \cite{TracyWidom}, one can write

\begin{equation}
Q_t = \frac{t}{4} + \frac{t^{1/3}}{2^{4/3}} \chi \,.
\end{equation}
where the  distribution of the random variable $\chi$ is given by

\begin{equation}
 {\rm Prob}\left(\chi \leq s \right) = 1 - F_2(-s) \,.
\end{equation}
The Tracy-Widom function $F_2(s)$ is the cumulative distribution of
the maximum eigenvalue $\lambda_{max}$ in the Gaussian Unitary
Ensemble (self-adjoint Hermitian matrices) {\it i.e.}

\begin{equation} 
   {\rm Prob}\left( \frac{\lambda_{\rm max} - \sqrt{2N}}{ (8N)^{-1/6}} \leq s
    \right)  =  F_2(s) \, ,
\end{equation}
where $N \gg 1$ represents here the size of the random matrix.  Exact
expressions for $F_2(s)$ can be found, for example, in
\cite{Kriecherbauer}.  This crucial relation between random matrix
theory and the asymmetric exclusion process, first established by
K. Johansson in 2000 \cite{Johansson}, has stimulated many works in
statistical mechanics and probability theory (see e.g.,
\cite{SpohnInde,SpohnCours,FerrariPatrick}).  The mathematical study
of the infinite system has grown into a subfield {\it per se} that
displays fascinating connections with random matrix theory,
combinatorics and representation theory.  We note that K. Johansson
did not use the Bethe Ansatz in his original work.  He studied a
discrete-time version of the TASEP in which the trajectories of the
particles were encoded in a waiting-time matrix, which specifies how
long a given particle stays at each given site. This integer-valued
matrix can be mapped via the Robinson-Schensted-Knuth correspondence
into a Young Tableau.  The value of the total current through the bond
(0,1) is linearly related to the length of the longest line of this
Young Tableau.  The statistics of the length of the longest line can
be found by using asymptotic analysis techniques {\it \`a la}
Tracy-Widom. In fact, a closely related question, the classical Ulam
problem of the longest increasing subsequence in a random permutation,
was solved a few years earlier by J. Baik, K.  Deift and K.
Johansson, using related methods \cite{Baik}.

Very recently, in a series of articles
\cite{Tracy1,Tracy2,Tracy3,Tracy4,Tracy5}, C. A. Tracy and H. Widom
have generalized Johansson's results to the partially asymmetric
exclusion process by deriving some integral formulas for the
probability distribution of an individual particle starting from the
step initial condition.  These expressions can be rewritten as a
single integral involving a Fredholm determinant that is amenable to
asymptotic analysis.  In particular, a limit theorem is proved for the
total current distribution.  Going from TASEP to ASEP is a crucial
progress because the weakly asymmetric process leads to a well-defined
continuous limit: the Kardar-Parisi-Zhang (KPZ) equation, a universal
model for surface growth. Indeed, a very important outgrowth of these
studies is the exact solution of the KPZ equation in one dimension.
The distribution of the height of the surface at any finite time is
now known exactly (starting from some special initial conditions),
solving a problem that remained open for 25 years: a description of
the historical development of these results and the contributions of
the various groups can be found in
\cite{SpohnCAIRNS,FerrariPatrick,SasamotoSpohnPRL,
  AmirCorwinQuastel,CLDR,Dotsenko010,ProlhacSpohnPRE,CalabDoussal,Imamura}.
One important feature to keep in mind is that the results (and
therefore the universality class) depend strongly on the chosen
initial conditions.  Lately, $n$-point correlation functions of the
height in KPZ have also been exactly derived by H. Spohn and
S. Prolhac \cite{SylvainMunich1,SylvainMunich2}.  For an overview of
these fascinating problems, we recommend the article by T. Kriecherbauer and J. Krug \cite{Kriecherbauer}, and the reviews by D.  Aldous and P. Diaconis \cite{AldousDiaconis} and I. Corwin
\cite{ReviewCorwin}.


\subsection{Hydrodynamic mean field approach}
\label{HMFT}

Though elegant and rigorous, the methods presented above cannot be
applied to systems much more complex than the TASEP. Typically, further
progress relies on a very successful approach, based on the mean field
approximation and a continuum limit, leading to PDE's for various
densities in the system. Known as the hydrodynamic limit, such
equations can be `derived' from the master equation
(\ref{Eq:Markov}). The strategy starts with the exact evolution
equation for $\rho_{i}(t) \equiv \langle \sigma_i \rangle_t =
\sum_{\cal C} \sigma_i P\left({\cal C},t \right) $. Differentiating
this $\rho_{i}(t)$ with respect to $t$ and using equation
(\ref{Eq:Markov}), we see that new operators appear:

\begin{equation}
{\cal O} ({\cal C}^{\prime}) = \sum_{\cal C} \sigma_i ({\cal C}) M({\cal C},
{\cal C}^{\prime}).
\end{equation}
In the case of ASEP, $M({\cal C},{\cal C}^{\prime })$ is
sufficiently simple that the sum over ${\cal C}$ can be easily
performed, leaving us with products of $\sigma$'s (associated with
configurations ${\cal C}^{\prime }$). Applying the mean field
approximation (equation (\ref{MF-2spins})), the right-hand side now consists
of products of $\rho_{i}(t)$ and $\rho_{i \pm 1}(t)$.

Taking the thermodynamic and continuum limit, we let $i,L \to \infty$
with finite $x = i/L$ and write $\rho_{i}(t) \simeq \rho (x,t)$. Next,
we expand $\rho_{i \pm 1}(t) \simeq \rho(x,t) \pm \varepsilon \partial_{x}\rho
 + (\varepsilon^{2}/2)\partial_{x}^{2}\rho + \ldots$, where $\varepsilon \equiv 1/L
\to 0$. The result is a hydrodynamic PDE for the particle density
\cite{LAKATOS2006}:

\begin{equation}
\fl {\partial \rho(x,t) \over \partial t} = \ve {\partial \over \partial
  x}\big[(q-p)\rho(1-\rho)\big] + {\ve^{2}\over 2}{\partial \over
  \partial x}\left[(1-\rho){\partial \over \partial
    x}\left[(p+q)\rho\right] + (p+q)\rho {\partial \rho \over \partial
    x}\right]. 
\label{INTERIOR}
\end{equation}
where $\varepsilon \equiv 1/L \to 0$. Note that even slow variations
in the hopping rates ($p_{i}\to p(x)$ and $q_{i}\to q(x)$) can be
straightforwardly incorporated \cite{SMOOTHPQ}. Analogous hydrodynamic
equations have also been derived for ASEPs with particles that occupy
more than one lattice site
\cite{SHAW2003,SCHOENHERR2004,SCHOENHERR2005}.  In addition to
spatially slowly varying hopping rates, this approach is useful for
exploring more complex models, e.g., a TASEP with particles that can
desorb and/or attach to every site (Section \ref{SEC:MOTORS}). Of
course, such equations are also the starting point for very successful
field theoretic approaches to dynamic critical phenomena
near-equilibrium \cite{DDBZ75,HHM74,HHM76,BJW76} as well as ASEPs
(even with interacting particles) in higher dimensions
\cite{JS86a,JS86b,LC86,JO96,OJ98,BJ99}.  Supplemented with noise
terms, these become Langevin equations for the density or
mangetisation fields. Then, fluctuations and correlations can be
computed within, typically, a perturbative framework.

Returning to equation (\ref{INTERIOR}), it is especially easy to analyse
in the steady-state limit where the resulting equation is an ODE in
$x$. But, the highest power of $\varepsilon$ multiplies the largest
derivative, so that we are dealing with `singularly perturbed'
differential equations \cite{LAKATOS2006,FREY2003,FREY2004}.  The
solutions for the steady-state mean field density $\rho(x)$ consist of
`outer solutions' that hold in the bulk, matched to `inner solutions'
corresponding to boundary layers of thickness $\varepsilon$ at the
open ends. For the ASEP, this approach correctly determines the phase
boundaries and the self-consistently determined steady-state current
$J$ (which arise both in the integration constant of equation
(\ref{INTERIOR}) and in the boundary conditions).  However, as
anticipated, the density profiles are not exactly determined, even
when $L\to \infty$. Nevertheless, for more realistic physical models
such as the ones to be discussed in the next section, such a
mean-field approach (or some of its variants) is often the only
available analytical method at our disposal.


\section{Biological and related applications of exclusion processes}
\label{TC}

In this section, we review a number of applications of stochastic
exclusion models to problems of transport in materials science, cell
biology, and biophysics. While quantitative models of these real-world
applications often require consideration of many microscopic details
(and their corresponding parameters), simple one-dimensional lattice
models nonetheless can be used to capture the dominant mechanisms at
play, providing a succint representation of the system. Moreover, these
types of models can be extended in a number of ways, and are amenable
to concise analytic solutions. As we will see, application of lattice
models to complex biophysical systems is aided by a few main
extensions and modifications to the TASEP presented above. These
modified dynamics are illustrated respectively, by figures
\ref{LARGE}-\ref{FB0}.

\begin{enumerate}

\item {\bf Longer-ranged interactions:} Objects represented by
  particles in an exclusion process may have molecular structure that
  carry longer-ranged particle-particle interactions. For example, to
  model ribosomes on an mRNA or cargo-carrying molecular motors, we
  should use large particles taking small discrete steps and introduce a rule beyond on-site exclusion. We may regard these as `extended particles' that occlude $\ell >1$ lattice sites 
(figure \ref{LARGE}).

\begin{figure}[h]
\begin{center}
\includegraphics[width=3.4in]{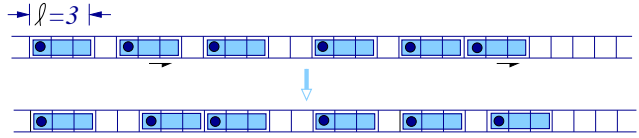}
\end{center}
\caption{An interior section of an asymmetric exclusion process 
with extended particles of size $\ell =3$. The individual particles 
occlude three of the lattice sites on which the hops occur.}
\label{LARGE}
\end{figure}

\item  {\bf Particle detachment and attachment:} Particles such
as molecular motors have finite processivity. That is, they can
spontaneously detach from their lattices before reaching their
destination. Conversely, particles in a bulk reservoir can also attach at random positions on the lattice. The coupling of the particle number to a bulk reservoir is analogous to the problem of Langmuir kinetics \cite{FREY2004,FREY2006} on a one-dimensional lattice, except that the particles are directionally driven on the substrate (figure \ref{DETACH}).

\begin{figure}[h]
\begin{center}
\includegraphics[width=3.1in]{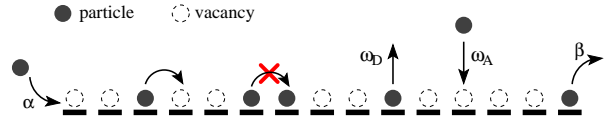}
\end{center}
\caption{A TASEP with Langmuir kinetics where particles can
  spontaneously detach and attach at every site with rates $\omega_{\rm
    D}$ and $\omega_{\rm A}$, respectively. Adapted from \cite{FREY2004G}.}
\label{DETACH}
\end{figure}

\item {\bf Multiple species:} Usually, biological transport
involves multiple interacting species in confined geometries, often one-dimensional in nature.  For example, these secondary species may represent particles that are co-transported or counter-transported with the primary particles, or, they may represent species that bind to pores and regulate primary particle transport \cite{WOOD2009}(figure \ref{TWO_SPECIES}).

\begin{figure}[h]
\begin{center}
\includegraphics[width=3.2in]{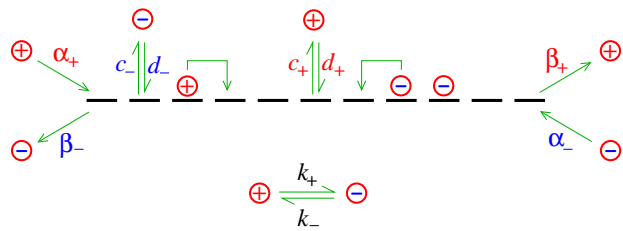}
\end{center}
\caption{A two-species exclusion model, adapted from
  \cite{MUHURI2008}, where $+$ and $-$ particles move in opposite
  directions but can interconvert with rate $k_{\pm}$. Attachment and
  detachment are also allowed with rates $c_{\pm}$ and $d_{\pm}$, respectively.}
\label{TWO_SPECIES}
\end{figure}

\item {\bf Partial exclusion \& coupled chains:} Often, the
strict one-dimensional nature of the exclusion dynamics can be
relaxed to account for particles that, while strongly repelling each other, can on occasion pass over each other.  This scenario might arise when pores are wide enough to allow particle passing.  Related extensions of single-chain exclusion processes are exclusion processes across multiple interacting chains (figure \ref{TWO_CHANNEL}).

\begin{figure}[h]
\begin{center}
\includegraphics[width=3.1in]{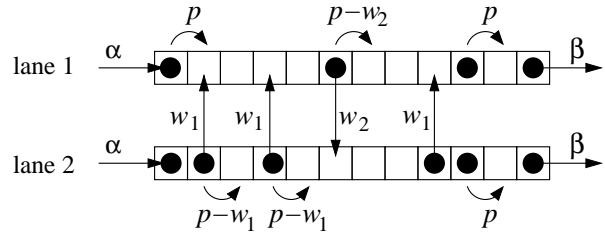}
\end{center}
\caption{Two coupled TASEP lattices with interchain hopping rates $w_{1}$ and $w_{2}$.
Adapted from \cite{PRONINA2006}.}
\label{TWO_CHANNEL}
\end{figure}

\item {\bf Internal states and nonexponentially distributed dwell times:} The physical mechanisms of how excluding particles are inserted, extracted, and hop in the interior of the lattice are often complex, involving many intermediate chemical steps 
\cite{INTERNAL2006,GARAI2009}. Therefore, the distribution of times between successive hops, even when unimpeded by exclusion interactions, is often non-exponential. Specific hopping time distributions can be incorporated into lattice particle models by explicitly evolving internal particle states (figure \ref{INTERNAL}).

\begin{figure}[h]
\begin{center}
\includegraphics[width=3.1in]{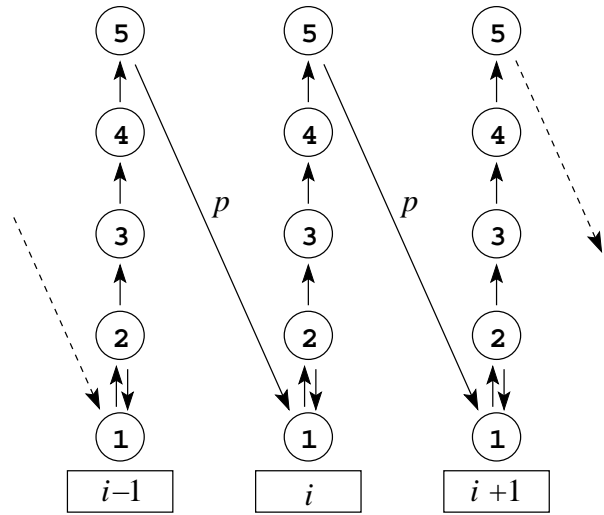}
\end{center}
\caption{Internal states (1-5) that determine the timing between particle 
  hops (adapted from \cite{GARAI2009}). Here, a linear sequence of
  Poisson processes generates a Gamma-distributed \cite{PAPOULIS}
  interhopping time distribution (dwell time). This model for the
  internal dynamics has been used to model mRNA translation by
  ribosome enzymes.}
\label{INTERNAL}
\end{figure}

\item {\bf Variable lattice lengths:} The lattices on which the
  exclusion processes occur may not be fixed in some settings. A
  dynamically varying length can arise in systems where growth of the
  lattice is coupled to the transport of particles to the
  dynamically-varying end of the lattice
  \cite{EVANS2007,READ2007,SUGDEN2007,HOUGH2009}.  While no exact
  solutions have been found, mean-field and hydrodynamic approaches
  have been successfully applied \cite{NOWAK2007}. In the continuum
  limit, this problem is analogous to the classic Stefan problem
  \cite{NOWAK2007,FOK2009}, except with nonlinear particle density
  dynamics (figure \ref{FB0}).

\begin{figure}[h]
\begin{center}
\includegraphics[width=4in]{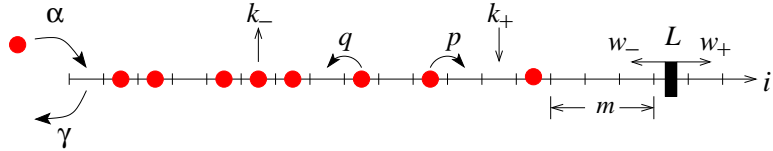}
\end{center}
\caption{Schematic of an ASEP (adapted from \cite{NOWAK2007}) with
  detachment and attachment rates $k_{-}$ and $k_{+}$, respectively, 
and a `confining' wall at site $L$. The wall has
  intrinsic dynamics described by hopping rates $w_{\pm}$.}
\label{FB0}
\end{figure}

\end{enumerate}

Not surprisingly, the techniques presented in the previous section are
typically not powerful enough for obtaining exact results for these
variants of the basic exclusion processes. Of course, for sufficiently
small systems (e.g., $L\lesssim 10$), high performance computers can
be used to diagonalize $\Bbb{M}$, but that method is not viable for
many physical or biological systems. There are also some cases,
including special cases of two-species exclusion and ASEP with
extended particles on a ring (to be discussed below, within their own
contexts), where exact solutions can be found for all $L$. Despite the
lack of progress in finding analytic results, there are two other
approaches which provide valuable insights into these more complex yet
more realistic systems. One is computer simulations, exploiting Monte
Carlo techniques on our lattice models. Once the stochastic rules of a
model are specified, then, this technique corresponds to implementing
equation (\ref{me}). In addition, molecular dynamics simulations with
off-lattice systems also proved to be very successful. The other
approach is based on designing good approximation schemes for
attacking exact equations (or expressions for specific
quantities). Mean field theory (MFT), or mean field
\emph{approximation}, is most often used. An example is equation
(\ref{MF-2spins}) for the TASEP.  However, systematically improving
these approximations it is not generally straightforward. Thus, we
will see that, for TASEP with extended objects, the substitution
(\ref{MF-2spins}) is very poor (when compared to simulation
results). Instead, a more sophisticated scheme, implemented by Gibbs,
\emph{et.al.} \cite{MGP68,MG69}, is much more successful at
predicting, say, the average current. In turn, when this scheme fails
to predict other quantities, another level of approximation
\cite{SETHNA2004} (in the spirit of the Bethe-Williams approximation
for the Ising model) is designed. Nevertheless, if the predictions are
in good agreement with simulations, the MFT provides some confidence
that we are `on the right track' towards our goal: understanding these
generalized ASEPs which have found wide applicability in modeling
complex biological processes.

\subsection{Pore Transport} 

Transport of atoms and small molecules arises in both man-made
structures and cell biological systems.  Materials such as zeolites
form networks of one-dimensional channels within which molecules, such
as light hydrocarbons, can pass through and/or react. A number of
authors have used exclusion processes as simple descriptions of
single-file or near single-file transport. It has long been known that
tracer diffusion in single-file channels follows subdiffusive dynamics
\cite{LEVITT1973,ALEXANDER1978,KARGER1992,BARKAI2010,PaulK}.  A
mean-square displacement of the form $\langle x^{2}\rangle \sim \sqrt{t}$
is found by considering equilibrium fluctuations of the density
around the tagged particle in an infinite system.  For steady-state
particle transport across finite pores, the lattice exclusion process
has proved a more useful starting point. Given the frequent
application of exclusion processes to pore transport, two important
points should be stressed.

First, the standard lattice exclusion process assumes that waiting
times are exponentially distributed and, other than lattice site
exclusion, are independent. This assumption can fail if, say, an
attractive interaction between particles is comparable to the
interaction between substrate (e.g., atoms that make up the pore
walls) and the particles. Here, concerted motion can arise and has
been shown to be important in particle transport. For example, Sholl
and Fichthorn have shown using molecular dynamics (MD) simulations how
concerted motions of clusters of water affects its overall transport
in molecular sieves \cite{SHOLL1997}.  Similarly, Sholl and Lee also
showed that concerted motions of clusters of CF$_{\rm 4}$ and Xe in
AlPO$_{4}$-5 and AlPO$_{4}$-31 zeolites, respectively, contribute
significantly to their overall mobilities \cite{SHOLL2000}. Concerted
motion has also been predicted to occur in transport across carbon
nanotubes \cite{LITHIUM}.  These concerted motions arise from
frustration due to a mismatch between particle-particle and
particle-pore interaction ranges, allowing for a lower barrier of
motion for bound pair of particles that for an isolated, individual
particle. Although concerted motion arises from geometrically
complicated arrangements of particles that form low energy pathways in
the high dimensional energy landscapes, if a small number of these
pathways support most of the probability flux, simplifying assumptions
can be made.  For example, coarse-grained treatments of
concerted motion were developed by Sholl and Lee \cite{SHOLL2000}.
Concerted motion has also been implemented within lattice models of
exclusion processes in a more draconian manner. Gabel, Krapivsky, and
Redner have formulated a `faciliated exclusion' process on a ring
whereby a particle hops forward to an empty site {\it only} if the
site behind it is also occupied \cite{GABEL2010}.  They find a maximal
current of $3-2\sqrt{2}$ which is less than the maximal current of
$1/4$ arising in the standard TASEP. A model that incorporates
concerted motion might be described by a facilitated exclusion model
where motion in either direction occurs faster if the particle is
adjacent to another one. The hopping rate might also be increased if
an isolated particle moves to a site that results in it having an
additional adjacent particle. If these accelerated steps occur much
faster than individual particle hops, then the dynamics will resemble
motion of pairs of particles.  It would be interesting to determine
how this model of near-concerted hops increase the overall particle
flux.

\begin{figure}[h]
\begin{center}
\includegraphics[width=2.9in]{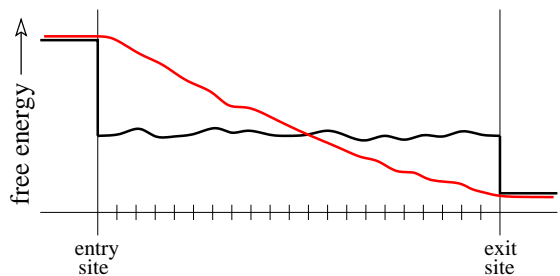}
\end{center}
\caption{Representative single-particle, one-dimensional free energy
  landscapes over which excluding particles travel.  The black curve
  free energy profile could represent a symmetric exclusion process,
  while the red landscape would describe an asymmetric exclusion
  processes. In the former, detailed balance is assumed to be violated
  at the entrance and exit sites and the particle flux is driven by
  differences between the chemical potentials of the two reservoirs.
}
\label{PROFILE}
\end{figure}

A second important point to emphasize is that different physical
systems are best modeled with varying degrees of asymmetry in the
exclusion processes.  Two extreme limits are the totally asymmetric
process, where particles only hop in one direction, and symmetric
exclusion, where particle hopping between any two adjacent sites obeys
detail-balance. In this case, detailed balance is violated only at the
two ends of the lattice. Net particle current arises only when there
exists an asymmetry in the injection and extraction rates at the two
ends. This latter scenario corresponds to boundary-driven exclusion
processes where differences in the chemical potential of the particles
in the two reservoirs drive the flux. Osmotic and pressure-driven
flows (when local equilibrium thermodynamics holds and particle
inertia is negligible) are examples of processes best described using
{\it symmetric} exclusion processes \cite{CHOU1998}. However, when
particles in the lattice are charged, and an external field is
applied, the internal jumps are asymmetric since there is a direct
force acting on the particles, breaking detailed balance.  A heuristic
delineation between asymmetric and symmetric (boundary-driven)
exclusion processes can be motivated by considering the {\it
  single-particle} free energy profiles.  Figure \ref{PROFILE} depicts
two hypothetical single-particle free energy profiles experienced by an isolated
particle under local thermodynamic equilibrium.

In biological systems, channels that transport water and neutral
(e.g., sugars) or charge-screened molecules, can be described by
symmetric exclusion, while motion of charged ions in single-file
channels would be best characterized by partially asymmetric
exclusions.  Modeling uncharged particles allows equal internal
hopping rates ($p = q$) with drive originating only at the boundaries
The steady-state current is easily found and is a function of the
asymmetry in the boundary injection and extraction rates
\cite{CHOU1998,KOLO_MEMBRANE}

\begin{equation}
J = {p(\alpha \beta -\gamma \delta) \over (L-1)(\alpha + \delta)(\beta + \gamma) 
+ p(\alpha +\beta +\gamma + \delta)}.
\end{equation}
The boundary injection and extraction rates are those defined in
figure \ref{fig:ASEPgeneral}(c). As expected, this expression for $J$ is
similar to that arising from simple diffusive flux in the $L\to \infty
$ limit.  However, as discussed, nontrivial current fluctuations, which cannot be
accounted for by simple diffusion, have also been worked out
\cite{Sylvain1,Sylvain2,Sylvain4,ROCHE,SANTOS}.

When particles are `charged', the level of asymmetry (the relative
difference between an ion hopping forward and hopping backwards) is
controlled by the magnitude of the externally applied `electric
field'.  Besides the exact solutions for the steady-state current
\cite{AlcarazRitt}, cumulants of the current in a weakly asymmetric
exclusion process have also been derived \cite{Sylvain3}, the slowest
dynamic relaxation mode computed \cite{deGier3}, and the tracer
diffusivity derived \cite{DMal}.

Specific biological realisations of driven transport include
ion transport across ion channels \cite{CHOU1998}, while transport
across nuclear pore complexes \cite{ZILMANNATURE,ZILMANPLOSCB2007},
and osmosis \cite{CHOU1999} are typically boundary-driven, or
ratcheted (see the section on molecular motors).

\begin{figure}[h]
\begin{center}
\includegraphics[width=5.1in]{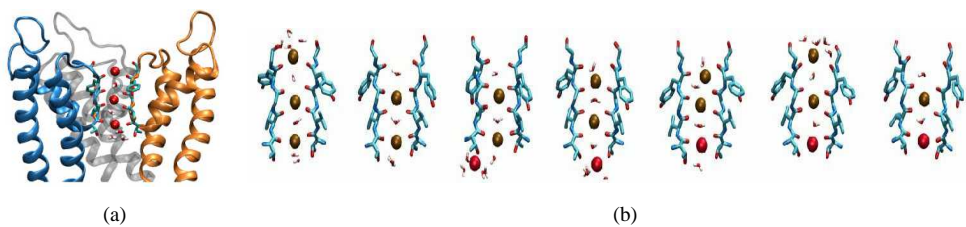}
\end{center}
\caption{(a) A ribbon figure of the Kv1.2 voltage-gated potassium ion
  channel embedded in a model lipid membrane. The pore structure
  clearly shows three dominant interior binding sites.  This potassium
  channel image was made with VMD and is owned by the Theoretical and
  Computational Biophysics Group, NIH Resource for Macromolecular
  Modeling and Bioinformatics, at the Beckman Institute, University of
  Illinois at Urbana-Champaign.  (b) A time series of a molecular
  dynamics simulation suggesting a concerted `knock-on' mechanism.
  Image adapted from \cite{SCHULTEN} with original labels removed, and
  used with their permission.}
\label{KCHANNEL}
\end{figure}

In 1998, the X-ray crystal structure of the K$^{+}$ ion channel was published~\footnote{This achievement was partially responsible for Peter Agre and Roderick McKinnonin being awarded the Nobel Prize in chemistry in 2003, ``for discoveries concerning channels in cell membranes."}, 
allowing a more detailed mechanistic understanding of the ion conduction and selectivity across a wide class of related, and 
biologically important, ion channels \cite{MACKINNON1998}. The crystal
structure showed three approximately in-line vestibules that can each
hold one K$^{+}$ ion. Representations of the voltage-gated Kv1.2
channel are shown in figure \ref{KCHANNEL}. Ion conduction can thus be
modeled by a simple three-site partially asymmetric exclusion process
provided the hopping rates at each occupation configuration are
physically motivated \cite{CHOU1999,ZILMAN2009,GWAN2007}. However,
note that there is evidence that ion dynamics within some ion channels
are `concerted' because the free energy barriers between ion binding
sites are small and ions entering an occupied channel can knock the
terminal ion out of the channel.  This `knock-on' mechanism
\cite{KNOCKON} has been motivated by molecular dynamics simulations
\cite{ROUX2001,SCHULTEN} and studied theoretically via a reduced
one-dimensional dynamical model \cite{KNOCKON2005}.

The symmetric exclusion process has also been used to consider osmotic
flow across membrane-spanning, single-file channels \cite{CHOU1999}.
Here, the solvent flux is driven by differences in the injection
rates at the two pore ends connected to two separate particle
reservoirs. Simple osmosis across a strictly semipermeable membrane
can be described by a single species symmetric exclusion process where
the injection rate of the solvent from one of the reservoirs is
reduced due to its smaller mole fraction arising from the presence of
solute in that reservoir \cite{CHOU1999}. The solvent current is
approximately linearly proportional to the difference in injection
rates at the two ends of the lattice, inversely proportional to the
length of the lattice, but with a weak suppression due to multiple
pore occupancy \cite{CHOU1998}.

Another transport channel in cells are nuclear pore complexes (NPC),
responsible for the selective shuttling of large molecules and
proteins (such as transcription factors, histones, ribosomal subunits)
through the double nuclear membrane. Exclusion processes have been
employed as theoretical frameworks for describing the efficiency and
selectivity NPC transport \cite{ZILMANPLOSCB2007,ZILMANPLOSCB2010}.

In all of the above applications, extensions exploiting multispecies
exclusion processes are often motivated. In biological systems ion
transport is typically `gated' by cofactors that bind to the pore,
either blocking ion transport, or inducing a conformational change in
the pore structure thereby affecting the entrance, exit, and internal
hopping rates \cite{WOOD2009}. The inclusion of additional species of
particles in the exclusion process has been used to describe
`transport factor' mediated nuclear pore transport
\cite{ZILMANNATURE,ZILMANPLOSCB2007,ZILMANPLOSCB2010}.  For osmosis,
solutes that are small enough to enter membrane pores may also
interfere with the transport of solvent particles through
channels. The solvent and solute species would have different entry,
exit, and internal hopping rates, describing their different
interactions with the pore. In \cite{CHOU1999}, a simple two-species,
three-site, partially asymmetric exclusion model was used to show how
solutes that can enter a pore and suppress solvent flux.  Moreover, it
was found that a small amount of slippage (passing of the solvent and
solute particles over each other) and a pore that was slightly
permeable to solute can very effectively shunt osmotic flow. When both
solute and solvent can pass through the channel (with equal forward
and backward hopping rates in the interior), an interesting
possibility arises whereby the flux of one of the species can drive
the other species from its low concentration reservoir to the high
concentration reservoir. Since the mechanism relies on entrainment of
particles that are driven up its chemical potential, slippage between
the pumping and convected particles destroys this entropic pumping
mechanism.  The efficiency of using a pumping particle that travels
from high chemical potential reservoir to low chemical potential
reservoir to pump the secondary particle was found using Monte-Carlo
simulations \cite{LOHSE1999}.

\begin{figure}[htb]
\begin{center}
\includegraphics[width=4in]{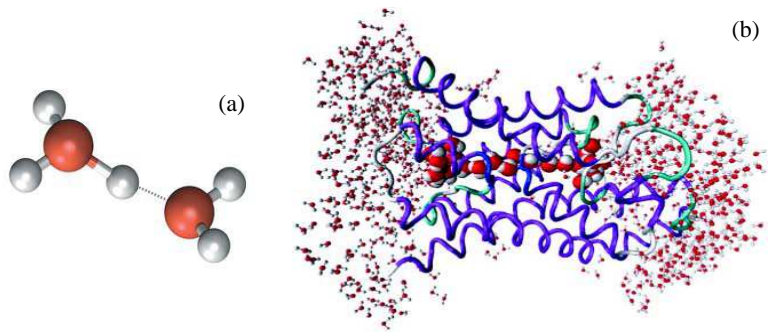}
\end{center}
\caption{(a) Schematic of the Grotthuss proton transfer mechanism.
  (b) An MD simulation showing the persistence of a `water wire'
  within a membrane-spanning Aquaporin-1 water channel protein
  \cite{ILAN2004}. The waters within the pore are shown magnified.}
\label{WATERWIRE0}
\end{figure}

An even more interesting application of multispecies exclusion
processes to biological transport arises when the transported ion is a
proton.  In addition to regulating ionic concentrations, pH regulation
is an important aspect of cellular function.  It has been known for
quite some time that the diffusivity of protons is about an order of
magnitude faster than that of small cations
\cite{GROTTHUSS1806,AGMON}. The physical mechanism invoked to explain
accelerated motion of protons is based on a proton transfer or
shuttling mechanism, analogous to electronic conduction. For a simple
ion to traverse a narrow ion channel, it must shed its hydration shell
and push any possible water molecules ahead of it within the pore. An
extra proton, however, can hop along an oxygen `backbone' of a line of
water molecules, transiently converting each water molecule it visits
into a hydronium ion H$_{3}$O$^{+}$. The Grotthuss mechanism of proton
conduction has been conjectured to occur across many narrow pores,
including those in gramicidin-A channels \cite{ROUX1996,VOTH2006},
proton transfer enzymes such as carbonic anhydrase \cite{CUI2003},
voltage gated proton channels such as Hv1 \cite{RAMSEYHV1}, Aquaporin
water channels \cite{GRUBMULLER2003,ILAN2004,HYDROXIDE2005}, and
carbon nanotubes \cite{DELLAGO2003}.  All of these structures have in
common the existence of a relatively stable water wire as shown on the
right of figure \ref{WATERWIRE0}.

\begin{figure}[htb]
\begin{center}
\includegraphics[width=2.6in]{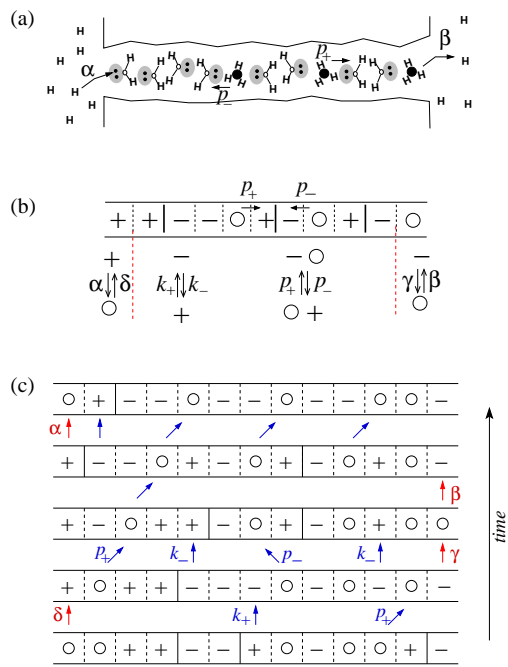}
\end{center}
\caption{(a) A cartoon of water-wire proton conduction. Hydrogen atoms
  and lone electron pairs of the water oxygen are shown. (b) A lattice
  model for the Grotthuss proton conduction mechanism. The symbols
  $\bigcirc, +,$ and $-$ correspond to hydronium ions, water with
  lone pair pointing to the left, and water with lone electron pair
  pointing to the right, respectively. Proton movement from left to
  right leaves the donor site in a $-$ configuration and can occur
  only if the receptor site is originally in the $+$
  configuration. The spontaneous left-right water flipping rates are
  $k_{\pm}$ and the proton forward and backward hopping rates
  (assuming a compatible configuration) are $p_{\pm}$. (c) A
  time-sequence of a trajectory of configurations. Figures adapted from 
  \cite{CHOU_WATER0,CHOU_WATER1}.}
\label{WATERWIRE1}
\end{figure}

The basic water wire mechanism can be mapped onto a three-species
exclusion model as shown in figure \ref{WATERWIRE1}.\footnote{We
  caution the reader that the usage of `$M$ species' in the
  literature is not uniform or unique. Thus, `three-species' here
  refer to three types of particles on the lattice, \emph{without}
  holes. Thus, in terms of states at each site, it is the same as the
  `two species' model describing solvent, solute, and holes at each
  site.} Unlike models where only one extra proton is allowed in the
channel at any given time \cite{SCHUMAKER2003}, the general exclusion
model allows each site to be in one of three states, protonated, water
lone pair electrons pointing to the left and to the right.  In this
particular application, no exact results are known, but mean-field
treatments have been used to extract qualitative phenomena
\cite{CHOU_WATER0,CHOU_WATER1}. For example, in order to sustain a
steady-state proton current, the lone-pair electrons of a water
molecule need to flip in order to relay successive protons.  Not only
was proton conduction found to be mediated by water flipping, but
nonlinear effects such as saturation at high voltages and even
negative differential resistances were exhibited by the model
\cite{CHOU_WATER0,CHOU_WATER1}.

The issue of concerted motions and precise definitions of the proton
carrier species also arises in more detailed considerations of proton
transport. There has been considerable effort devoted to identifying
the precise molecular species that solvates and relays the protons
\cite{AGMON2007}.  Moreover, there is some evidence that proton
dynamics in water wire conduction may be concerted \cite{CUI2003}.
Nonetheless, the three-species exclusion model and its potential
extensions are fruitful ways of understanding the gross mechanisms in
proton conduction.

\subsection{Simple models of molecular motors}
\label{SEC:MOTORS}

Perhaps the simplest models of active biological transport are those
of isolated molecular motors that move along one-dimensional tracks
such as actin, microtubules, DNA, or RNA. Motors are enzymes such as
dynein, kinesin, and myosin that hydrolyze molecules such as ATP or
GTP, turning the free energy released to directed motion along their
one-dimensional substrate \cite{CHOWDHURY2005,CHOWDHURY2008}.
Motoring is necessary for sustaining cell functions such as mediating
cell swimming and motility and intracellular transport of
biomolecules, particularly at length scales where diffusion is not
efficient, or where spatial specificity is required. Due to the
importance of molecular motors in intracellular transport and cell
motility, there is an enormous literature on the detailed structure
and chemo-mechanical transduction mechanisms of molecular motors.

From the theoretical physics point of view, molecular motors are
useful examples of non-equilibrium systems. Indeed, such motors
typically operate far from equilibrium, in a regime where the usual
thermodynamic laws do not apply.  A molecular motor is kept far from
equilibrium by a coupling with some external `agent' (e.g., a chemical
reaction, or a load-force): under certain conditions, work can be
extracted, although the motor operates in a medium with constant
(body) temperature.  We emphasize that there is no contradiction with
thermodynamics: the system is far from equilibrium and the motor
simply plays the role of a transducer between the energy put in by the
agent (e.g., chemical energy) and the mechanical work extracted.
Molecular motors have been described theoretically either by
continuous ratchet models or by models based on master equations on a
discrete space, which are similar to exclusion-type systems, to which
some of the exact analytical techniques described above can be
applied.  Using an extension of the discrete two-state motor model,
Lau {\it et al.} \cite{LLM007} investigated theoretically the
violations of Einstein and Onsager relations and calculated the
efficiency for a single processive motor in \cite{LLM007}.
Furthermore, it can be shown that the fluctuation relations (such as
the Gallavotti-Cohen theorem, the Jarzysnki-Crooks relations) play the
role of a general organizing principle.  Indeed, cellular motors are
systems of molecular size which operate with a small number of
molecules, and for these reasons undergo large thermal fluctuations.
The fluctuation relations impose general constraints on the function
of these nanomachines that go beyond classical thermodynamics. They
provide a way to better understand the non-equilibrium energetics of
molecular motors and to map out various operating regimes
\cite{LLM007,LLM008,LLM009}.

Significant effort in the investigation into molecular motors has
focussed on identifying and understanding the molecular mechanics and
the coupling of molecular motion with a chemical reaction such as ATP
hydrolysis.  Form these studies, complex descriptions of molecular
motors have been developed, including a somewhat artificial
classification of motors employing `power stroke' or `Brownian
ratchet' mechanisms. This classification refers to how detailed
balance is violated {\it within} the large motor molecule or
enzyme. If certain internal degrees of freedom in a molecule are made
inaccessible at the right times, a net flux along these states can
arise, ratcheting the motion. If a particular transition is strongly
coupled to specific chemical step such as the hydrolysis of an ATP
molecule bound in a pocket of the motor molecule, the motion has been
described as a power stroke motor. It has been shown that this
distinction is quantitative, rather than qualitative \cite{HWANG2002}.
From a stochastic processes point of view, when the dynamics of the
microscopic internal motor degrees of freedom are represented by,
e.g., a Markov chain, power stroke and ratchet mechanisms can be
distinguished by where detailed balance is violated in the cycle. For
example, if the relative amount of violation occurs duriong
transitions across states that are directly coupled to motion against
a load, the motor will be ``ratchet-like''. 

If the absicca in figure \ref{PROFILE} represents sequential internal
molecular states, a purely ratchet mechanism is naturally represented
by a state evolving along the flat energy profile.  In this case, a
steady-state probability flux, or molecular motion, arises from the
absorbing and emitting boundary conditions that ratchet the overall
motion.  In this stochastic picture, the power-stroke/Brownian ratchet
distinction is mathematically recapitulated by state-space boundary
conditions and by the relative amount of convection through
state-space.  How much a motor utilizes a power-stroke mechanisms
would be described by the Peclet number within the framework of
single-particle convection-diffusion or Fokker-Planck type equations.
For a more detailed delineation of regimes of mechanical-chemical
coupling, see the review by Astumian \cite{ASTUMIAN2010}.

In many cellular contexts, molecular motors are crowded and interact
with each other and exhibit collective behavior. Examples include
connected myosin motors in muscle, multiple motors and motor types on
cellular filaments and microtubules (often carrying large cargo such
as vesicles), and motors that process DNA and RNA.  To model such
systems, the details of how an isolated motor generates force and
moves may be best subsumed into a single parameter representing the
mean time between successive motor displacements. The internal
dynamics, whether power-stroke or Brownian ratchet, moves the motor
one step against a load or resistance at random times. These times are
drawn from a distribution of hopping times that is determined by the
underlying, internal stochastic process.

\begin{figure}[h]
\begin{center}
\includegraphics[width=4.2in]{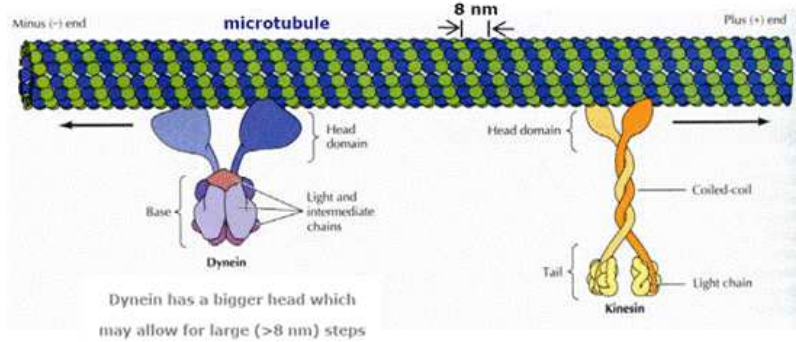}
\end{center}
\caption{A schematic of two types of molecular motors moving along a
  cellular microtubule (adapted from \cite{COOPER2007}). In this
  picture, kinesin moves to the right while dynein moves to the
  left. Each motor can be attached to large cargoes such as vesicles
  or other filaments. Note that the microtubule is constructed of
  twisted lanes of repeated molecular subunits.}
\label{MOTORS}
\end{figure}

One can simplify the modeling of molecular motors by assuming that the
motor stepping time is exponentially distributed with an
inverse mean that defines the hopping rate. Each motor hops along a
one-dimensional track and can exclude other motors. Typically,
concerted motions are also neglected in this application.  That is, a
motor is not allowed to push another one in front of it, moving both
motors ahead simultaneously. In the extreme limit, multiple motors are
coupled with, e.g., elastic elements and have been extensively
modeled using simple Fokker-Planck equations that incorporate
mechanical and thermal forces
\cite{JULICHER1995,AJDARI1997,VILFAN1998,BADOUAL2002,STUKALIN2005,FISHER2007,
  BRUGUES2009}. Weaker interactions that do not bound the distance
between motors can be modeled using the excluding particle picture by
assigning different rules for the hopping of two adjacent particles,
analogous to the facilitated diffusion models of Gabel, Krapivsky, and
Redner \cite{GABEL2010}. Given the assumptions discussed above,
exclusion processes can be directly used to model collections of
motors moving on one-dimensional tracks. Additional effects particular
to applications in biomolecular motors include detachment and
attachment kinetics and different types of motors that move in
opposing directions.

Molecular motors `walk' along filaments but have a finite
`processivity' since they can spontaneously detach. Thus, they have a
distribution of run lengths. Conversely, motor molecules diffusing in
the `bulk' cytoplasmic space can attach to interior locations of the
lattice. Detachment violates particle conservation on the lattice and
has been studied using mean field models, hydrodynamic approximations
\cite{SANTEN2003,JUHASZ2004}, and Monte-Carlo simulations
\cite{FREY2004,FREY2003,MIRIN2003,GREULICH2009}.  The detachment and
possible reattachment of particles on a lattice have been extensively
studied in the context of gas adsorption isotherms or `Langmuir
kinetics' (see \cite{GAST} and references within).  While many models
of Langmuir isotherms exist, previous studies considered only passive,
undriven particles. In the new work combining Langmuir kinetics with
driven exclusion processes, analytic progress can be made by
considering the infinite site, continuum hydrodynamic limit.  If
attachment and/or detachment occurs, the hopping rates must be rescaled by
the number of lattice sites such that the rates $\omega_{\rm A,D}$
(cf. figure \ref{DETACH}) at each site are inversely proportional to
the number of sites $L$. If this is not done, particles would only
occupy a small region near the injection end of a long lattice. To
arrive at a nontrivial structure, the attachment and detachment rates
must be decreased so that the cumulative probability of desorption is
a length-independent constant.

As we mentioned in section \ref{KM}, continuum hydrodynamic equations
allow more complicated models to be approximately treated. This has
been the case for one-dimensional exclusion processes with Langmuir
kinetics.  In the steady-state limit, Parmeggiani {\it et al.} find
\cite{FREY2004}

\begin{equation}
{\varepsilon \over 2}{\partial^{2} \rho(x) \over \partial x^{2}} 
+ (2\rho -1){\partial \rho(x) \over \partial x} +\Omega_{\rm A}(1-\rho(x))
-\Omega_{\rm D}\rho(x) = 0,
\label{HYDRO}
\end{equation}
where $\Omega_{\rm A,D} = \omega_{\rm A,D}L \equiv \omega_{\rm
  A,D}/\varepsilon$ represents appropriately rescaled detachment and
attachment rates that in this simple model are {\it independent} of $L$.
A detailed asymptotic analysis of equation (\ref{HYDRO}) was performed
and a phase diagram as a function of four parameters (the injection
and extraction rates at the ends of the lattice, and the adsorption
and desorption rates) was derived \cite{FREY2003,FREY2004}. They find
a rich phase diagram with coexisting low and high-density phases
separated by boundaries induced by the Langmuir kinetics. Langmuir
kinetics have also been investigated in the presence of bottlenecks
\cite{FREY2006BOTTLENECK}. Klumpp and Lipowsky \cite{KLUMPP2003} have
considered asymmetric exclusion in tube-like structures such as those
seen in axons or in filapodia \cite{ALBERTS,LIPOWSKY2006}. In this
simplified geometry, the diffusion of motors within the tube can be
explicitly modeled \cite{KLUMPP2003}.

Since motors are used to carry cargoes across the different regions
within a cell, they travel on directed filaments and
microtubules. Different motors are used to carry cargo in one
direction versus the other. For example, kinesins travel along
microtubules in the `+' direction, while dynein travels in the `-'
direction, as shown in figure \ref{MOTORS}. This scenario can be modeled
using a two-species exclusion process with Langmuir kinetics
overtaking. Moreover, microtubules are composed of multiple twisted
filamentous tracks. Not only can motors travel in opposite directions
on the same track, but filaments may be oppositely-directed within
confined spaces such as axons. Motors traveling in opposite directions
on the same filament or on a nearby parallel filament can be modeled
using coupled chains of exclusion processes. Many groups have explored
the dynamical properties of two-species and two-lane asymmetric
exclusion processes (as shown in figure \ref{TWO_CHANNEL})
\cite{KOLO1997,PRONINA2007,TSEKOURAS2008B,DU2010,DENNIJS2007}. With
the proper biophysical identification of the parameters, results from
these studies of interacting lattices should provide illuminating
descriptions of more complex scenarios of moleculer motor-mediated
transport.

\subsection{mRNA translation and protein production}

\begin{figure}[h]
\begin{center}
\includegraphics[width=5in]{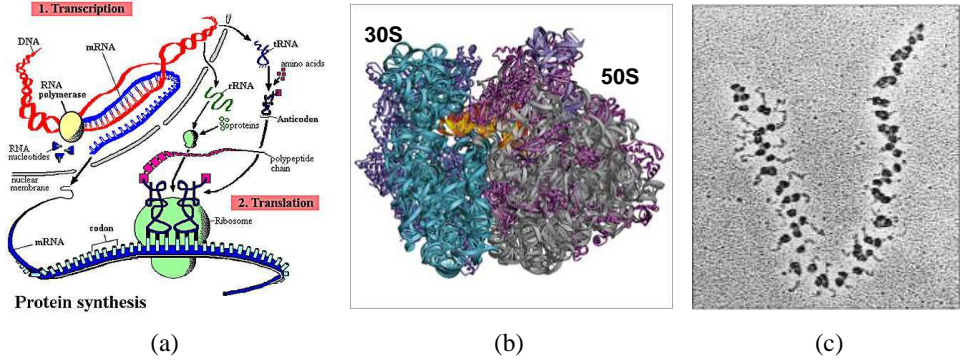}
\end{center}
\caption{(a) A cartoon of the `central dogma' in biology where mRNA is
  synthesized from cellular DNA and transported to the cytoplasm. The
  cytoplasmic mRNAs are then translated by ribosomes into polypeptides
  which may then finally be folded and processed into functioning
  proteins. (b) Crystal structure of both subunits of bacterial
  ribosome.  (c) An electron micrograph of multiple ribosomes
  translating a single mRNA.}
\label{TRANSLATION}
\end{figure}

One very special case of interacting motors moving along a
one-dimensional track arises in cell biology. In all cells, proteins
are synthesized by translation of messenger RNA (mRNA) as 
schematised in figure \ref{TRANSLATION}. Complex
ribosome enzymes (shown in figure \ref{TRANSLATION}(b)) unidirectionally scan the
mRNA polymer, reading triplets of nucleotides and adding the
corresponding amino acid to a growing polypeptide chain. Typically,
many ribosomes are simultaneously scanning different parts of the
mRNA.  An electron micrograph of such a {\it polysome} is shown in
figure \ref{TRANSLATION}(c).

The ribosomes are actually comprised of two subunits, each made up
mostly of RNA and a few proteins.  Not only do these `ribozymes'
catalyze the successive addition of the specific amino acid as they
scan the mRNA, their unidirectional movement from the 5' end to the 3'
end, (where the 5' denotes the end of the mRNA that has the fifth
carbon of the sugar ring of the ribose as the terminus) constitutes a
highly driven process.  Therefore, each ribosome is also a molecular
motor that rarely backtracks as it moves forward. The fuel providing
the free energy necessary for codon recognition and unidirectional
movement is supplied in part by the hydrolysis of GTP
\cite{RIBOSOMEENERGETICS}. Quantitatively, mRNA translation is
different from typical molecular motors in that ribosome processivity
is very high, allowing one to reasonably neglect detachments except at
the termination end. Moreover, there are no known issues with
`concerted motions' or `facilitated exclusion'. If one ribosome
prematurely pushes the one ahead of it, one would expect many
polypeptides to be improperly synthesised.

Thus, at first glance, the TASEP seems to be the perfect model for
this translation process, with the particles being ribosomes and a
site being a codon -- a triplet of nucleotides. On closer examination,
it is clear that protein production is much more complicated. Many biophysical
features relevant to mRNA translation are missing from the basic
TASEP. The desire to have more `realistic' models of protein
production has motivated the development of various extensions to the
basic TASEP. In this subsection, we will discuss only two efforts to
generalize TASEP: allowing particles of size $\ell>1$ and incorporating
inhomogeneous, site-dependent hopping rates. In each case, we will
describe the cell biology which motivates the modifications.

Although the fundamental step size of ribosomes is a codon, the large
size of each ribosome (20nm for prokaryotic ribosomes and 25-30nm for
eukaryotic ribosomes) means that they each cover approximately 10
codons along the mRNA chain and exclude each other at this
range. Therefore, TASEPs comprised of extended objects, occluding
$\ell>1$ lattice sites have therefore been developed, dating back to the
late 1960's \cite{MGP68,MG69}. Although exact results for such a
generalized TASEP on a ring are available \cite{ALCARAZ1999}, the
problem with open boundaries remains unsolved. Instead, various MFT's
have been successful at capturing many important properties for
modeling mRNA translation. Let us devote the next paragraphs to this
system.

First, consider a ring of $L$ sites filled with $N$ particles, and
note that every configuration with particles of extent $\ell$ can be
matched with a configuration with $N$ point particles ($\ell =1$) on a
ring of $L-N(\ell -1)$ sites. This mapping is easily understood by
regarding a configuration, $\mathcal{C}$, as clusters of adjoining
vacancies followed by clusters of adjoining particles (regardless of
their sizes).  Therefore, the stationary distribution, $P^{*}\left(
\mathcal{C}\right)$, is again flat and independent of $\ell$ for all
$p,q$. Of course, the sum of the particle density ($\rho \equiv N/L$)
and the hole density ($\rho_{\rm hole}$) now satisfy 
$\ell \rho +\rho_{\rm hole}=1$, but with $\rho$ lying in a limited interval
$\left[0,1/\ell \right]$.  With $P^{*}\propto 1$, finding the probability
of particle-hole pairs is straightforward, resulting in an exact
expression \cite{SHAW2003} for the current density relation, $J(\rho)
=\rho \rho_{\rm hole}/(\rho +\rho_{\rm hole}-1/L)$, in the TASEP
case. To lowest order in $1/L$, the formula
\begin{equation}
J=\frac{\rho \left(1-\ell \rho \right) }{1-(\ell -1)\rho }  
\label{J-ring}
\end{equation}
was known as early as 1968 \cite{MGP68,MG69}, and leads to a maximal
current of $(1+\sqrt{\ell})^{-2}$ associated with the optimal density
of $\hat{\rho}=(\ell +\sqrt{\ell })^{-1}$. Note that, though $J(\rho)$
is no longer symmetric about $\hat{\rho}$, the particle-hole symmetry
($\rho \leftrightarrow \rho_{\rm hole}$) is preserved. This invariance
is expected on physical grounds, as the current arises only from
exchanges of particle-hole pairs. Though the stationary measure is
trivial and $G(r)$, the expectation value of two particles separated
by $r$ sites, can be written formally as a sum of products of
binomials. Of course, period-$\ell$ structures are to be expected in
$G\left(r\right)$. Despite the conceptual simplicity of this problem,
remarkably intricate patterns emerge \cite{GBBM11}, especially near
complete filling, $L\cong N \ell$ \cite{DZunpublished}. Such
structures are completely absent from the $\ell =1$ case (whether in
periodic or open TASEP), showing us that, even in seemingly trivial
situations, statistics of extended objects can produce surprises.

Turning to the open boundary TASEP, we must first specify how a particle
enters/exits the lattice. One possibility is `complete entry,
incremental exit' \cite {LAKATOS2003}, where a particle may enter
completely provided the first $\ell $ sites are empty, while it may exit
moving one site at a time. In the NESS, exact expressions for the
current like (\ref{CourantCorrelTASEP}) can still be written. When
both $i$ and $i+\ell $ are within $\left[1,L\right] $, we have $J=\langle
\sigma _i(1-\sigma _{i+\ell })\rangle $. From the `incremental exit' rule,
we have $J=\langle \sigma _{L-\ell }\rangle =\ldots =\langle \sigma
_{L-1}\rangle =\beta \left\langle \sigma _L\right\rangle $, leading to
a simple profile next to the exit. The major challenge comes from the
`complete entry' condition: $J=\alpha \langle \prod_{k=1}^{\ell} (1-\sigma
_k)\rangle $. Since this problem can no longer be solved exactly, many
conclusions can only be drawn from simulation studies or imposing mean
field approximations. For the latter, the na\"{i}ve replacement of
averages of products of $\sigma $ by products of $\left\langle \sigma
\right\rangle $ (e.g., $\langle \sigma _i\sigma _{i+\ell }\rangle
\rightarrow \langle \sigma _i\rangle \langle \sigma _{i+\ell }\rangle $)
leads to extremely poor predictions. Gibbs,
\emph{et.al. }\cite{MGP68,MG69} took into account some of the effects
of exclusion at a distance and approximated $J=\langle \sigma
_i(1-\sigma _{i+\ell })\rangle$ by

\begin{equation}
J=\frac{\rho _i\bar{\rho}_i}{\rho _{i+\ell }+\bar{\rho}_i}  \label{MGPcurrent}
\end{equation}
where $\bar{\rho}_i\equiv 1-\sum_{k =i+1}^{i+\ell }\rho_k $ is an
effective hole density in the $\ell$ sites following $i$. For the ring,
the profile is flat and $\rho_i=\rho $, so that (\ref{MGPcurrent})
reduces to (\ref{J-ring}). Supplemented with the appropriate boundary
equations, (\ref{MGPcurrent}) can be regarded as a recursion relation
for the profile. The successes of this approach include predicting a
phase diagram that is the same as the one in figure
\ref{fig-DIAGPHASE}(c), except that the boundaries of the maximal
current phase are now at $\alpha, \beta
=\left(1+\sqrt{\ell}\right)^{-1}$. Simulations largely confirm such
predictions \cite{SHAW2003,LAKATOS2003}, suggesting that the
correlations neglected by the scheme of Gibbs, \emph{et al.}
\cite{MGP68,MG69} are indeed small. On the other hand, for more
sensitive quantities like the profile, this MFT is less successful,
especially for the high density phase ($\beta \ll \alpha \sim 1$). To
produce better predictions, Shaw, \emph{et al.}  \cite{SETHNA2004}
introduced a more sophisticated MFT, taking into account some pair
correlations. So far, no higher level of MFT have been attempted.

The second modification to the basic TASEP we consider here is
site-dependent hopping rates. Since the translation of mRNA into
polypeptides depends on the sequence of nucleotides, the hopping rates
of the ribosome TASEP particles can vary dramatically as a function of
its position on the lattice. The local hopping rates depend on the
effective abundance of the specific amino-acid-charged tRNA that
participates in each elongation step at each site.  One of the first
treatments of TASEPs on a nonuniform lattice considered a single
defect, or slow hopping site\footnote{If an isolated site had faster
  hopping than all its neighbors, the average current is hardly
  affected, though there are noticable changes to the profile. Thus,
  we focus on slow sites.}, in the middle of an asymptotically long
lattice. Kolomeisky derived the expected steady-state particle flux
across a lattice with a single slow hopping site by ignoring
particle-particle correlations across the defect \cite{KOLO1998}. He
self-consistently matched the exit rate from the first half of the
chain with the injection rate of the second half of the chain. This
approach indicated that each long uniform region between slow sites
can approximately be treated as separate, but self-consistently
connected, uniform exclusion processes.  Later work by Chou and
Lakatos developed a more refined method of connecting the current
between sections separated by interior defects.  Their method
generalizes mean-field theory by explicitly enumerating the
configurations of the sites straddling the inhomogeneity, and
self-consistently matching the current across this segment with the
asymptotically exact steady-state currents across the rest of the
homogeneous lattice.  Steady-state particle currents across localized
defects can be accurately computed \cite{CHOU2004}.  Using this
approach, the synergistic effects of two nearby slow hopping sites
were analyzed. It was shown that two defects decreased the current
more when they are placed close to each other. The decrease in current
approaches that of a single defect as the distance between two defects
diverges since the dynamical `interaction' from overlap of density
boundary layers vanishes.

\begin{figure}[h]
\begin{center}
\includegraphics[height=2in]{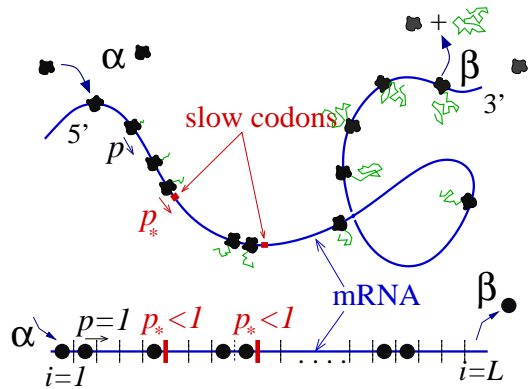}
\end{center}
\caption{mRNA translation with fixed slow codons, or bottlenecks
  across which ribosome motion is slower ($p^{*} < p$) than across other
  codons. This local slowing down can arise from a limited supply of
  the appropriate amino acid-charged tRNA corresponding to the slow
  site.}
\label{DEFECT}
\end{figure}

Another method for approximately treating inhomogeneous hopping rates
was developed by Shaw {\it et al.} \cite{SETHNA2004} where the
original mean-field equations of Gibbs {\it et al.}  \cite{MGP68,MG69}
were generalized to include site-dependent elongation rates.  Dong,
Schmittman, and Zia \cite{DONG2007A,DONG2007B} systematically analyzed
the effects of defects on the steady-state current of extended
particles that occupy multiple ($w > 1$) sites.  They used a
combination of self-consistent mean field theory and Monte-Carlo
simulations to show the importance of the location of the defect,
especially if the defect is placed close entry or exit sites where
they may possibly `interact' with the end-induced density boundary
layers.

As mentioned in section \ref{SSASEP} hydrodynamic MFT developed for
analyzing systems with slowly varying hopping rates.  Continuum
equations, such as (\ref{INTERIOR}), that incorporate spatially
varying hopping rate functions $p(x)$ and $q(x)$ have been derived for
the mean field ribosome density $\rho(x)$
\cite{SHAW2003,SCHOENHERR2004,SCHOENHERR2005,LAKATOS2006}.  For
single-site particles, these hydrodynamic equations can be integrated
once to arrive at singular nonlinear first order differential equation
where the integration constant is the steady-state particle current.
Using singular perturbation theory, analytic solutions for density
profiles have been found for very specific hopping rate functions
$p(x)$ and $q(x)$ \cite{LAKATOS2006,SMOOTHPQ}.

The stochastic process underlying the models of mRNA translation
described above all assume an exponentially-distributed waiting time
between elongation events. Dwell time distributions would be difficult
to incorporate into simple ODEs for the mean-field particle density or
continuum hydrodynamic equations. Although different dwell time
distributions would not qualitatively affect steady-state ribosome
current and protein production rate, to model them explicitly requires
incorporation of intermediate biochemical steps involved in
initiation, elongation, and termination of the individual ribosomes.
These steps include binding of an amino-acid charged transfer RNA
(tRNA) to one of the binding sites, hydrolysis,
etc. \cite{CHEN2011}. Models that include more complex elongation
steps have also been developed by a number of researchers
\cite{CHOWDHURY2007,CHOWDHURY2009A,CHOWDHURY2009B,GARAI2009,PAOLO2009,INTERNAL2006,INTERNAL2010}.
The main qualitative result of these studies is that the standard
current phase diagram is shifted due to a varying {\it effective}
elongation (internal hopping) rate.  Since the internal hopping rate
depends on the details of their models, it cannot be
nondimensionalized and the standard phase diagram is roughly
reproduced with $\alpha_{\rm eff}/p_{\rm eff}$ and $\beta_{\rm
  eff}/p_{\rm eff}$ as the tuning variable. Here $\alpha_{\rm eff}$,
$\beta_{\rm eff}$, and $p_{\rm eff}$ are are effective rates that
might be estimated as the inverse of the mean of the associated dwell
time distributions.

\begin{figure}[h]
\begin{center}
\includegraphics[width=4in]{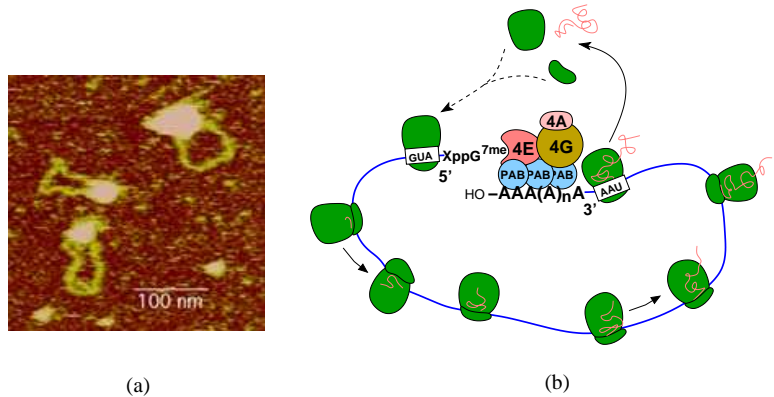}
\end{center}
\caption{(a) AFM image of circularized mRNA. (b) Schematic of a model
  where ends of mRNA are sticky (the poly-A tails are known to bind to
  initiation factors \cite{SACHS1990,WELLS1998,POLYA}), increasing the
  probability of loop formation.}
\label{CIRCLE}
\end{figure}

Finally, the mRNA translation process, much like molecular
motor-facilitated intracellular transport, occurs in complex,
spatially heterogeneous environments and involves many molecular
players that initiate and terminate the process.  For example, certain
initiation factors that prime the initiation site for ribosomal entry
have been shown to bind to the polyadenylated tails of mRNA, thereby
forming circularized RNA. One hypothesis is that mRNA circularization
facilitates the recycling of ribosomes. In circularized mRNA,
ribosomes that detach from the termination site after completing
translation have a shorter distance to diffuse before rebinding to the
initiation site.  A model for this effect was developed in
\cite{CHOU2003}, where an effective injection rate $\alpha_{\rm eff}$
was self-consistently found by balancing the steady-state ribosome
concentration at the initiation site with the diffusive ribosome flux
emanating from the nearly termination site. In this model, the
diffusive feedback tends to increase the protein production rate,
especially when overall ribosome concentrations are low. One can also
imagine a strong feedback if factors regulating translation initiation
were themselves products of translation. Newly produced cofactors can
readily maintain the initiation rate $\alpha$ at a high value.

An even more important aspect of mRNA translation is that ribosomes
and initiation factors \cite{WILLETT2010}, and mRNAs \cite{ER_RNA2006}
can be actively localized to endoplasmic reticular (ER) membranes and
compartments, depending on what types of proteins they code for, and
where these proteins are needed.  In confined cellular spaces, the
supply of ribosomes and initiation factors may be limited. Moreover,
there are many different mRNA copies that compete with each other for
ribosomes. This global competition has been modeled by Cook {\it et al.}
\cite{COOK2009A,COOK2009B} who defined an effective initiation rate
$\alpha_{\rm eff}(N)$ which is a monotonically increasing function of
the free ribosome concentration. They considered mRNAs of different
length (but of identical initiation, elongation, and termination
rates) and found that steady-state protein production for different
length mRNA's were comparable, but that their ribosome loading levels
exhibit varying levels of sensitivity on the total ribosome mass.

\subsection{Free boundary problems and filament length control}

Our final example of a class of biological application of exclusion
processes involves changes in the length of the underlying
one-dimensional substrate.  A dynamically varying lattice length
arises in at least three different cellular contexts: growth of
filaments such as hyphae and cellular microtubules, replication forks,
and mRNA secondary structure. in each of these examples, there is a
`moving boundary' whose dynamics depends on transport within the
domain bounded by the boundary.

\begin{figure}[h]
\begin{center}
\includegraphics[width=3in]{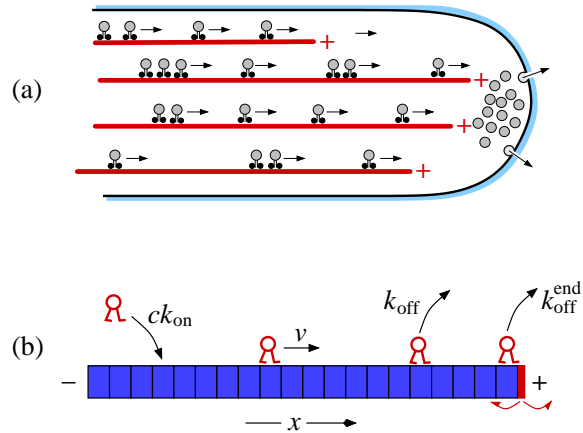}
\end{center}
\caption{(a) Schematic of multiple TASEPs a dynamically growing
  tips. The particles are motors carrying building blocks of the
  underlying filaments \cite{EVANS2007}. (b) A model of microtubule
  length control where particles reaching the tip depolymerises the
  last subunit \cite{HOUGH2009}. Figures are adapted from those in
  \cite{EVANS2007} and \cite{HOUGH2009}.}
\label{FB}
\end{figure}

An analogy can be made with the classic free boundary problem arising
in continuum physics. In the `Stefan problem' a diffusive quantity
mediates the growth of an interface bounding the dynamics. This
diffusing quantity may be heat which melts away water-ice interface
\cite{CARSLAW}, or a particulate systems that deposits and makes the
interface grow \cite{FOK2009}. In systems relevant to cell biology,
the dynamics (i) occur in confined, often one-dimensional geometries,
and (ii) occur on a fluctuating mesoscopic or molecular scale. In
confined geometries, particle-particle exclusion become important.  By
contrast, this exclusion is absent when the transported field is say,
heat. On small scales, statistical fluctuations of the interface, and
its coupling with the fluctuating particle field can also be
important.

A stochastic one-dimensional moving boundary problem can also be
described within an exclusion process where the number of sites is
allowed to fluctuate.  Such models have been described in
\cite{NOWAK2007,EVANS2007,READ2007,NOWAK2007,SUGDEN2007}, and
\cite{HOUGH2009}. In \cite{NOWAK2007}, a fluctuating wall is coupled
to an asymmetric exclusion process as shown in figure \ref{FB0}. The
wall particle acts like a piston and defines one boundary of the
lattice and has its own intrinsic forward and backward hopping rates.
Particles impinge on the wall and have a certain detachment rate.
Using a moving frame version of asymptotic matching \cite{CHOU2004}
the expected position and fluctuations of the wall were accurately
computed.  The wall was found to either extend the lattice
indefinitely, compress the particles and fall off the lattice at the
injection site, or find an equilibrium lattice length.

In terms of specific biological applications, Evans and Sugden
\cite{EVANS2007} and Sugden {\it et al.} \cite{READ2007} consider a
free boundary TASEP as a model for hyphae growth in fungi (cf. figure
\ref{FB}(a)). The hypothesis is that kinesin motors carry cargo and
move unidirectionally toward the tip. The delivered cargo can then
extend the tip by incrementing the number of lattice sites of the
TASEP. In this case, there is no confining wall, and the velocity tip
extension is proportional to the number of arriving TASEP
particles. Even though the tip is always growing with a fixed
velocity, the particle density profiles can acquire different
structure, including a jammed, high density phase.

Another application of dynamic-boundary TASEPs involving molecular
motors was described by Hough {\it et al.} \cite{HOUGH2009}.  In their
model of microtubule length control, kinesin-8 motors move along a
microtubule and {\it depolymerize} the tip when it is reached.  The
tip also has an intrinsic degradation and assembly rate, the latter
depending on the concentration of subunits in the bulk.  Therefore,
this model is similar to a TASEP with a confining wall
\cite{NOWAK2007} representing the end of the lattice. However, in this
model, the kinesis motors also undergo Langmuir kinetics by attaching
to, and detaching from the microtubule lattice. Hough {\it et al.}
find regimes where microtubule length-dependent depolymerization rates
arise, as well as how the behavior depends on bulk motor
concentrations \cite{HOUGH2009}.

A number of potentially new models remain relatively unexplored.  For
example, molecular motors often encounter an obstacle. As the helicase
motor separates DNA strands for transcription, it must break base pair
bonds and push the replication fork forward
\cite{BETTERTON2005,BETTERTON2005JPC,GARAI2008,MANOSAS2010}.  The fork
may be represented by a confining wall that tends to reseal the single
strands of DNA. Similarly, mRNA translation by ribosome `motors'
typically occur in the presence of hairpins that must be separated
before the ribosomes can progress.  In both of these applications the
motor-wall interaction may be considered passive (ratchet-like) or
active (forced separation) \cite{MANOSAS2010}.  These limits are again
related to the issue of concerted motion or faciliated exclusion
mentioned earlier. The motor can actively advance the replication
fork, whereby the forward motions of the leading particle and the wall
occur simultaneously in a concerted fashion.  Alternatively, the
opening of the fork can be thermally activated, temporarily allowing
the leading motor to ratchet the wall. Sometimes, due to the
particular geometry, a motor may not need to move against a barrier,
but may pass through it. An interesting realisation arises in mRNA
translation where ribosome may jump over hairpins, effectively making
hops to an empty site far away on the lattice. This kind of
`short-cut' model has been studied using mean-field theory and
simulations by Kim {\it et al.} \cite{KIM2011} and have been found to
exhibit coexistence of empty and jammed phases.  The consequences of
these rich mechanisms on cell biology remain to be explored.


\section{Summary and outlook}
\label{sum}

Non-equilibrium statistical mechanics is the study of stochastic
processes for a system involving many interacting degrees of
freedom. As such, it concerns essentially the entire spectrum of
natural phenomena, ranging over all length scales and relevant for all
areas in science and engineering.  While a small encyclopedia would be
needed to adequately cover NESM, reviews dealing with more specialised
topics are abundant in the literature. In this article, we attempt a
compromise by providing (a) a bird's eye view of NESM, and (b) a
`bridge' between the two different approaches to the subject. For the
former, we review some of the fundamental issues associated with NESM
and why the techniques (and assumptions) of equilibrium statistical
mechanics cannot be easily generalized. Using the language of
stochastic processes, we base our discussions on master equations and
focus on systems which violate detailed balance (or time
reversal). One serious consequence is that, given a set of such rates,
even the stationary distribution, $P^{*}$ -- a NESS analog of the
Boltzmann factor $e^{-\beta {\cal H}}$ -- is generally unknown.  There
is an additional challenge: the presence of persistent {\em
  probability currents} (forming closed loops, of course), $K^{*}$,
much like the steady electric current loops in mangetostatics. By
contrast, $K^{*}\equiv 0$ if the transition rates obey detailed
balance, in analog to electrostatics. We emphasized that these
currents play important roles in a NESS, as they lead to persistent
fluxes of physically observable quantitities (e.g., mass, energy,
etc.) as well as constant entropy production in the surrounding
medium.

Naturally, our `bird's eye view' is both limited and colored. For
example, we offer only a brief glance on fluctuation theorems and
symmetry relations, a topic of NESM which has witnessed considerable
progress over the past two decades.  Even more generally, our focus on
master equations implicitly assumes that the dynamics can be
decomposed into a series of Markov processes with exponentially
distributed waiting times in each configuration.  Age-dependent
processes described by more complex waiting time distributions of each
configuration may often be better studied using theories of branching
processes.  Moreover, master equation approaches are not very suited
for studying the statistics of individual trajectories or
realisations. Understanding properties of individual trajectories in
configuration space is important, for example, in problems of
inference and recontruction of the stochastic rules from data. These
attributes are better analysed using techniques of, e.g., stochastic
calculus or path integrals. Therefore, a comprehensive overview of the
foundations of NESM and stochastic processes would include concepts
and formalism from other disciplines such as probability theory,
statistics, and control theory - a task far beyond our scope.

To meet the challenges posed by NESM, a wide range of approaches have
been developed. On the one hand, we have model systems, sufficiently
simple that rigorous mathematical analysis is feasible. At the other
extreme are models which account for many ingredients deemed essential
for characterizing the large variety of phenomena found in
nature. Understandably, those working on either end of this spectrum
are not typically familiar with the progress at the other end. A
historic example appeared in the late 1960's when Spitzer
\cite{Spitzer70} and Gibbs, {\em et.al.}  \cite{MGP68,MG69}
investigated independently the same stochastic process, but were
unaware of each others' studies.
%
%
In this article, we attempt to bridge this divide by providing brief
reviews of a limited set of topics from both ends.

For the exclusion process, we presented many exact results, along with
two complementary techniques to obtain them. The stationary
distribution for a system with periodic boundary conditions (i.e.,
ASEP on a ring) was known to be trivial \cite{Spitzer70}. On the other
hand, a rich variety of dynamic properties, even for SEP, have been
discovered over the last four decades. With open boundaries, particles
may enter or exit the lattice at both ends, so that the total
occupation becomes a dynamic variable. As a result, this system poses
much more of a challenge. Little was known until the 1990's, when the
exact $P^{*}$ was found \cite{DeDoMuk,DEHP,SchutzDomany} through an
ingenious recognition of a matrix representation.  For simplicity, we
presented the details of this method only for the extreme case of
TASEP, in which particles enter/exit with rate $\alpha/\beta $
(section \ref{TASEP-MR}). Once $P^{*}$ is known, it is straightforward
to compute the exact {\em particle} current, $J$, and the mean density
profile $\rho_{i}$. A non-trivial phase diagram in the
$\alpha$-$\beta$ plane emerged, with three distinct phases separated
by both continuous and discontinuous transitions (section
\ref{TASEP-PD}). In addition, algebraic singularities are always
present in one of the phases. Such phenomena are novel indeed, since
the conventional wisdom from EQSM predicts no phase transitions (i.e.,
long-range order), let alone generic algebraic singularities, for
systems in one dimension, with short-ranged interactions! Beyond NESS,
another powerful method -- the Bethe Ansatz -- have been applied
successfully to obtain time-dependent properties (section
\ref{BA}). Analyzing the spectrum and the eigenvectors (Bethe wave
functions) of the full Markov matrix, other physically interesting
quantities in a NESS can be also computed. Examples include the
fluctuations of the current (around the mean $J$), encoded in the
large deviation function, $G\left( j\right) $. The knowledge of
dynamic properties also allows us to explore, in principle, ASEPs on
{\em infinite} lattices, for which the initial configuration must be
specified and the system properties after long times are of
interest. In practice, however, these problems are attacked by another
technique, fascinatingly related to the theory of random matrices,
representations and combinatorics (section \ref{Infinite}).  These
powerful methods have yielded mathematically elegant and physically
comprehensible results, rendering exclusion models extremely appealing
for modeling real systems.

Apart from exact solutions, we also presented an important
approximation scheme -- the mean-field theory. Based on physically
reasonable and intuitive arguments, we consider coarse-grained
densities as continuum fields, obeying hydrodynamic equations, e.g.,
the density $\rho\left(x,t\right)$ in ASEP. Remarkably, such a MFT
predicted some aspects of TASEP exactly (e.g., the current density
relation $J=\rho\left(1-\rho \right)$).  Though neither rigorous nor
a systematic expansion (such as perturbation theory with a small
parameter), MFT has provided valuable insights into behaviour which we
cannot compute exactly. This is especially true when appropriate noise
terms are added, so that we can access not only the (possibly
inhomogeneous) stationary profiles but also the fluctuations and
correlations near them.  A good example is the average power spectrum
associated with $N(t)$, the total mass in an open
TASEP. Intimately related to the time-dependent correlation of the
currents at the two ends, this quantity has eluded efforts to find an
exact expression, despite our extensive knowledge about the full
Markov matrix.  Instead, starting with a MFT and expanding around a
flat profile in the high/low density phase, many interesting
properties of this power spectrum can be reasonably well understood
\cite{ASZ07,CookZia10}.
%
%
Undoubtedly, there are many other physically interesting quantities
(in ASEP and other `exactly solvable systems') for which theoretical
predictions can be obtained only from mean field approximations. Of
course, as we consider modeling transport phenomena in biological
systems, more realistic features must be included. Then, MFT typically
offers the only route towards some quantitative understanding.

We next turned to specific generalizations of the ASEP which are
motivated by problems in biological transport. While the exactly
solvable ASEP consists of a single lattice of fixed length with at
most one particle occupying each site and hopping along from site to
site with the same rate, the variety of transport phenomena in cell
biology calls for different ways to relax these constraints. Thus, in
the first model for translation in protein synthesis, Gibbs, {\em
  et.al.} already introduced two features \cite{MGP68,MG69} absent
from the standard ASEP.  Representing ribosomes/codons by
particles/sites, and being aware that ribosomes are large molecules
covering $O(10)$ codons, they considered ASEP with {\em
  extended objects}.  They were also mindful that the sequence of
codons in a typical mRNA consists of different codons, leading to a
ribosome elongating (moving along the mRNA) with possibly different
rates. The result is an ASEP with {\em inhomogeneous hopping rates}.
Beyond these two aspects, we know that there are typically thousands
of copies of many different mRNA's (synthesizing different proteins)
within a single cell. Now, they compete for the same pool of
ribosomes. To account for such competition, we should study multiple
TASEPs, with different lengths, `interacting' with each other through
a common pool of particles. Within the cell, the ribosomes presumably
diffuse around, leading to possibly more complex pathways of this
`recycling of ribosomes.'

Of course, ribosome motion along mRNA is only one specific example of
molecular motors. Therefore, the TASEP is also suited for modeling
cargo transport by molecular motors `walking' along cellular
microtubules. In this case, many essential biological features are
also absent from the standard TASEP model. Motors are known to detach
from the microtubule and reattach at other points, so that Langmuir
kinetics \cite{GAST} should be introduced. The cargoes they carry are
typically much larger than their step size, leading us again to
long-ranged particle-particle interactions. There are many lanes on a
microtubule, so that we should include multiple TASEPs in parallel,
with particles transfering from one lattice to another, much like
vehicular traffic on a multilane highway. Molecular motors come in
many varieties (dynein, kinesin, etc.) which move in opposite
directions and at different speeds.  Consequently, a proper model
would consist of several particle species hopping in different
directions. The variety of speeds is the result of complex, multi-step
chemical reactions, so that the dwell times are not necessarily
distributed according to simple exponentials. To account for such
details, particles with internal states can be used necessary.  This
level of complexity is also sufficient for modeling another important
biological process: transport through channels on membranes
(pores). Various ions, atoms and molecules are driven in both
directions, often `squeezing by' each other. Finally, microtubles grow
and shrink, a process modeled by a dynamic $L$, the length of our
lattice. Typically, the associated rates are governed by the densities
of particles at the tip, leading us to an entirely new dimension in
mathematical complexity.


Finally, let us provide an outlook of NESM beyond the topics presented in the sections here. 
In the realm of exactly solvable models, ASEP is just
one of many. Not surprisingly, all but a few are one-dimensional
systems.  For example, we noted in section \ref{BAapplications} the
zero-range process. Also introduced by Spitzer \cite{Spitzer70}, it is
a closely related model for mass transport. Multiple occupancy is
allowed at each site, while some of the particles hop to the next
site. Thus, this process is well suited to describe passengers at bus
stops, with some of them being transported to the next
stop. Especially interesting is the presence of a `condensation
transition,' where a macroscopic number of particles occupy a single
site as the overall particle density on a ring is increased beyond a
critical value. Much progress has emerged, especially in the last two
decades (see \cite{EvansHanney05} for a review). Another notable
example of transport is the ABC model \cite{EKKM98,CDE03}, mentioned
at the end of section \ref{General}. In general, it also displays
transitions of a non-equilibrium nature, admitting long-range order
despite evolving with only short-ranged dynamics.

Apart from transport models, exact results are also known for systems
with no conservation law. A good example is the kinetic Ising chain
\cite{Glauber63}, coupled to two thermal reservoirs at different
temperatures ($T > T^{\prime }$). As a result, it settles down to a
NESS, with generally unknown $P^{*}$. Depending on the details of the
coupling, exact results are nevertheless available. In particular, if
every other spin is updated by a heat-bath dynamics associated with
$T$ and $T^{\prime }$, then \emph{all} multi-spin correlations are
known exactly \cite{RaczZia94,SchmittmannS02}.  Remarkably, even the
full time dependence of these correlations can be obtained exactly, so
that the full $P({\cal C},t)$ can be displayed explicitly as well
\cite{MSZ05}
\footnote{Indeed, due to the simplicity of heat-bath dynamics, some
  exact results are known even if the rates are \emph{time dependent}
  \cite{Aliev09}.}. As a result, it is possible to compute the energy
flow \emph{through} the system (from the hotter bath to the cooler
one) in the NESS, as well as the entropy production associated with
the two baths \cite{RaczZia94}. Similar exact results are available in
a more common form of imposing two baths, namely, joining two
semi-finite chains (coupled to $T$ and $T^{\prime }$) at a single
point. Since the energy flows from the hotter bath, across this
junction, to the cooler side, we may ask for say, (a) the power
injected to the latter and (b) the profile of how this injected energy
is lost to the colder bath. Only the average of the latter is known
\cite{Lavrentovich10}, but the large deviation function of the total
injected power can be computed exactly
\cite{FaragoPitard07,FaragoPitard08}. These are just some examples of
other exactly solvable systems which evolve with detailed balance
violating dynamics.

In the realm of potential applications, exclusion processes extend
well beyond the examples in biological transport presented here. In
the general area of `soft condensed matter,' the exclusion mechanism
arises in many other systems, such as motion of confined colloids
\cite{WEI2000,CHAMP2010}.  Further afield, the process of surface
growth in a particular two-dimensional system can be mapped to an ASEP
\cite{Rost81,Kriecherbauer}.  On larger scales, exclusion processes
lends themselves naturally to modeling traffic flow
\cite{TRAFFIC1,TRAFFIC2,TRAFFIC3} and service queues
\cite{QUEUES}. For each of these applications, though ASEP and its
variants may not be sufficiently `realistic', they nonetheless provide
a succinct physical picture, some insights from mean-field analysis
and a precise language on which sophisticated mathematical techniques
can be applied.  Along with the overall improvement of various
technologies (in e.g., nanoscience, renewable energy), we expect that
there will be many opportunities for exclusion processes to play a
role, both in modeling newly discovered phenomena and in shaping
directions of further research.

Broadening our outlook from exactly solvable models and potential
applications to NESM in general, the vistas are expansive. It is
beyond our scope to provide an exhaustive list of such systems, which
would include problems in aging and branching processes, directed
percolation, dynamic networks, earthquake prediction, epidemics
spreading, granular materials, financial markets, persistence
phenomena, population dynamics, reaction diffusion, self organized
critically, etc. On the purely theoretical front, many fundamental
issues await further exploration. For example, the implications of
probability currents, beyond the computation of physical fluxes, may
be far reaching. If we pursue the analog with electromagnetism, we
could ask if these currents can be linked to a form of `magnetic
fields,' if there is an underlying gauge theory, and if these concepts
are constructive. Perhaps these ideas will lead us to a meaningful
framework for \emph{all} stationary states, characterized by the pair
of distributions $\left\{ P^{*},K^{*}\right\} $, which encompasses the
very successful Boltzmann-Gibbs picture for the equilibrium states. In
particular, in attempting to describe systems which affect, and are
affected by, their environment (through e.g., entropy production) NESS
represents a significant increase of complexity from EQSS. Of course,
the hope is that an overarching theory for NESM, from the full
dynamics to predicting NESS from a given set of rates, will emerge in
the near future. Such a theory should help us reach the ultimate goal
for say, biology -- which would be the ability to predict the form and
behaviour of a living organism, based only on its intrinsic DNA
sequence and the external environment it finds itself. For the latter,
we have in mind both sources of input (e.g., light, air, food,
stimulations) and output (e.g., waste disposal, work,
reproduction). In the absence of such interactions with the
environment, an isolated DNA will evolve to an equilibrium state --
probably just an inert macromolecule. To fully understand the physics
of life, we believe, a firm grasp of non-equilibrium statistical
mechanics is absolutely vital.

\section{Acknowledgments}
We are indebted to many colleagues for enlightening discussions on
many topics. In particular, we wish to thank C. Arita, J. J. Dong,
M. R. Evans, C. V. Finkielstein, O. Golinelli, G. Lakatos, S. Mallick,
S. Prolhac, B.  Schmittmann, U. Seifert, and B. Shargel for their
collaborations and continuing support in our endeavours.  The authors
also thank F. W. Crawford and A. Levine for helpful comments on the
manuscript. One of us (RKPZ) is grateful for the hospitality of
H. W. Diehl in the Universit\"{a}t Duisburg-Essen and H.  Orland at
CEA-Saclay, where some of this work was carried out. This research is
supported in part by the Alexander von Humboldt Foundation, and grants
from the Army Research office through grant 58386MA (TC) and the US
National Science Foundation, DMR-0705152 (RKPZ), DMR-1005417 (RKPZ),
DMS-1032131 (TC), ARO-58386MA (TC), and DMS-1021818 (TC).

\newpage

\bibliography{references4}

\end{document}